\documentclass[a4paper,10pt]{article}
\pdfoutput=1
\usepackage[utf8]{inputenc}
\usepackage{amsmath, amssymb, graphicx,color}

\usepackage[natbibapa]{apacite}

\usepackage[top=3cm, bottom=3cm, left=3cm, right=3cm]{geometry}
\definecolor{webblue}{rgb}{0, 0, 0.5} 
\usepackage[pdfpagelabels,  
plainpages=true,
bookmarks=true,
bookmarksnumbered=true,
hypertexnames=true,
breaklinks=true,%
linkcolor=webblue,]{hyperref}
\hypersetup{
colorlinks  = true,
linkcolor = webblue,
urlcolor = webblue,
citecolor = webblue,
pdfauthor   = {John F. Rudge},
pdfsubject = {},
pdftitle    = {Creep},
pdfkeywords = {},
}

\renewcommand{\d}{\mathrm{d}}

\newcommand {\dd}[2]{\frac{\d #1}{\d #2}}

\newcommand {\pdd}[2]{\frac{\partial #1}{\partial #2}}

\newcommand{\n}{\mathbf{n}}
\newcommand{\x}{\mathbf{x}}
\renewcommand{\t}{\mathbf{t}}
\newcommand{\uuline}[1]{\underline{\underline{#1}}}
\newcommand{\tensor}[1]{\uuline{\boldsymbol{#1}}}

\newcommand{\srate}{\dot{\tensor{e}}}
\newcommand{\ssrate}{\dot{e}}
\newcommand{\stress}{\tensor{\sigma}^s}
\newcommand{\sbars}{\overline{\tensor{\sigma}}^s}

\newcommand{\sbarl}{\overline{\tensor{\sigma}}^l}

\newcommand{\sbar}{\overline{\tensor{\sigma}}}
\newcommand{\ssbar}{\overline{\sigma}}
\newcommand{\I}{\tensor{I}}
\newcommand{\g}{\tensor{\gamma}}
\newcommand{\Pl}{P^l}
\newcommand{\Vc}{V_\text{cell}}

\newcommand{\nablas}{\nabla_\perp}
\newcommand{\logcap}{\kappa(\theta)}

\newcommand{\etashort}{\eta_{0,\text{shorted}}}

\begin{document}
\title{The viscosities of partially molten materials undergoing diffusion 
creep}
\author{John F. Rudge\\
\small{Bullard Laboratories, Department of Earth Sciences, 
University  of 
 Cambridge}}


\maketitle

\begin{abstract}
Partially molten materials resist shearing and compaction. This resistance 
is described by a fourth-rank effective viscosity tensor. When the tensor 
is isotropic, two scalars determine the resistance: an effective shear and 
an effective bulk viscosity. Here, calculations are presented of the 
effective viscosity tensor during diffusion creep for a 2D tiling of 
hexagonal unit cells and a 3D tessellation of tetrakaidecahedrons 
(truncated octahedrons). The geometry of the melt is determined by 
assuming textural equilibrium. The viscosity tensor for the 2D tiling is 
isotropic, but that for the 3D tessellation is anisotropic. Two parameters 
control the effect of melt on the viscosity tensor: the porosity and the 
dihedral angle. Calculations for both Nabarro-Herring (volume diffusion) 
and Coble (surface diffusion) creep are presented. For Nabarro-Herring 
creep the bulk viscosity becomes singular as the porosity vanishes. This 
singularity is logarithmic, a weaker singularity than typically assumed in 
geodynamic 
models. The presence of a small amount of melt (0.1\% porosity) causes the 
effective shear viscosity to approximately halve. For Coble creep, 
previous modelling work has argued that a very small amount of melt may 
lead to a substantial, factor of 5, drop in the shear viscosity. Here, a 
much smaller, factor of 1.4, drop is obtained for tetrakaidecahedrons. 
Owing to a Cauchy relation symmetry, the Coble creep bulk viscosity is a 
constant multiple of the shear viscosity when melt is present.   
\end{abstract}

\section{Introduction}

Dynamical models are often highly dependent on assumptions about the 
rheology of the material being deformed. In many situations in the Earth 
Sciences, this rheology is poorly known, and this is particularly true of 
the polycrystalline rocks in the partially molten regions of the Earth's 
mantle. There are two main approaches to making progress towards a better 
understanding of rheology: One approach is to perform deformation 
experiments on materials in the laboratory, and parametrise the results of 
these experiments into empirical laws (e.g. \citet{Mei2002}). A second 
approach is to produce models of the microscale physics and upscale these 
models to produce rheological laws suitable for use at larger scales 
\citep[e.g.][]{Cooper1984,Takei2009}. This study follows the second 
approach.

One of the remarkable features of the Earth's mantle is its ability to 
flow on long time scales, despite being solid for the most part. This flow 
is only possible because the solid is fairly close to its melting 
temperature, which enables creep by the diffusive transport of matter in 
the solid phase. 
Diffusion creep comes in two main types: Nabarro-Herring creep 
\citep{Nabarro1948,Herring1950}, where diffusion takes place through the 
bodies of grains, and Coble creep \citep{Coble1963}, where diffusion takes 
place along the boundaries between grains. Due to the different activation 
enthalpies and grain-size dependencies of the two diffusion mechanisms, 
Nabarro-Herring creep dominates for large grains and high temperatures; 
Coble creep for small grains and low temperatures.

A pure solid material resists shear, and that resistance to shear is 
characterised by a shear viscosity, which for diffusion creep can be 
calculated from simple mathematical models of the diffusive transport of 
matter \citep{Herring1950,Coble1963,Lifshitz1963}. A partially molten 
material allows an additional mode of deformation, compaction, whereby 
grains are packed closer together and melt is expelled under an isotropic 
stress. The resistance to this kind of deformation can be described in 
terms of an effective bulk viscosity \citep{McKenzie1984}, which can also 
be determined using microscale models of diffusion 
\citep{Arzt1983,Cocks1990,Cocks1996,Takei2009}.

In partially molten materials undergoing diffusion creep, the presence of 
melt causes a reduction in the effective shear viscosity of the material 
compared to the melt-free situation. One reason for this is that diffusion 
in the melt phase is typically much faster than in the solid phase 
\citep{Cooper1984,Takei2009}. Since the rate of diffusion creep is 
dependent on the rate at which material can be transported from one part 
of a grain to another, the presence of fast melt pathways speeds up the 
overall rate of creep for a given stress. How large this effect is depends 
crucially on the geometry of the melt at the grain scale. The simplest 
model of melt geometry at the grain scale is that of textural equilibrium, 
a state which minimises surface energy. Calculations of such melt 
geometries have recently been presented for a tessellation of 
tetrakaidecahedral unit cells \citep{Rudge2018}. The aim of this 
manuscript is to explore the effect of melt on creep viscosities using 
these geometries. 

This work can be seen as a direct extension of the work by 
\citet{Takei2009}. That study presented a detailed account of the effect 
of melt on viscosity during Coble creep, but made a series of 
approximations to the geometry (spherical grains with circular contact 
patches) to allow analytical solutions to be obtained for the relevant 
diffusion problems. Here, the diffusion problems are solved numerically by 
the finite element method using the full geometries. Moreover, in this 
study, both Nabarro-Herring creep and Coble creep are considered. 

The manuscript is organised as follows. The next section presents an 
overview of the governing equations of diffusion creep, both for 
Nabarro-Herring creep and Coble creep. Section \ref{sec:results} gives 
results of the creep calculations for the textural equilibrium geometries. 
To build insight, for each type of diffusion creep the case of a 2D 
hexagonal array of grains is considered first before examining the 3D 
case. Discussion and conclusions sections follow that compare these 
results to those in the wider literature. Detailed mathematical derivations 
are given in the appendices. 

\section{Governing equations}\label{sec:governing}

The basic governing equations of diffusion creep are well-established and 
can be found in many previous studies (e.g. 
\citet{Nabarro1948,Herring1950,Lifshitz1963,Cocks1990,Cocks1996,Takei2009}
). These equations establish the relationship between the macroscale 
stress and strain tensors, which can be written in terms of a fourth-rank 
effective viscosity tensor. The presentation of the equations given below 
largely follows \citet{Lifshitz1963} in expressing all relationships in 
tensor form, and follows \citet{Cocks1990} and \citet{Takei2009} in its 
treatment of the melt phase.

\subsection{Nabarro-Herring (volume diffusion) 
creep} \label{sec:nhsetup}

Let the concentration of vacancies (number of vacancies per unit volume) 
in the 
solid grain be $c$. Fick's law 
describes the motion of vacancies within the grain, with flux
\begin{equation}
 \mathbf{j}_\text{v} = - D_\text{v} \nabla c, \label{eq:fluxdef}
\end{equation}
where $D_\text{v}$ is the diffusivity of vacancies. There is a 
corresponding flux $\mathbf{j} = - \mathbf{j}_\text{v}$ of atoms. 
Conservation of vacancies within the grain ($\nabla \cdot 
\mathbf{j}_\text{v}=0$, assuming 
a 
quasi-static approximation) yields Laplace's equation for the 
concentration of vacancies 
\begin{equation}
 \nabla^2 c = 0. \label{eq:laplace}
\end{equation}
The grain changes shape because of the flux of vacancies to the 
boundaries of the grain, which are assumed to act as perfect sources and 
sinks 
for vacancies. The shape change is described by the plating rate 
$\dot{r}$, 
which quantifies the rate at which a boundary grows, and is given by
\begin{equation}
 \dot{r} = - \Omega \mathbf{j}_\text{v} \cdot \mathbf \n 
\label{eq:plating}
\end{equation}
where $\Omega$ is the atomic volume, and $\n$ is the normal outward to the 
grain. For the regular 
tessellations of identical grains considered here, we will assume grain 
boundaries 
are perpendicular to lines connecting neighbouring grain centres. Addition 
or loss of 
material at the grain boundaries leads to a change of shape of the grain, 
which manifests as a macroscopic strain rate $\srate$. To maintain 
compatability between grains during deformation, the 
assumption will be made that 
the polyhedral grain remains polyhedral (all grain boundaries remain 
flat). Thus the plating rate $\dot{r}$ will be a 
constant on each planar 
face of the 
grain \citep{Cocks1990}. Coordinates are chosen such that the centre of 
mass of the 
grain is the origin, and it will be assumed this point is fixed. 
  The 
plating rate $\dot{r}$ on 
the planar grain boundaries can then be related to the strain rate tensor 
$\srate$
by
\begin{equation}
 \n \cdot \srate \cdot \n = \frac{\dot{r}}{\x \cdot \n}, \label{eq:strain}
\end{equation}
 where $\x \cdot \n$ is the perpendicular distance of the plane from 
the centre of the grain, and $\x$ is the position vector of a point on the 
plane.
Gradients in vacancy concentration are created because the grain is under 
stress, and this changes the equilibrium concentration of vacancies at the 
grain boundaries, which depends on the normal stress on those boundaries. 
Unlike the plating rate, the normal stress varies over each grain 
boundary.
A linearised relationship between the concentration of vacancies and the 
normal stress is given by \citep{Herring1950}
\begin{equation}
 c = c_0 \left( 1 + \frac{\Omega}{k T}  \n \cdot \stress \cdot \n \right) 
\text{ on }S, \label{eq:convac}
\end{equation}
where $S$ is the surface of the grain, $c_0$ is the equilibrium 
concentration of vacancies, $k$ is the Boltzmann constant, and $T$ is 
temperature. For a partially molten system, the surface of the grain can 
be divided into two types: a section of grain--grain contact 
($S_\text{ss}$) 
and a section of grain--liquid contact ($S_\text{sl}$). On areas of 
contact 
with the liquid it will be assumed that
\begin{equation}
\stress \cdot \n = - \Pl \n \text{ on } S_\text{sl}, \label{eq:pressure}
\end{equation}
 where $\Pl$ is the liquid pressure (surface tension across the 
solid--liquid interface is neglected). Implicit in this boundary condition 
is an assumption that all the melt pores are connected and at the same 
pressure. Conservation of momentum inside the 
grain is 
\begin{equation}
 \nabla \cdot \stress = \mathbf{0}. \label{eq:conmom}
\end{equation}
It will be assumed that grains can slide freely at grain-boundaries 
\citep{Lifshitz1963,Raj1971}, and 
thus the corresponding shear stresses are relaxed,
\begin{equation}
 \t \cdot \stress \cdot \n = 0 \text{ on } S_{ss}, \label{eq:slip}
\end{equation}
where $\t$ is a vector tangential to the boundary plane. The main quantity 
of interest is the mean stress inside the grain, defined by
\begin{equation}
 \sbar^s \equiv \frac{1}{V_\text{s}} \int \stress \; \d V,
\end{equation}
where $V_\text{s}$ is the volume of the grain. The ultimate aim of the 
creep 
calculations is to relate this mean stress $\sbar^s$ to the macroscopic 
strain rate $\srate$. This can be done without explicitly solving for the 
stress inside the grain, because the mean stress inside the grain can be 
related to the 
boundary tractions $\stress \cdot \n$ by
\begin{equation}
 \int \stress \; \d V = \int \stress + \left(\nabla \cdot \stress \right) 
\x  
\; \d V = \int \nabla \cdot \left(\stress \x \right) \; \d V  = \int \x \, 
\stress \cdot \n \; \d S
\end{equation}
where the first equality exploits the conservation of 
momentum \eqref{eq:conmom}, and the final equality exploits the divergence 
theorem and the symmetry of the stress tensor. Furthermore, 
because the boundaries are assumed to be free-slipping \eqref{eq:slip}, 
the 
boundary tractions are purely normal, and can be written $\stress \cdot \n 
= (\n \cdot \stress \cdot \n) \n$. Hence,
\begin{equation}
 \sbar^s = 
\frac{1}{V_\text{s}} 
\int (\n \cdot \stress \cdot \n)\, \x \, \n \; \d S \label{eq:sbar2}
\end{equation}
 In turn, $\n \cdot \stress \cdot 
\n$ can be related directly to the concentration of vacancies through 
\eqref{eq:convac}.

\subsubsection{Scaling}

The equations above can be simplified by scaling to produce dimensionless 
governing equations. All lengths can be scaled on a characteristic length 
scale $d$ (a measure of grain size). It is useful to introduce the 
self-diffusion coefficient $D = D_v c_0 \Omega$, and scale all times with 
the diffusive timescale $\tau = d^2 / D$. A natural scale for stresses 
(and thus pressure) is given from \eqref{eq:convac} as $P_0 = k T / 
\Omega$. It 
follows that a natural scale for viscosities is $P_0 \tau = k T d^2 / D 
\Omega$, 
which is the classic scaling of Nabarro-Herring creep, with viscosities 
proportional to the square of grain size. For the concentration of 
vacancies 
it 
is useful to introduce a scaled variable as
\begin{equation}
 c^\prime = \frac{c}{c_0} - 1 + P^{l\prime}
\end{equation}
where $P^{l\prime}= \Pl/P_0$ is the scaled liquid pressure. The 
dimensionless governing equations are then
\begin{gather}
 \nabla^2 c = 0 \text{ in }V_\text{s}, \label{eq:nh1} \\
c = 0 \text{ on } S_\text{sl}, \label{eq:nh2} \\
\pdd{c}{n} = \left(\n \cdot \srate \cdot \n \right) \x \cdot \n 
\text{ on } S_\text{ss}, \label{eq:nh3}\\
\sbars = - \Pl \I + \frac{1}{V_\text{s}} \int c \,\x\, \n \; \d 
S. \label{eq:nh4}
\end{gather}
where primes have been dropped on the dimensionless variables for ease of 
reading. The Laplace's equation of \eqref{eq:nh1} follows directly from 
\eqref{eq:laplace}. The Dirichlet boundary condition on the 
solid--liquid interfaces \eqref{eq:nh2} follows from combining 
\eqref{eq:convac} and \eqref{eq:pressure}. The Neumann 
boundary condition \eqref{eq:nh3} results from combining 
\eqref{eq:fluxdef}, \eqref{eq:plating}, and \eqref{eq:strain}. The mean 
stress expression \eqref{eq:nh4} results from \eqref{eq:sbar2} and 
\eqref{eq:convac}, and $\I$ is the identity tensor. 
Given a desired macroscopic strain rate $\srate$, \eqref{eq:nh1}, 
\eqref{eq:nh2}, and \eqref{eq:nh3} can be solved to determine $c$, which 
can then be substituted into \eqref{eq:nh4} to determine the mean stress 
that 
needs to be applied to produce that strain rate.

Another way of writing \eqref{eq:nh4} is in terms of the total stress 
tensor,
\begin{equation}
 \sbar = \phi \sbarl + (1-\phi) \sbars,
\end{equation}
where $\phi$ is the porosity (volume fraction of melt). $\phi \equiv 
V_\text{l} 
/ V_\text{cell}$ where $V_\text{l}$ is the volume of liquid melt and 
$V_\text{cell}$ is the volume of a unit cell (when considering a 
tessellation of unit cells). In the calculations that follow it will be 
 assumed that as porosity varies, the volume of the cell remains fixed. 
\eqref{eq:nh4} can be written in terms of total stress as
\begin{equation}
 \sbar = - \Pl \I + \frac{1}{V_\text{cell}} \int c \,\x\, \n \; \d 
S. 
\end{equation}
Since \eqref{eq:nh1}, \eqref{eq:nh2}, \eqref{eq:nh3} are linear equations, 
the vacancy concentration 
$c$ is linearly related to the strain rate tensor $\srate$. As a 
consequence, the results of the calculations can be written in suffix 
notation as
\begin{equation}
\ssbar_{ij} = -\Pl \delta_{ij} + C_{ijkl} \ssrate_{kl}
\end{equation}
where $\delta_{ij}$ is the Kronecker delta and $C_{ijkl}$ is the 
viscosity tensor. It is possible to write a single statement of the 
problem for this viscosity tensor. The linear relationship between $c$ and 
$\srate$ can be written as $c = \gamma_{ij} \ssrate_{ij}$ where 
$\gamma_{ij}$ is a symmetric second rank tensor that satisfies
\begin{gather}
 \nabla^2 \gamma_{ij} = 0 \text{ in }V_s, \label{eq:g1} \\
\gamma_{ij} = 0 \text{ on } S_\text{sl}, \label{eq:g2} \\
\pdd{\gamma_{ij}}{n} = x_p n_p n_i n_j 
\text{ on } S_\text{ss}, \label{eq:g3}\\
C_{ijkl} = \frac{1}{\Vc} \int \gamma_{kl}  x_i n _j\; \d 
S. \label{eq:g4}
\end{gather}

%
%

\subsection{Coble (grain-boundary diffusion) creep}

The governing equations for Coble creep are very similar to those outlined 
above for Nabarro-Herring creep. The only differences that arise from 
those in section \ref{sec:nhsetup} are in changes 
to equations \eqref{eq:fluxdef}, \eqref{eq:laplace}, and 
\eqref{eq:plating}. 
In Coble creep, diffusion only transports matter on the boundaries of the 
grains. The corresponding Fick's law becomes
\begin{equation}
 \mathbf{j}_\text{v} = - D^\text{gb}_\text{v} \nablas c, 
\label{eq:fluxdef_c}
\end{equation}
where $D^\text{gb}_\text{v}$ is the diffusivity of vacancies 
along the grain boundary, and $\nabla_\perp$ is the perpendicular gradient 
operator, defined by
\begin{equation}
 \nablas \equiv (\tensor{I} - \n \n) \cdot \nabla,
\end{equation}
where $\n$ is the outward normal to the grain. Conservation of mass 
relates the divergence of the flux to the plating 
rate
\begin{equation}
\tfrac{1}{2} \Omega \delta \nablas \cdot \mathbf{j}_\text{v} = 
\dot{r}, 
\label{eq:coblemass}
\end{equation}
where $\delta$ is the grain boundary thickness. The factor of 1/2 arises 
because each grain boundary borders two grains. Note that some authors, 
such as 
\citet{Takei2009}, use the symbol $\delta$ to denote the grain-boundary 
half-width. Here $\delta$ denotes the full-width, as used by 
\cite{Cocks1990} 
and \citet{Raj1971}. As before, the plating rate $\dot{r}$ is constant 
on each planar face of the grain. Combining 
\eqref{eq:coblemass} and \eqref{eq:fluxdef_c} leads to a Poisson 
equation for the concentration of vacancies
\begin{equation}
- \tfrac{1}{2}  \Omega \delta D^\text{gb}_\text{v} \nablas^2 c = 
\dot{r} \text{ on } S_\text{ss}  \label{eq:coble_main}.
\end{equation}

An additional subtlety arises in Coble creep, in that boundary conditions 
also need to be specified on the grain edges. When melt is present on the 
grain edges, this condition is simply that $c$ is constant on the edges. 
When melt is not present, $c$ must be continuous across the grain edges, 
and flux must be conserved, which implies that
\begin{equation}
 \sum_m \boldsymbol{\nu}_m \cdot \nablas c = 0   \text{ on }\Gamma,
\end{equation}
where $\Gamma$ represents the contact line where grain boundaries meet, 
$m$ is 
an index identifying each grain boundary, and $\boldsymbol{\nu}_m$ is the 
outward-pointing co-normal to each surface at the contact line (i.e. 
$\boldsymbol{\nu}_m$ is perpendicular to both the normal to the surface 
and a 
vector in the direction of the contact line).

\subsubsection{Scaling}

As for Nabarro-Herring creep, the governing equations can be simplified by 
scaling. The scaling for length and stresses are the same, but 
\eqref{eq:coble_main} motivates a slightly different choice of timescale, 
$\tau = d^3 / \delta D^\text{gb}$ where  $D^\text{gb} = 
D^\text{gb}_\text{v} c_0 
\Omega$ is the self-diffusion coefficient for grain-boundary diffusion. 
The 
natural viscosity scale is then $k T d^3 / \delta D^\text{gb} \Omega$, 
proportional to the third power of grain-size. The dimensionless equations 
are
\begin{gather}
-  \nablas^2 c = 2 \left(\n \cdot \srate \cdot \n \right) \x \cdot \n 
\text{ 
on }S_\text{ss}, \label{eq:cb1}\\
c = 0\text{ on }S_\text{sl}, \label{eq:cb2}\\
\sbar = - \Pl \I + \frac{1}{V_\text{cell}} \int c \,\x\, \n \; \d 
S. \label{eq:cb3}
\end{gather}
As for Nabarro-Herring creep, we may write $c = \gamma_{ij} \ssrate_{ij}$ 
to get a single statement of the problem for the viscosity tensor as
\begin{gather}
 -\nablas^2 \gamma_{ij} = 2 x_p n_p n_i n_j \text{ on }S_\text{ss}, 
\label{eq:gc1} \\
\gamma_{ij} = 0 \text{ on }S_\text{sl}, \\
\sum_m \boldsymbol{\nu}_m \cdot \nablas \gamma_{ij} = 0 \text{ on 
}\Gamma,\\
C_{ijkl} = \frac{1}{\Vc} \int \gamma_{kl}  x_i n _j\; \d 
S. \label{eq:gc2}
\end{gather}

\section{Results}\label{sec:results}

\subsection{Nabarro-Herring (volume diffusion) creep}

\subsubsection{2D: Tiling of hexagons}

The simplest and most natural two-dimensional geometry one can consider is 
a tiling of hexagonal grains. The hexagonal symmetry demands that the 
fourth-rank viscosity tensor is isotropic, and hence can be described by 
just two numbers: the shear viscosity $\eta$ and the bulk viscosity 
$\zeta$. 

In the absence of melt, only shear is possible, with shear viscosity given 
in dimensional form as
\begin{equation}
 \eta_0 = \frac{k T d^2}{36 D \Omega}, \label{eq:nh_hex_eta0}
\end{equation}
where $d$ is the perpendicular distance between opposite sides of the 
hexagon, and the subscript 0 is used to denote the absence of melt. The 
dimensional factors in \eqref{eq:nh_hex_eta0} are obtained by scaling, and 
the numerical prefactor of 1/36 results from a calculation. Remarkably, 
this numerical prefactor can be obtained from a simple geometrical 
integral, without the need to explicitly solve Laplace's equation (see 
appendix \ref{sec:nhpure} for the derivation). \eqref{eq:nh_hex_eta0} is 
identical to the formula attributed to Gibbs by \citet{Raj1971}. A 
slightly 
different numerical prefactor for hexagonal grains of 1/30.2 was obtained 
by \citet{Beere1976}. The 1/36 factor given here is exact, and 
\citet{Beere1976}'s results only differ due to approximations he made to 
the geometry in order to get an analytic expression for the solution of 
Laplace's equation. The 1/36 prefactor is not that different from that 
for a circle of diameter $d$, which is 1/32.

\begin{figure}
\centering
 \includegraphics[width=\columnwidth]{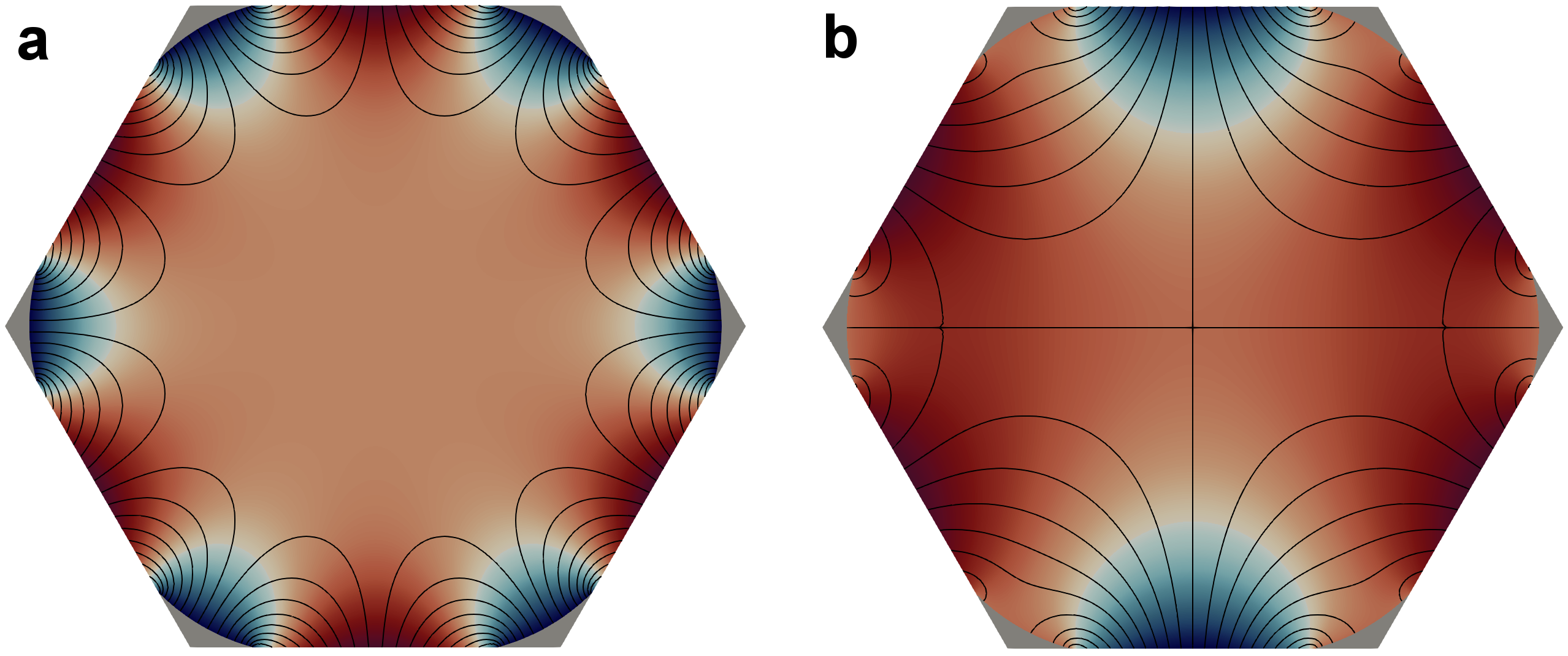}
\caption{Illustration of the hexagonal unit cell undergoing a) isotropic 
bulk 
deformation and b) pure shear. Regions of melt are shown in grey. The 
colour scale inside the 
solid grain is used to show concentration of vacancies. Black lines show 
the paths of vacancy flux. Under bulk compression material is transported 
from the grain--grain boundaries to the melt pores, resulting in shrinking 
of the melt pores and overall compaction.}
\label{fig:hex_grain}
\end{figure}

Our principal interest here is the effect of melt on the viscosity, which 
requires assumptions about how the melt is organised at the grain scale. 
For the hexagonal tiling we assume that melt lies at the vertices of the 
hexagons, and takes the form of a texturally equilibrated geometry 
(minimum 
surface energy). This implies that the solid--melt interfaces are simply 
arcs of circles, which meet the solid--solid (grain--grain) contacts at 
the 
dihedral angle (\citet{German2009}). There are thus two parameters that 
describe the effect of melt: the porosity (volume fraction of melt) $\phi$ 
and the dihedral angle $\theta$.

\begin{figure}
 \includegraphics[width=\columnwidth]{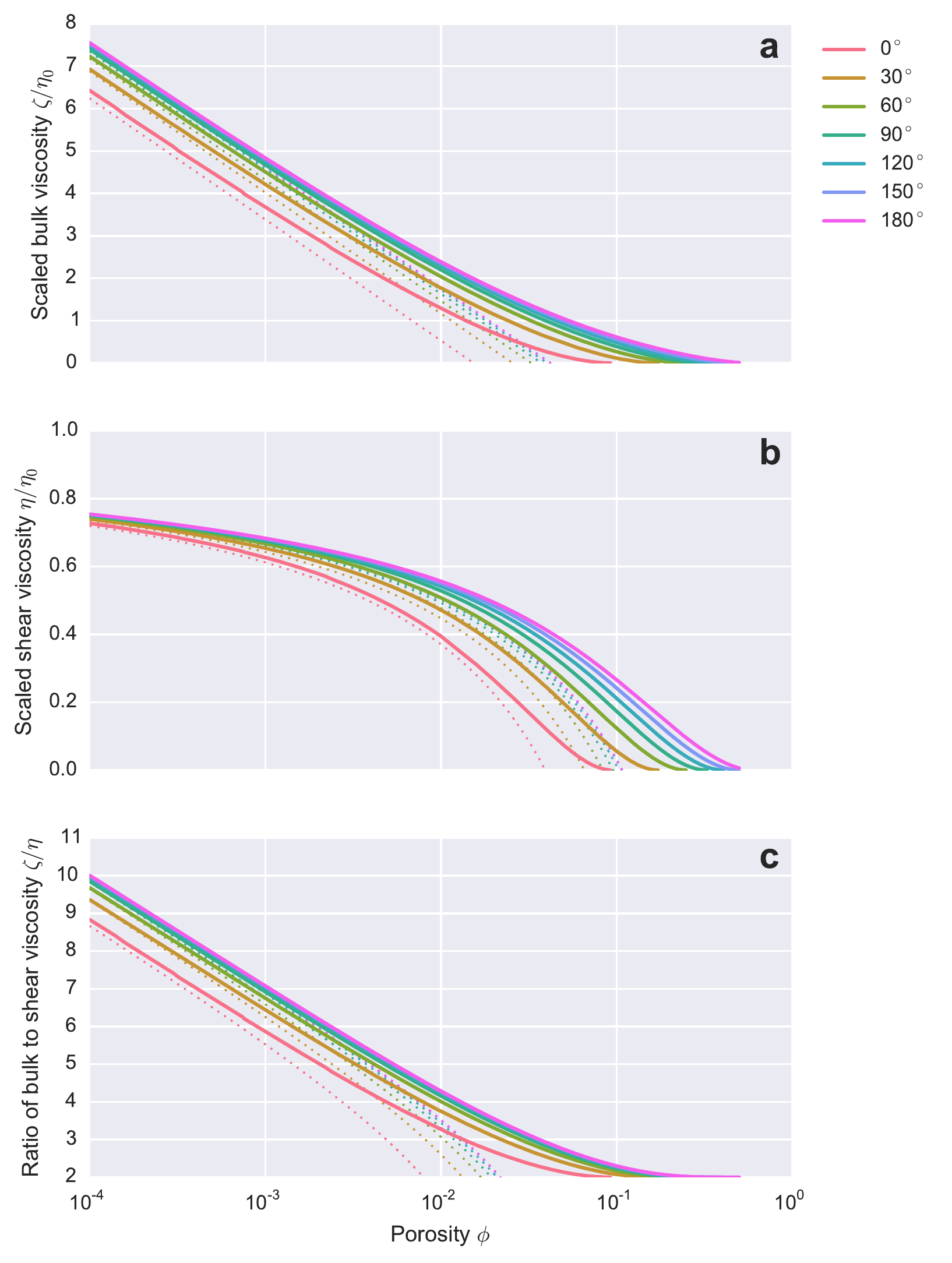}
\caption{Bulk and shear viscosities for Nabarro-Herring creep of hexagonal 
grains with melt. Solid lines show the finite element calculations for 
different dihedral angles as indicated in the legend. Dotted lines show 
the 
asymptotic results for small porosity. Porosity is shown on a logarithmic 
scale on the horizontal axis. On the vertical axes are shown a) scaled 
bulk 
viscosity $\zeta/\eta_0$, b) scaled shear viscosity $\eta/\eta_0$, and c) 
the ratio of bulk to shear viscosity $\zeta/\eta$.}
\label{fig:nh_hex_summary}
\end{figure}

The geometry of the hexagonal unit cell with melt present is illustrated 
in 
Figure \ref{fig:hex_grain}. The governing equations in 
\eqref{eq:g1}-\eqref{eq:g4} have been solved numerically using the FEniCS 
software \citep{Logg2010,Logg2012} to obtain the bulk and shear 
viscosities, presented in Figure \ref{fig:nh_hex_summary}. Such finite 
element calculations have been performed previously for circular pores by 
\citet{Cocks1996}, and the calculations here extend Cocks's analysis to a 
wider range of pore shapes.

\begin{figure}
\includegraphics[width=\columnwidth]{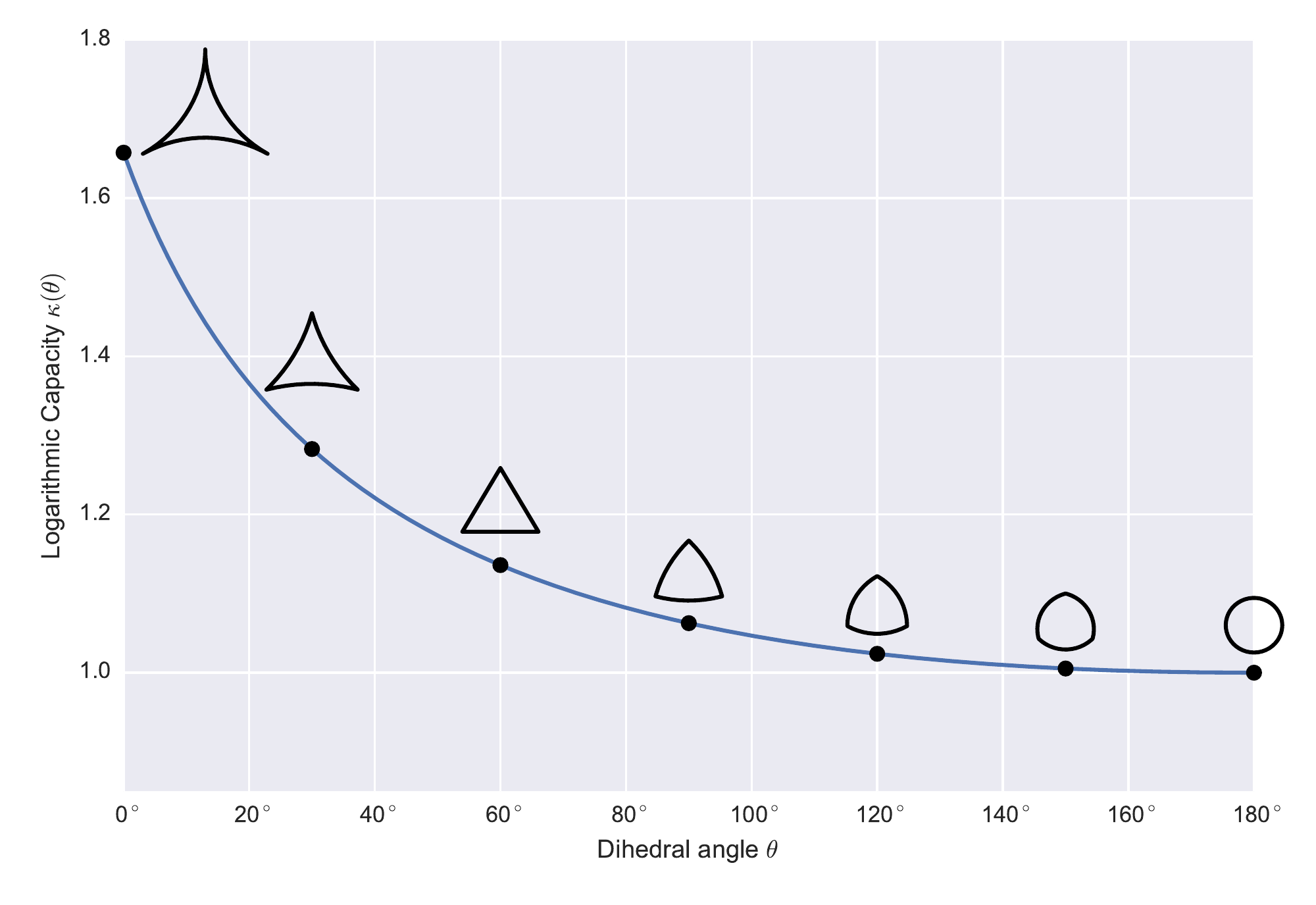}
\caption{Logarithmic capacity $\logcap$ as a function of dihedral angle 
$\theta$. 
$\logcap$ varies from 1.6574 at zero dihedral angle to 1.0 at $180^\circ$. 
Also 
shown are the pore shapes for $\theta = 0^\circ$, $30^\circ$, $60^\circ$ 
(an equilateral triangle), $90^\circ$, $120^\circ$, $150^\circ$, and 
$180^\circ$ (a circle). The scaling is such that each pore has the same 
area, equal to $\pi$.}
\label{fig:logcap}
\end{figure}

Figure \ref{fig:nh_hex_summary}a shows the bulk viscosity (scaled by the 
reference shear viscosity $\eta_0$ in the absence of melt), which becomes 
singular as the porosity vanishes. As apparent from the plot, this 
singularity is logarithmic in the porosity. This singularity arises 
because very small melt pores act like point sources/sinks of vacancies at 
the vertices of the hexagon. In two-dimensions, point-source solutions of 
Laplace's equations are proportional to the logarithm of distance away 
from the source. Thus at very small porosities, the vacancy concentration 
grows logarithmically approaching the vertices of the hexagon; this 
manifests in a logarithmic singularity in the bulk viscosity.  The 
behaviour at 
small porosity can be obtained formally by matched asymptotics (appendix 
\ref{sec:nh_bulk_asm}) as 
\begin{equation}
 \frac{\zeta}{\eta_0} \sim - \frac{9 \sqrt{3}}{4 \pi} \log \phi - 3.92922 
- 
\frac{9 \sqrt{3}}{2 \pi} \log \logcap,
\end{equation}
which is plotted as the dotted lines in Figure \ref{fig:nh_hex_summary}. 
The 
leading order term, proportional to $\log \phi$, is independent of the 
shape of melt pores. Hence all curves in Figure \ref{fig:nh_hex_summary}a 
approach the same slope as porosity tends towards zero. The melt pore 
shape 
affects the next order, constant term, which is expressed in terms of the 
logarithmic capacity $\logcap$ of the pore shape, which in turn depends on 
the dihedral angle $\theta$. Figure \ref{fig:logcap} 
plots the 
logarithmic capacity $\logcap$ calculated using the 
\texttt{capacity} routine of the \textit{Schwarz-Christoffel Toolbox} 
\citep{Driscoll2002,Driscoll2002a}. For a fixed porosity, the bulk 
viscosity 
increases with increasing dihedral angle. The full numerical solutions 
start to depart significantly from the asymptotic solutions once $\phi 
\gtrsim 
0.3\%$. The bulk viscosity vanishes once all the grain boundaries have 
been 
wetted, which happens for larger porosities at larger dihedral angles.

The shear viscosity (again scaled by the reference shear viscosity 
$\eta_0$ 
in the absence of melt) is plotted in Figure \ref{fig:nh_hex_summary}b. 
The 
shear viscosity is not singular as the porosity vanishes, and tends to 
$\eta_0$. However, just a small amount of melt can cause a notable drop in 
the shear viscosity -- for a 0.1\% porosity, the shear viscosity is around 
63 to 68\% of the melt-free value, depending on the dihedral angle. The 
approach of the shear viscosity to $\eta_0$ as porosity vanishes is 
proportional to the reciprocal of the 
logarithm of $\phi$. Matched asymptotics (appendix \ref{sec:nh_shear_asm}) 
yields the 
small-porosity expression, 
\begin{equation}
 \frac{\eta}{\eta_0} \sim 1 + \frac{2 \pi F}{\sqrt{3} \left( \log 
(\epsilon 
\logcap) + 2 \pi R \right)} \label{eq:shearasym_main}
\end{equation}
where $F = 0.319889078$ and $R=0.150237305$ are numerical constants, and 
$\epsilon$ is related to $\phi$ by
\begin{equation}
 \epsilon = \left(\frac{\sqrt{3} \phi}{4 \pi} \right)^{1/2}. 
\label{eq:epstopor_maintext}
\end{equation}
Similar to the bulk viscosity, at higher porosities the shear viscosity
vanishes on wetting of the grain boundaries. Figure 
\ref{fig:nh_hex_summary}c shows the ratio $\zeta/\eta$ of bulk to shear 
viscosity, which also demonstrates a logarithmic singularity at small 
porosities. At larger porosities the ratio $\zeta/\eta$ tends to 2.

%

\subsubsection{3D: Tessellation of tetrakaidecahedrons}

A natural generalisation of the tiling of hexagonal grains in 2D is a 
tessellation of tetrakaidecahedrons (truncated octahedrons) in 3D. The 
tetrakaidecahedron has 14 faces, 6 of which are squares, 8 of which are 
hexagons. In the tessellation, three grains meet along grain edges, and 
four 
grains meet at the vertices. The tetrakaidecahedron has cubic symmetry. 
Unlike in the 2D case of hexagons, this symmetry does not lead to isotropy 
of the 
fourth-rank viscosity tensor. However, it does reduce the fourth rank 
tensor to three independent components, which can expressed in Voigt 
notation as
\begin{equation}
\left(\begin{matrix}
      C_{1111} & C_{1122} & 
C_{1133} & C_{1123} & C_{1113} 
& C_{1112}\\
C_{2211}       & C_{2222} & 
C_{2233} & C_{2223}
& C_{2213} & C_{2212}\\
C_{3311}     & C_{3322}& C_{3333} & C_{3323} & C_{3313} 
& C_{3312}\\
C_{2311}  & C_{2322}  & C_{2333} & C_{2323} & C_{2313} & C_{2312}\\
C_{1311} & C_{1322} & C_{1333} & C_{1323} & C_{1313} & C_{1312}\\
C_{1211} & C_{1222} & C_{1233} & C_{1223} & C_{1213} & C_{1212}\\
     \end{matrix}\right) =  \left(\begin{matrix}
      \zeta + \tfrac{4}{3} \eta_1 & \zeta-\tfrac{2}{3} \eta_1 & 
\zeta-\tfrac{2}{3} \eta_1 & 0 & 0 
& 0\\
      \zeta-\tfrac{2}{3} \eta_1 & \zeta + \tfrac{4}{3} \eta_1 & 
\zeta-\tfrac{2}{3} \eta_1 & 0 
& 0 & 0\\
      \zeta-\tfrac{2}{3} \eta_1 & \zeta-\tfrac{2}{3} \eta_1 & \zeta 
+ \tfrac{4}{3} \eta_1 & 0 & 0 
& 0\\
      0 & 0 & 0 & \eta_2 & 0 & 0\\
      0 & 0 & 0 & 0 & \eta_2 & 0\\
      0 & 0 & 0 & 0 & 0 & \eta_2\\
     \end{matrix}\right), \label{eq:Voigt}\end{equation}
for a bulk viscosity $\zeta$ and two shear viscosities $\eta_1$ and 
$\eta_2$. Here we will define the co-ordinates of the tetrakaidecahedron 
such that the $x$, $y$, and $z$ axes are aligned with the normals to the 
square faces. 

In the absence of melt, only shear deformation is permitted, and the two 
shear viscosities can be calculated by a geometric integral (appendix 
\ref{sec:nhpure}) over the grain as 
\begin{equation}
 \eta_{10} = \frac{23}{1536} \frac{k T d^2}{D \Omega}, \quad \quad 
\eta_{20} = \frac{31}{1536} \frac{k T d^2}{D \Omega}, 
\label{eq:nh_tet_visc}
\end{equation}
where $d$ is the distance between opposite square faces of the 
tetrakaidecahedron, and zeros are again used to indicate the absence of 
melt. The anisotropy is apparent, and can expressed by the Zener ratio 
$A\equiv\eta_2/\eta_1 = 31/23 \approx 1.35$. Shear is easiest when the 
principal axes of shear are aligned with the square faces. The 
Voigt-average over all  of the viscosities in 
\eqref{eq:nh_tet_visc} is
\begin{equation}
 \overline{\eta}_0 \equiv\frac{2}{5} \eta_{10} + \frac{3}{5} \eta_{20} = 
\frac{139}{7680} \frac{k T d^2}{D \Omega},
\end{equation}
and will be used as a reference quantity in what follows. The 
Voigt-average is 
obtained by averaging the viscosity tensor $C_{ijkl}$ over all possible 
orientations of the grain. The numerical 
prefactor $139/7680 \approx 0.0181$ is reasonably similar to the numerical 
prefactor for a sphere of diameter $d$, which is $1/40 = 0.025$ 
\citep{Herring1950}. The prefactor $139/7680$ is also similar to another 
calculation for the tetrakaidecahedron by \citet{Shah1998}, which yielded 
a 
prefactor of $1/54 = 0.0185$, but that study did not consider the 
inherent anisotropy of the shape. 

\begin{figure}
\centering
\includegraphics[width=0.8\columnwidth]{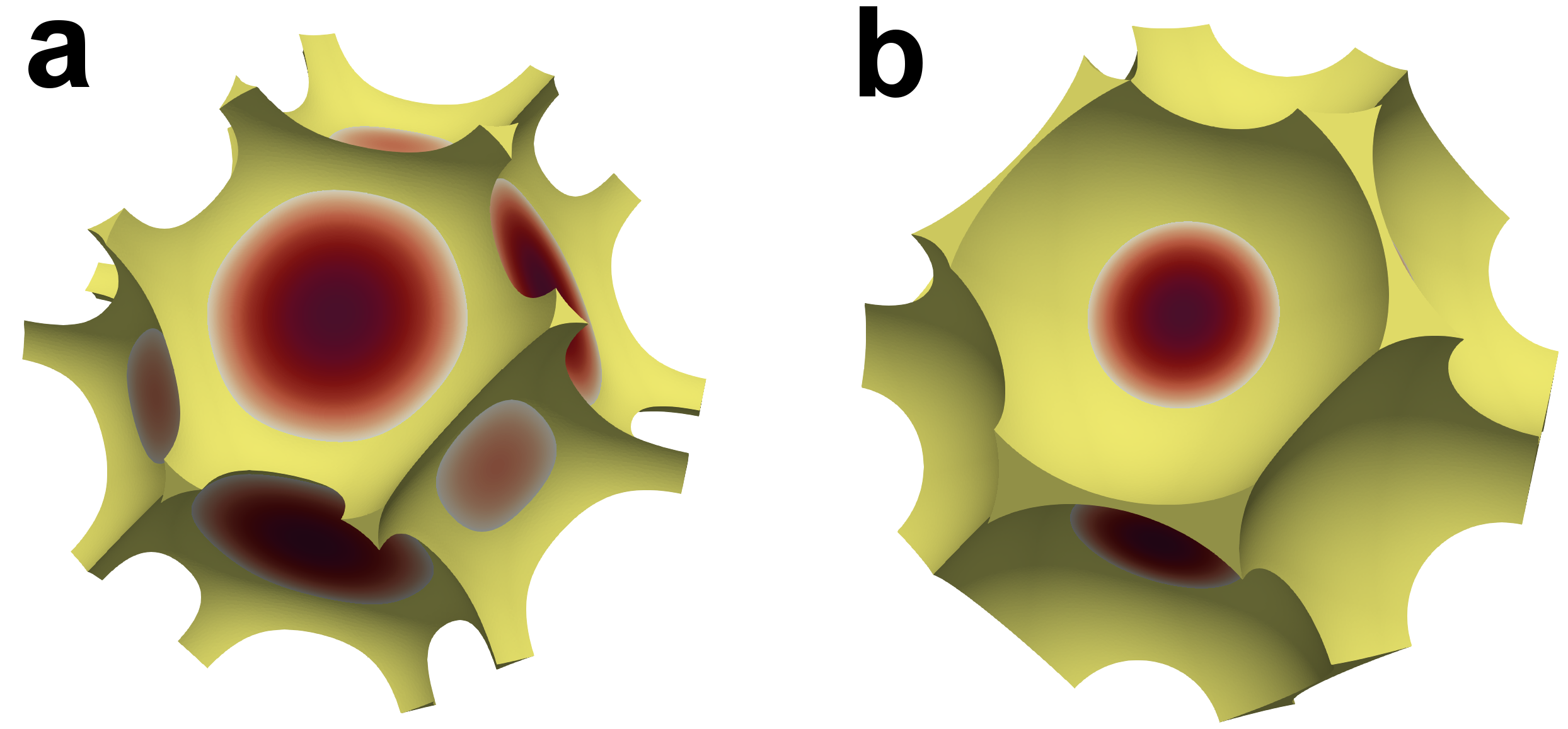}
\caption{Examples of two melt topologies around tetrakaidecahedral grains. 
Both examples have a dihedral angle of 30$^\circ$ \citep{Rudge2018}. a) 
shows the connected 
or ``c'' topology for a porosity of 0.03. b) shows the square-wetted or 
``s'' topology for a porosity of 0.12. The melt network is shown in 
yellow, 
and the grain faces are in a colour scale showing vacancy concentration 
during bulk deformation for Coble creep.}
\label{fig:melt_topo}
\end{figure}

\begin{figure}
 \includegraphics[width=\columnwidth]{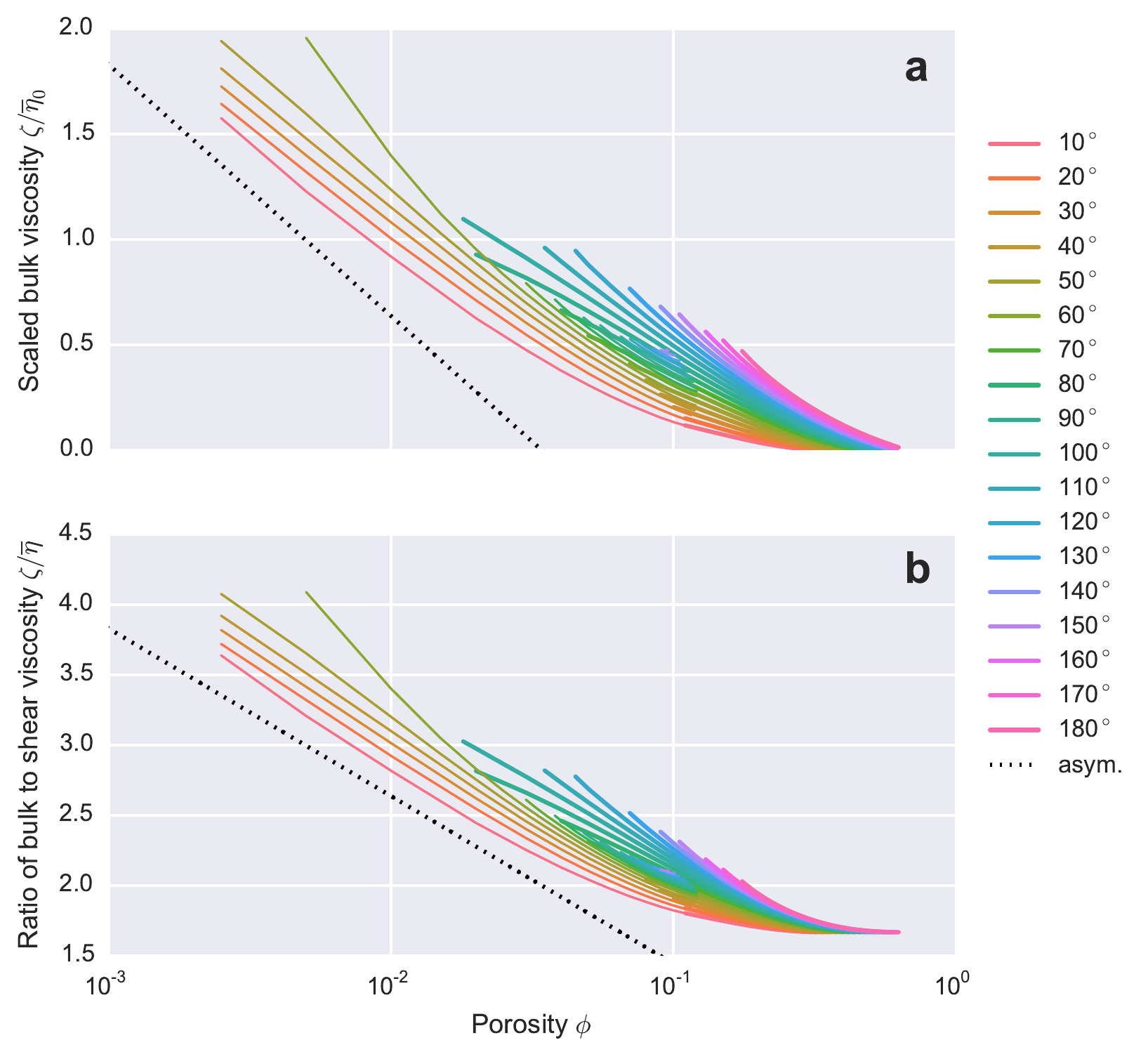}
\caption{a) Scaled bulk viscosity and b) scaled ratio of bulk to shear as 
a 
function of porosity for a tessellation of tetrakaidecahedral unit cells 
undergoing Nabarro-Herring creep. 
Thin lines show ``c'' topologies (connected along the grain edges, Figure 
\ref{fig:melt_topo}a), thick 
lines show ``s'' topologies (where the square faces are wetted, Figure 
\ref{fig:melt_topo}b). The dashed 
black line shows the expected slope from the asymptotic analysis for small 
porosities (the intercept of the line has been chosen arbitrarily).}
\label{fig:herring_bulk}
\end{figure}

\begin{figure}
 \includegraphics[width=\columnwidth]{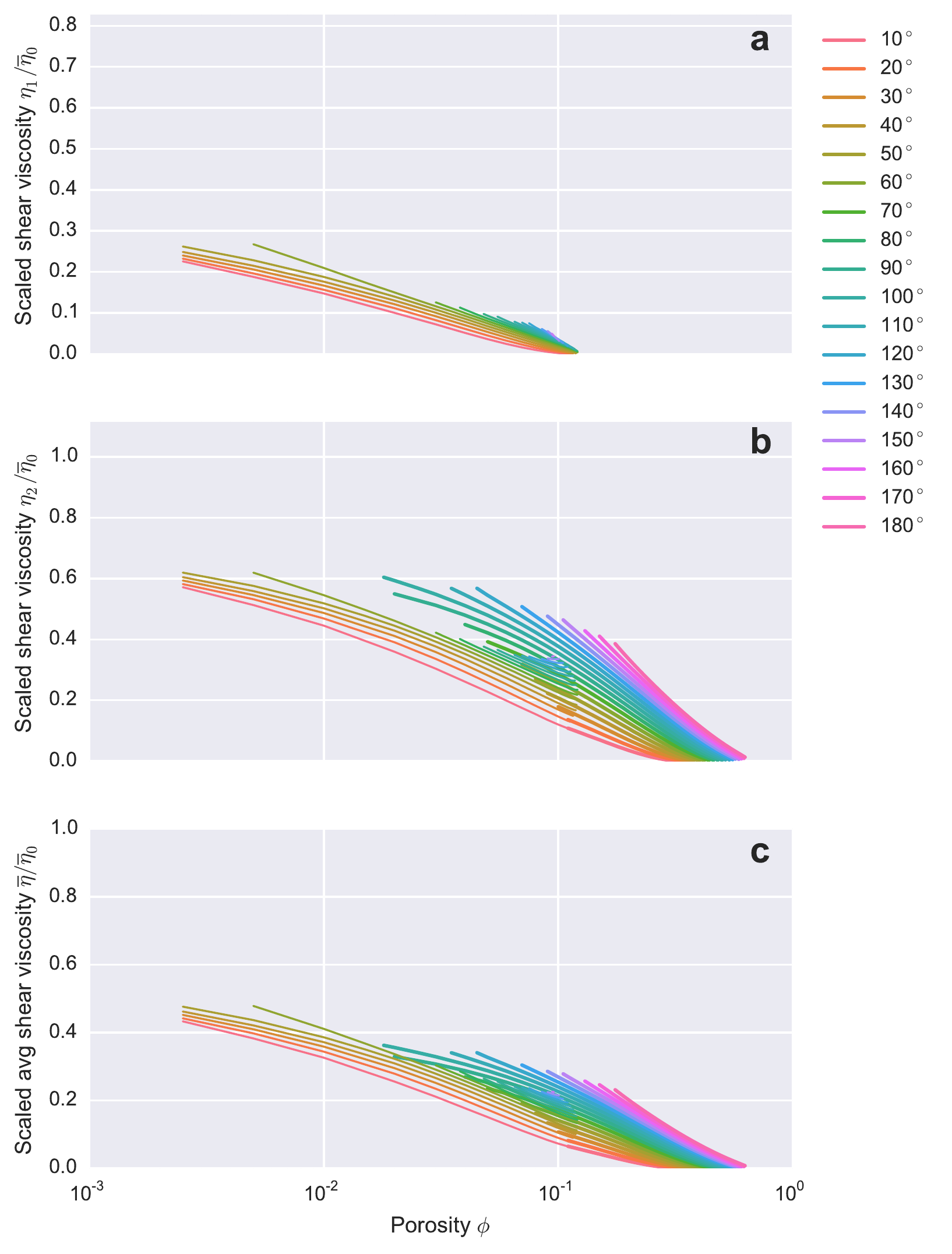}
\caption{Scaled shear viscosities as a function of porosity for a 
tessellation of tetrakaidecahedral unit cells undergoing Nabarro-Herring 
creep. a) 
$\eta_1/\overline{\eta}_0$, b) $\eta_2/\overline{\eta}_0$, and c) 
Voigt-average $\overline{\eta}/\overline{\eta}_0$. Lines styles are the 
same as Figure \ref{fig:herring_bulk} corresponding to the different 
topologies of the melt network. }
\label{fig:herring_shear}
\end{figure}

In three-dimensions, determining the textural equilibrium melt geometries 
is 
rather more involved than in 2D. Such calculations have recently been 
presented 
in \citet{Rudge2018}, and examples of these melt geometries are shown in 
Figure \ref{fig:melt_topo}. Figures \ref{fig:herring_bulk} and 
\ref{fig:herring_shear} plot the bulk and shear viscosities for a 
tessellation of tetrakaidecahedral unit cells with melt present, using the 
calculated geometries of \citet{Rudge2018}. The viscosities are again a 
function of two parameters: the porosity $\phi$ and the dihedral angle 
$\theta$. As discussed in \citet{Rudge2018}, there are different 
topologies 
of the melt network depending on these two parameters. When the dihedral 
angle is less than $60^\circ$ and the porosity small, melt lies along the 
triple lines where three grains meet, and forms a connected network. For 
small porosities, and dihedral angles greater than $60^\circ$, melt 
resides 
in isolated pockets and does not form a connected network. The effect of 
melt on 
viscosity has only been calculated here for the cases where the melt forms 
a 
connected 
network (so that it is possible to expel melt by compaction). 

The topology that has melt connected along the grain edges (termed ``c'' 
for connected) is shown as thin lines on Figures \ref{fig:herring_bulk} 
and 
\ref{fig:herring_shear}. At higher porosities, melt starts to wet the 
square faces of the tetrakaidecahedrons, forming a different melt topology 
(termed ``s'' for square-wetted). This topology is shown as the thicker 
lines on Figures \ref{fig:herring_bulk} and \ref{fig:herring_shear}. 

For dihedral angles less than $60^\circ$, the ``c'' topologies have a 
logarithmic singularity in the bulk viscosity. The reason for this 
logarithmic singularity is similar to the 2D case, with melt tubules 
acting as line sources/sinks of vacancies along grain edges when the 
porosity is small. An asymptotic 
analysis (appendix \ref{sec:leadingorder}) yields the small porosity 
behaviour
\begin{equation}
 \frac{\zeta}{\overline{\eta}_0} \sim - \frac{160 \sqrt{2}}{139 \pi} \log 
\phi, \label{eq:leadingorderbulk}
\end{equation}
which is plotted as the black dotted line in Figure 
\ref{fig:herring_bulk}. 
The next term in the asymptotic analysis is a constant term, and depends 
on the 
details of the melt geometry. This geometry is nontrivial to determine in 
3D for an infinitesimal amount of melt. Nevertheless, the leading order 
term 
in \eqref{eq:leadingorderbulk} seems to provide a good match to the slope 
of the numerical results at low porosity. The slope in 3D (-0.518) is 
rather less than the slope in 2D (-1.240). More broadly, the predicted 
variations in the ratio of bulk to shear viscosity are relatively modest, 
unless 
the porosity becomes extremely small: $\zeta/\eta$ varies from around 4.0 
at a porosity of $\phi \sim 10^{-3}$ to a limiting value of $5/3 \approx 
1.67$ at higher porosities. 

Figure \ref{fig:herring_shear} plot the shear viscosities, both the 
individual 
$\eta_1$ and $\eta_2$, along with the Voigt-average 
$\overline{\eta}=\tfrac{2}{5} \eta_1 + \tfrac{3}{5} \eta_2$. One notable 
feature of this plot is the vanishing of $\eta_1$ when wetting of the 
square faces occurs. The presence of melt along the square faces allows 
easy shearing in pure shear with principal 
axes aligned with the normals to the square faces. The general trend of 
shear 
viscosity 
against porosity for small porosity is similar to that seen in 2D. 
However, there is a greater overall reduction in the shear viscosity with 
porosity. For example, even a small porosity of $0.1\%$ is expected to 
have 
a Voigt-average shear viscosity approximately half that of the melt-free 
value.


\subsection{Coble (grain-boundary diffusion) creep}

\subsubsection{2D: Tiling of hexagons}

\begin{figure}
 \includegraphics[width=\columnwidth]{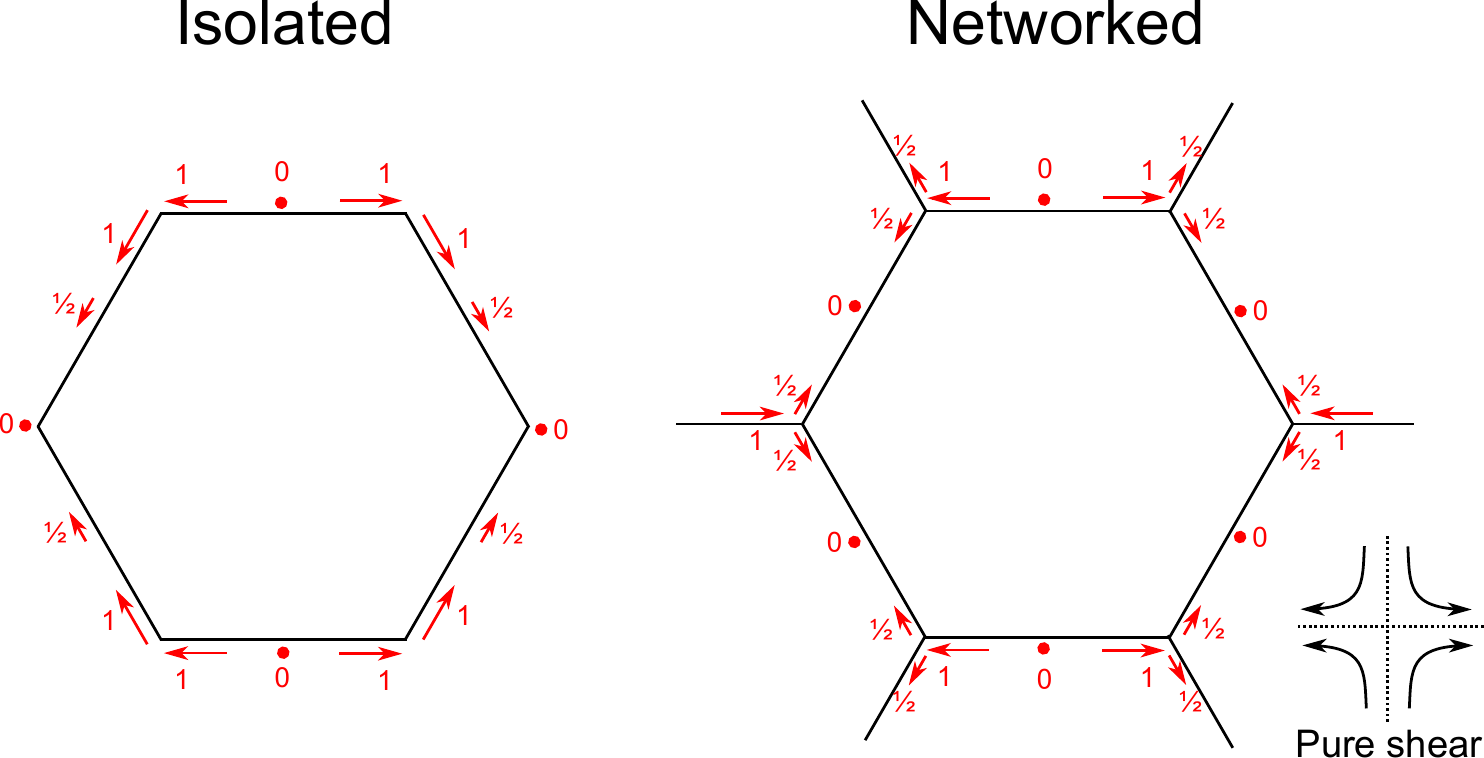}
\caption{Two distinct conceptual models for Coble creep in the absence of 
melt 
for hexagonal grains. On the left is the isolated assumption, where grain 
boundary diffusion is restricted to an individual grain. On the right is 
the 
networked assumption, where matter can be transported from one grain to 
another 
through the network of grain boundaries. Pure shear is applied, and shown 
in red 
with arrows are the corresponding fluxes of atoms at different locations.}
\label{fig:hex_coble_pure}
\end{figure}

Analysis of Coble creep is generally simpler than that of Nabarro-Herring 
creep, because diffusion takes place over manifolds of dimension one 
lower 
than that of the space (e.g. over lines in 2D, planes in 3D). Coble creep 
in 
the absence of melt for hexagonal grains is illustrated in 
Figure \ref{fig:hex_coble_pure}. An additional subtlety arises for Coble 
creep, in 
that the shear viscosity one obtains depends crucially on assumptions 
about 
what happens at the triple points where three grains meet. Two 
possible 
assumptions are shown in Figure \ref{fig:hex_coble_pure}: In the first 
case, 
the 
grain is considered isolated, and grain-boundary diffusion is restricted 
to 
an 
individual grain (this is the assumption made by \citet{Takei2009}). In 
the 
second case, it is assumed that all grain boundaries are connected such 
that 
matter can be transported from one grain to neighbouring grains through 
the 
network of grain boundaries (this is the assumption made by 
\citet{Spingarn1978}). When considering a tessellation of identical 
grains, it is more reasonable to make the networked assumption as there is 
no physical process that would be expected to restrict diffusion to an 
individual grain.

If the grain is assumed isolated, the shear viscosity is given by 
(appendix \ref{sec:coble_hex_pure})
\begin{equation}
 \eta_{0, \text{isolated}} = \frac{7}{432} \frac{k T d^3}{\delta 
D^\text{gb} 
\Omega},
\end{equation}
where $d$ is the distance between opposite sides of the hexagon. The 
numerical 
prefactor of $7/432 \approx 0.0162$ is very similar to that for a circle 
of 
diameter $d$, which is $1/64 \approx 0.0156$ \citep{Takei2009}. When the 
grain-boundaries are networked, the viscosity is significantly less, and 
is 
given by
\begin{equation}
  \eta_{0, \text{networked}} = \frac{1}{144} \frac{k T d^3}{\delta 
D^\text{gb} 
\Omega}, \label{eq:networked_eta}
\end{equation}
as found by \citet{Spingarn1978}. The 
viscosity 
for the networked case is lower by a factor of $3/7 \approx 0.43$. Under 
the 
networked assumption, each vertex of the hexagon is identical, and 
therefore 
must be at the same chemical potential (note that each of the triple 
junctions 
in Figure \ref{fig:hex_coble_pure} has the same pattern of fluxes). The 
vertices 
are not identical in the isolated case, where the four vertices at the top 
and 
bottom of the hexagon in Figure \ref{fig:hex_coble_pure} are at a 
different 
chemical potential to the two on the far left and right. The networked 
case is weaker 
because there is a shorter effective diffusion distance from one part of 
the 
hexagon to another, provided by the ``short-circuiting'' at the triple 
points.


\citet{Takei2009} and \citet{Holtzman2016} have argued that the presence 
of a very small amount of 
melt 
can radically reduce the shear viscosity. They argue this is due to the 
``short-circuiting'' nature of the melt, which acts as a fast path for 
diffusion. 
Whether such a reduction is seen for the hexagonal configuration here 
depends 
on whether the ``isolated'' or ``networked'' assumption is made for the 
situation without melt. In fact, adding an infinitesimal amount of melt at 
the triple junctions of the networked example makes no difference, because 
all 
triple junctions are already at the same chemical potential. Hence in this 
case 
there is no weakening effect of a small amount of melt. However, if the 
situation without melt is assumed isolated, then there is a drop by a 
factor of 
$3/7$ 
when melt is introduced.

\begin{figure}
 \includegraphics[width=\columnwidth]{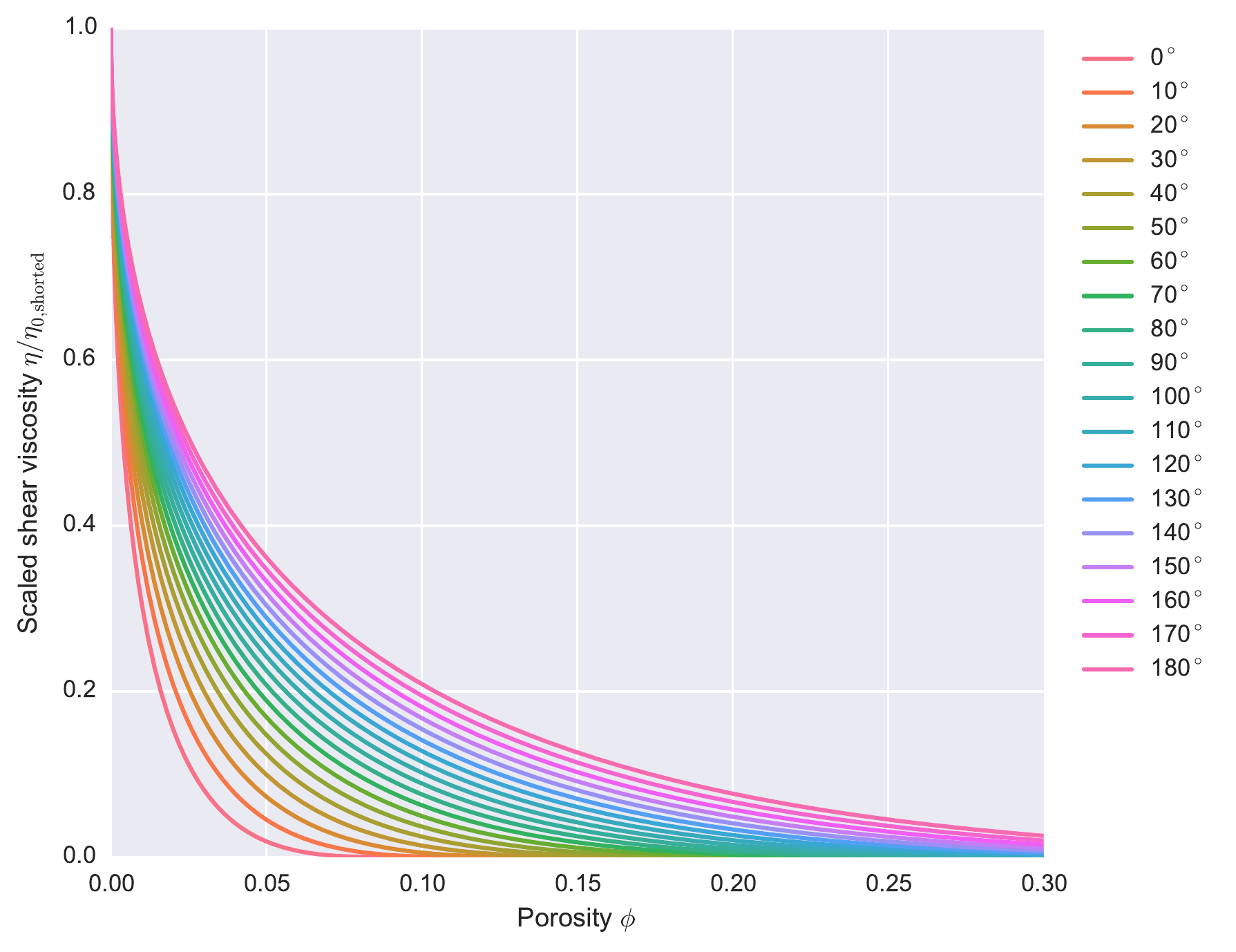}
\caption{Scaled shear viscosity plotted against porosity for hexagonal 
grains 
undergoing Coble creep.}
\label{fig:coble_hex}
\end{figure}

The effect of melt pores on a hexagonal array of grains has been discussed 
by 
\citet{Cocks1990}, who find for the same geometry of melt pores as 
depicted 
in Figure \ref{fig:hex_grain} that
\begin{equation}
 \frac{\eta}{\etashort} = \left(\frac{A_\text{ss}}{A_\text{cell}} \right)^3
\end{equation}
where $A_\text{ss}/A_\text{cell}$ is the fraction of the boundary of the 
unit 
cell that is grain--grain 
contact ($A_\text{ss}=A_\text{cell}$ with no melt present, $A_\text{ss}=0$ 
when 
complete wetting occurs). 
$\etashort$ is the viscosity in the presence of an infinitesimal amount of 
melt, 
equal to the networked result given in \eqref{eq:networked_eta}. The 
fraction 
$A_\text{ss}/A_\text{cell}$ is related to porosity $\phi$ and dihedral 
angle $\theta$ by an 
expression 
of the form
\begin{equation}
 \frac{A_\text{ss}}{A_\text{cell}} = 1 - \sqrt{\frac{\phi}{\phi_d(\theta)}}
\end{equation}
where $\phi_d(\theta)$ is the porosity at which disaggregation occurs 
(i.e. complete wetting of the grain boundaries). $\phi_d(\theta)$ as a 
function of dihedral angle is given explicitly 
in 
appendix \ref{sec:hex_melt}. Hence
\begin{equation}
  \frac{\eta}{\etashort} = \left(1 - \sqrt{\frac{\phi}{\phi_d(\theta)}} 
\right)^3, \label{eq:2dCobleresult}
\end{equation}
which is plotted in Figure \ref{fig:coble_hex}.
Once melt is present, bulk deformation is possible. Owing to a symmetry of 
the 
governing equations known as the Cauchy relation, bulk and shear 
viscosity 
are 
linked by $\zeta = 2\eta$ in 2D (see appendix \ref{sec:coble_melt}). As 
consequence, both bulk and shear viscosity show the same behaviour with 
porosity for Coble creep.

%


\subsubsection{3D: Tessellation of tetrakaidecahedrons}

\begin{figure}
 \includegraphics[width=\columnwidth]{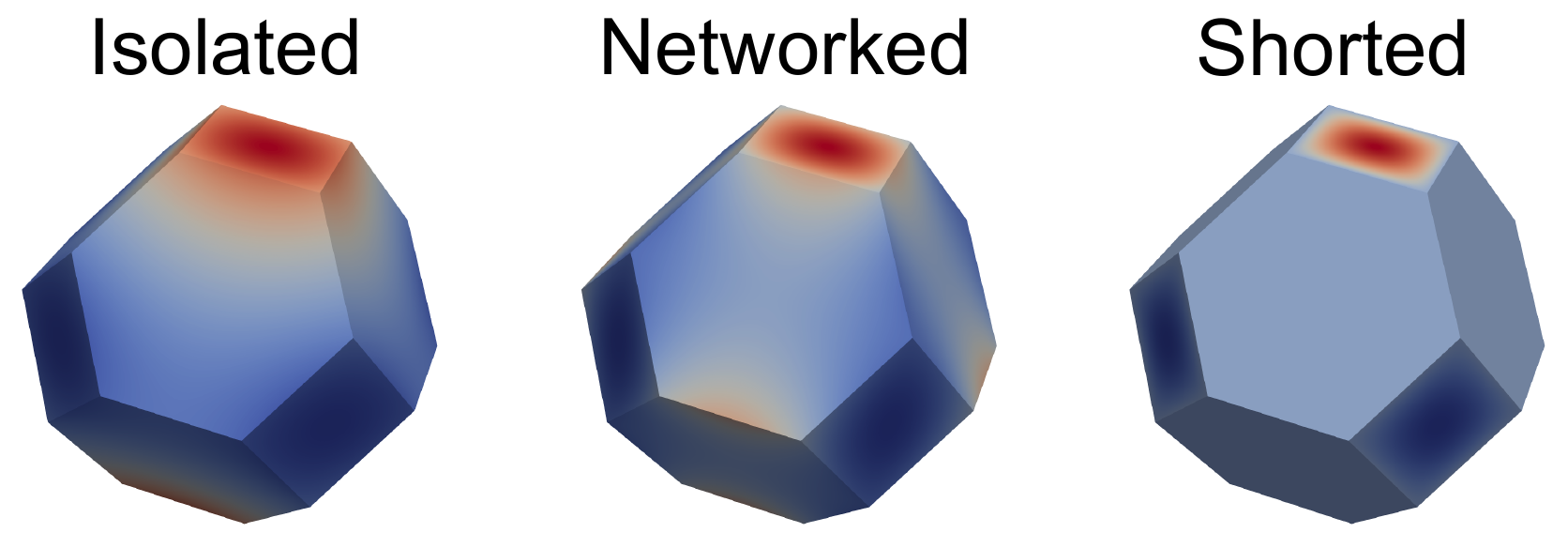}
\caption{Vacancy concentrations under pure shear for Coble creep of 
tetrakaidecahedrons with three different assumptions (isolated, networked, 
or 
shorted). The principal axes of shear are aligned with the square faces of 
the 
tetrakaidecahedron.}
\label{fig:tet_pics}
\end{figure}

As in 2D, the viscosity of tetrakaidecahedrons in 3D 
depends on assumptions about the behaviour at the triple lines where three 
grains meet. Figure \ref{fig:tet_pics} depicts three possibilities: 1, 
Isolated, 
where grain-boundary diffusion is restricted to an individual grain; 2, 
Networked, where the grain shown is one grain of an infinite tessellation 
of 
grains, and grain-boundary diffusion can move matter from one grain to 
neighbouring grains; 3, Shorted, where an infinitesimal amount of melt is 
assumed to lie along the grain edges forming a fast path for diffusion. 
The 
corresponding viscosities for these three situations are given in 
Table \ref{tab:tetvisc}. The numerical prefactor for the  Voigt-average 
viscosity 
for the isolated case, 0.0063471, is fairly similar to that for a sphere 
of 
diameter $d$, which is $1/120 \approx  0.0083333$ \citep{Takei2009}. The 
anisotropy 
in viscosity is larger than seen for Nabarro-Herring creep, with Zener 
ratios 
$\eta_2/\eta_1$
given by 1.78, 2.84, and 4.91 for the three cases. 

\begin{table}
 \begin{center}
\begin{tabular}{llll}\hline
 & $\eta_{10}$ & $\eta_{20}$ & $\overline{\eta}_0$\\ \hline
isolated & 0.0043178 & 0.0077000 & 0.0063471\\
networked & 0.0012242 & 0.0034715 & 0.0025726\\
shorted & 0.0005491 & 0.0026965 & 0.0018376 \\ \hline
 \end{tabular}
 \end{center}
\caption{Shear viscosities of tetrakaidecahedrons for Coble creep under 
three different assumptions (isolated, networked, shorted). The numerical 
prefactors above should be multiplied by $k T d^3/\delta D^\text{gb}$ to 
give 
the dimensional viscosity.}
\label{tab:tetvisc}
\end{table}

As in 2D, the networked configuration has a significantly reduced 
viscosity 
compared to the isolated configuration (in terms of the Voigt-average, the 
viscosity is reduced by a factor of 0.40, a similar factor to that seen 
for 
the 
hexagonal case). A key difference is that in 3D, unlike in 2D, the shorted 
and 
networked configurations are not identical. The shorted configuration 
reduces 
the viscosity by a factor of 0.71 compared to the networked configuration, 
and is a factor of 0.29 lower than the isolated configuration. 

\begin{figure}
 \includegraphics[width=\columnwidth]{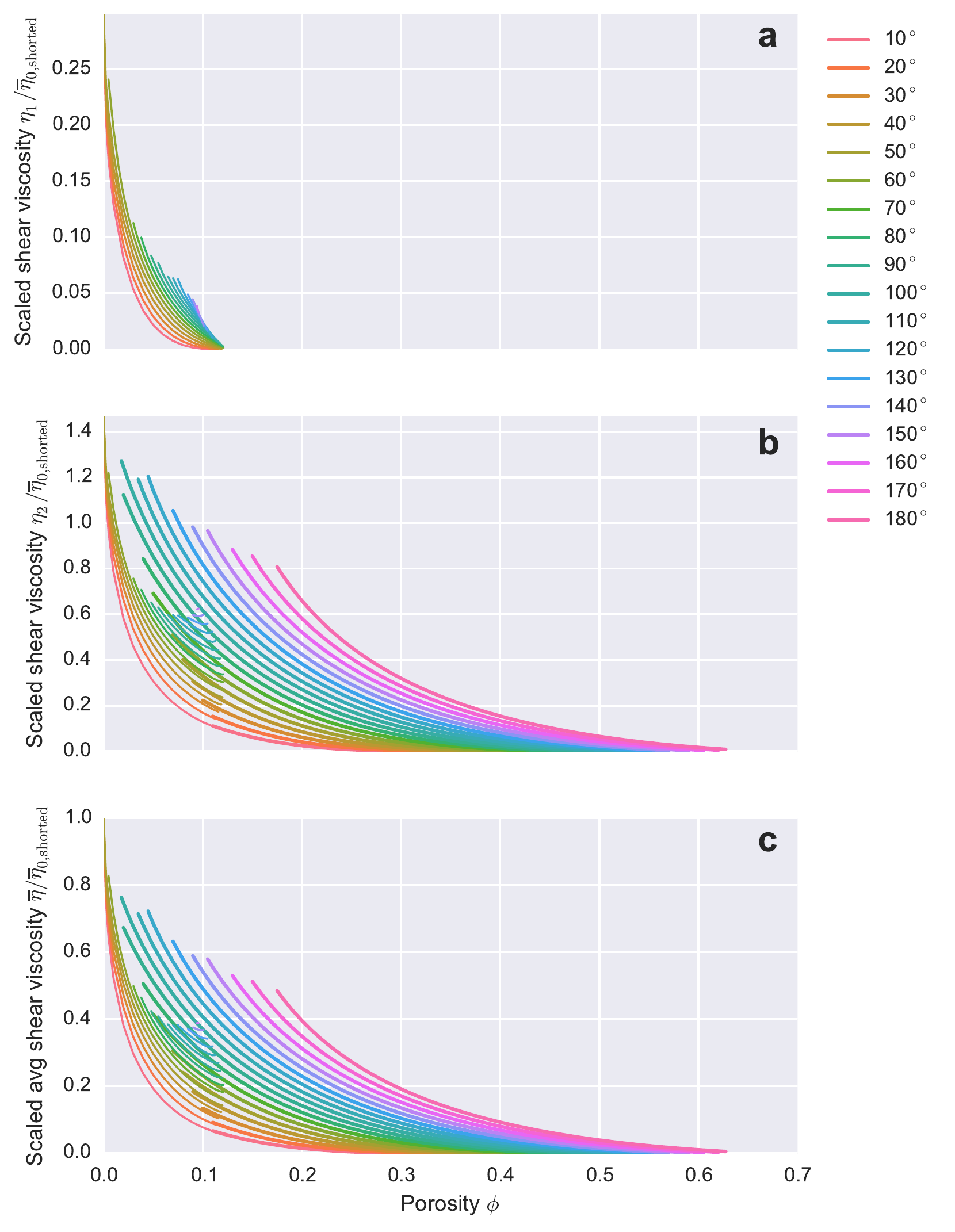}
\caption{Scaled shear viscosities plotted against porosity for Coble creep 
of tetrakaidecahedrons. Line styles are as in Figure 
\ref{fig:herring_bulk}. a) 
shows 
$\eta_1/\overline{\eta}_0$, which vanishes when the square faces become 
wetted. 
b) shows $\eta_2/\overline{\eta}_0$, and c) shows the Voigt-average 
$\overline{\eta}/\overline{\eta}_0$.  The scaling has used the ``shorted'' 
reference value from Figure \ref{tab:tetvisc}. Note that this plot only 
shows values with melt present ($\phi>0$), and not the potential jump in 
shear viscosity at the onset of melting.}
\label{fig:coble_summary}
\end{figure}

The effect of melt on the viscosities is plotted in 
Figure \ref{fig:coble_summary}. The anisotropy is clear, and arises 
largely 
from 
the smaller contact areas of the square faces compared to the hexagonal 
faces. As for Nabarro-Herring creep, the shear viscosity $\eta_1$ vanishes 
when 
the square faces become wetted. Owing to the 
Cauchy relation symmetry, the bulk viscosity is a constant multiple of the 
Voigt-average shear viscosity, given by $\zeta = \tfrac{5}{3} 
\overline{\eta}$ (appendix \ref{sec:coble_melt}).

The main behaviour of viscosity as a function of porosity in 3D for 
Coble creep can be understood from the simple theory developed by 
\citet{Cooper1984,Cooper1986} and \citet{Takei2009}. In the study by 
\citet{Takei2009} 
the grain geometry is simplified to be a sphere with circular contact 
patches. With these simplifications, the shear viscosity with melt present 
can be 
related to the area of solid--solid contact by
\begin{equation}
\frac{\overline{\eta}}{\etashort} \approx 
\left(\frac{A_\text{ss}}{A_\text{cell}} \right)^2, \label{eq:approxCoble}
\end{equation}
where $A_\text{ss}/A_\text{cell}$ is the fraction of the boundary of the 
unit cell that is grain--grain contact. The exact relationship between 
shear viscosity and contact areas for the 
more complex geometry considered here is given in 
appendix \ref{sec:coble_melt}, and is a weighted sum of squares of the 
individual contact areas. The expression above does not take account of 
the 
differences between the different contacts (i.e. the differences in both 
shape and area of the square 
and hexagonal contacts, and the differences in distances of the contacts 
from the grain 
center). Nevertheless, \eqref{eq:approxCoble} provides a useful 
approximation. In 
turn, the contact areas can be approximately related to porosity by
\begin{equation}
 \frac{A_\text{ss}}{A_\text{cell}} \approx 1 -  
\sqrt{\frac{\phi}{\phi_d (\theta)}}
\end{equation}
 where $\phi_d(\theta)$ is a function of dihedral angle 
\citep{vonBargen1986, Rudge2018}. This is a good 
approximation for a tube-like melt geometry (see (B14) of 
\citet{Rudge2018}). Hence for $\phi>0$ \citep{Takei2009}
\begin{equation}
\frac{\overline{\eta}}{\etashort} \approx \left(1 -  
\sqrt{\frac{\phi}{\phi_d(\theta)}}  \right)^2, \label{eq:approx_form}
\end{equation}
which differs from the 2D result in \eqref{eq:2dCobleresult} only in the 
exponent of 2 rather than 3.
\begin{figure}
 \includegraphics[width=\columnwidth]{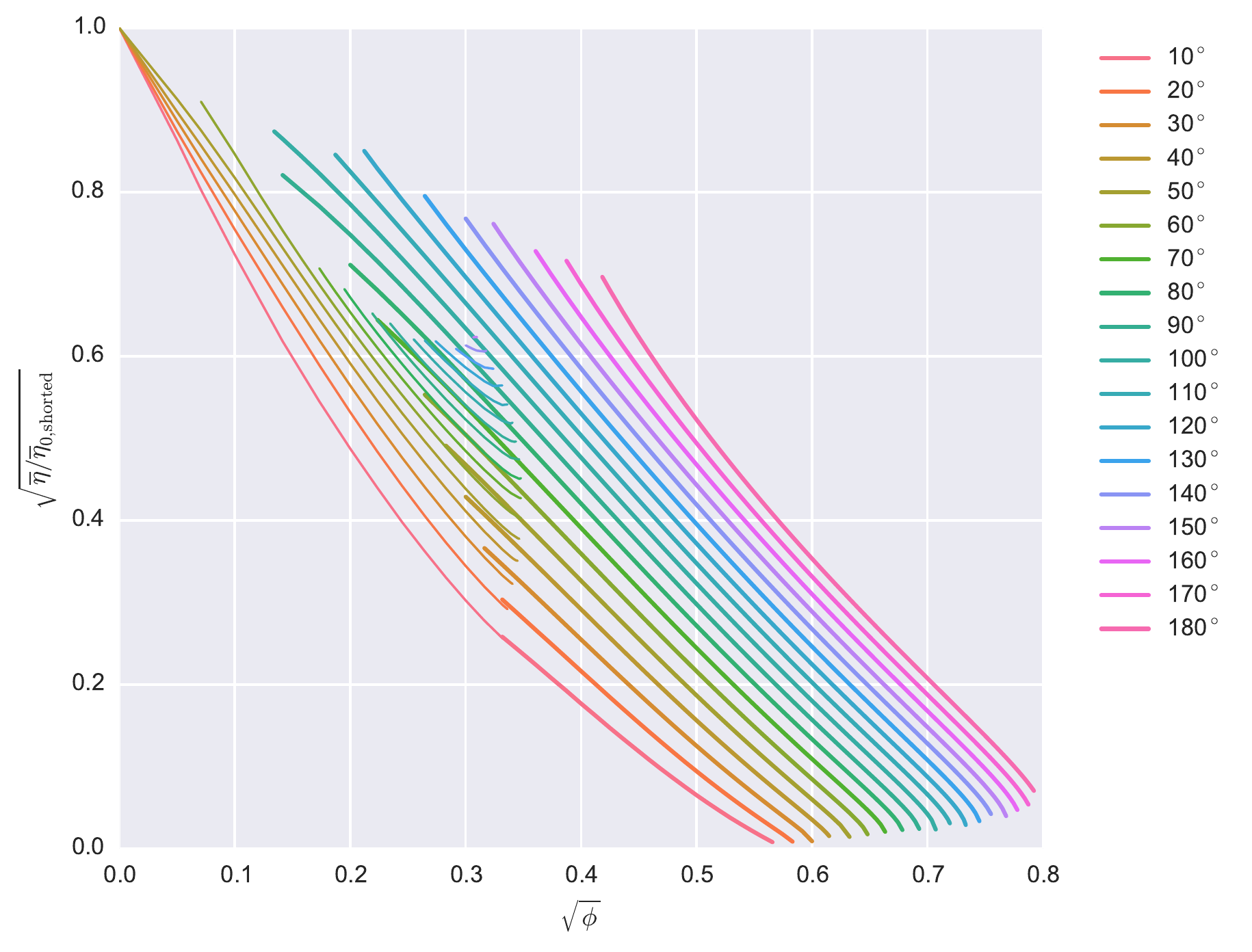}
\caption{A plot of the same data for Coble creep of tetrakaidecahedrons as 
in  
Figure \ref{fig:coble_summary}c, except axes now show the square root of 
scaled, 
Voigt-average shear viscosity against the square root of porosity to 
reveal 
a 
near-linear trend.}
\label{fig:coble_tetra_app}
\end{figure}

The broad validity of the approximation in \eqref{eq:approx_form} is 
illustrated in Figure \ref{fig:coble_tetra_app}, which plots the square 
root 
of 
the scaled shear viscosity against the square root of porosity. On this 
plot, 
an expression of the form \eqref{eq:approx_form} should give a 
straight 
line. Certainly for low porosities and connected topologies, the 
approximation appears to be good. For a $40^\circ$ dihedral angle typical 
of 
basaltic melts, and porosities less than 10\%, this implies a $\phi_d$ of 
around 0.24 (just slightly larger than $\phi_d$ of $2.3^{-2}\approx 0.19$ 
suggested by \citet{Takei2009} for the olivine--basalt system). Unlike the 
2D case, $\phi_d$ now does not represent exactly the porosity at which 
disaggregation occurs, because the relationship in \eqref{eq:approx_form} 
is only approximate. Instead $\phi_d$ represents where disaggregation 
would be expected to occur if one extrapolates the approximate linear 
relationship seen at low porosities in Figure \ref{fig:coble_tetra_app} to 
higher porosities.

The full calculations show that the shear viscosity $\eta_2$ is 
proportional to the areas of the hexagonal contacts squared, and $\eta_1$ 
the same for the square contacts (appendix \ref{sec:coble_melt}). Thus the 
main trends in Figure \ref{fig:coble_summary}a and b mimic those in 
Figures 16 
and 17 of \citet{Rudge2018} which plot the relevant areas against 
porosity. 

\begin{figure}
 \includegraphics[width=\columnwidth]{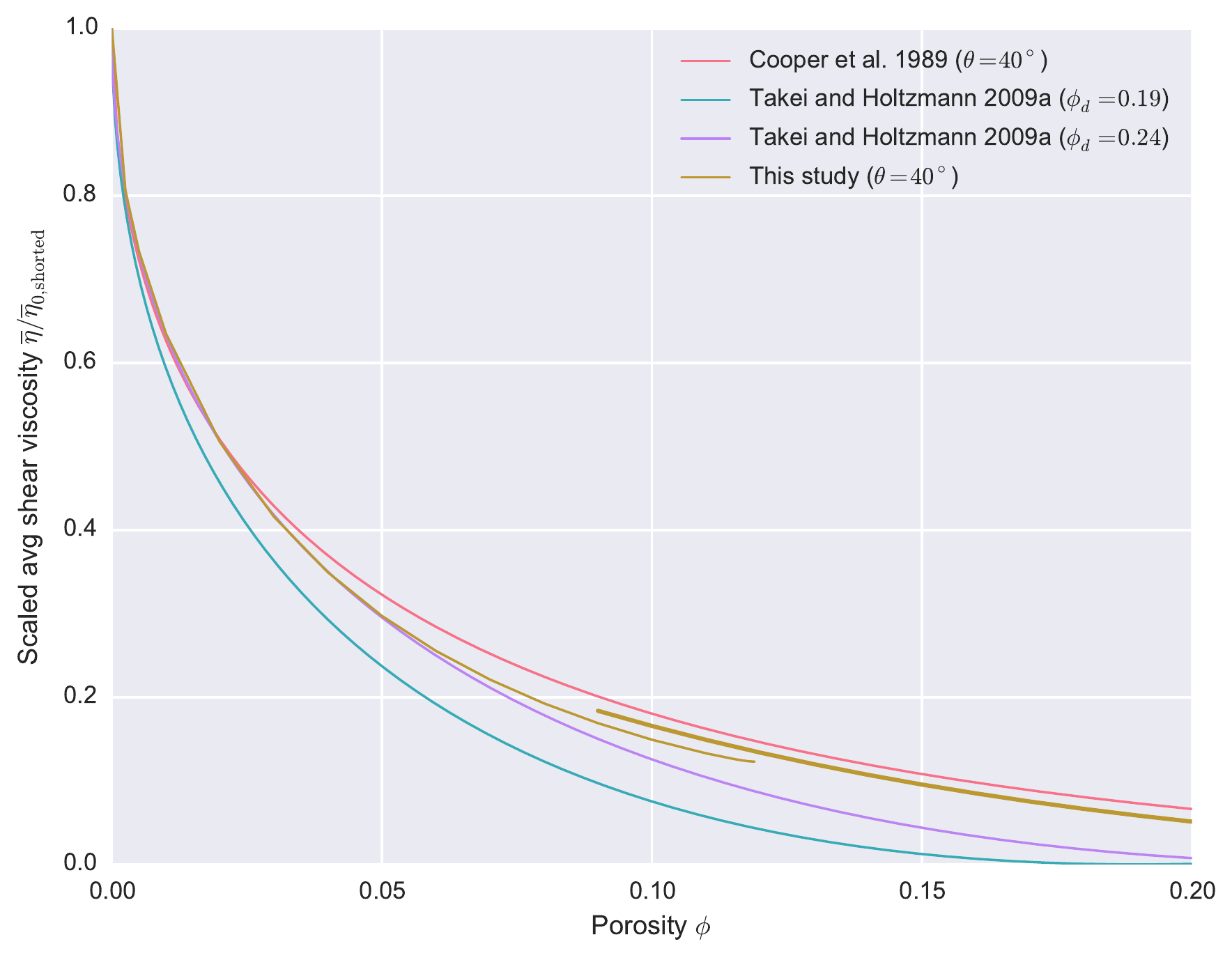}
\caption{Scaled average shear viscosity against porosity for Coble creep 
of tetrakaidecahedrons as in Figure \ref{fig:coble_summary}c, but just 
showing the results for a dihedral angle of 40$^\circ$. Also shown are 
three
alternative model curves. The first is given 
by equations 5, 6, and 7 of \citet{Cooper1989} for a dihedral angle of 
40$^\circ$. The other two curves show the
approximate law given by \citet{Takei2009} and \eqref{eq:approx_form} of 
this manuscript, with two choices of $\phi_d$.}
\label{fig:cooper_comparison}
\end{figure}

The broad agreement of the Coble creep calculations presented here with 
those of previous studies is illustrated in Figure 
\ref{fig:cooper_comparison}. Shown in the figure is the calculated 
variation in average shear viscosity for a dihedral angle of $40^\circ$, 
along with model curves from the studies of \citet{Takei2009} and 
\citet{Cooper1989}. The curves marked \citet{Takei2009} are given by 
equation \eqref{eq:approx_form}, where two values of $\phi_d$ are shown. 
The first ($\phi_d=0.19$) is that directly suggested by 
\citet{Takei2009} as appropriate for olivine-basalt; the second 
($\phi_d=0.24$) is the value which better fits the creep calculations 
presented here for a dihedral angle of $40^\circ$ at small porosities. 
With the appropriate choice of $\phi_d$ the law given in 
\eqref{eq:approx_form} clearly provides a very good approximation to the 
numerical calculations up to porosities of about $\phi \sim 0.1$, as can 
also be seen in the near linear nature of the curves in Figure 
\ref{fig:coble_tetra_app} up to $\sqrt{\phi} \sim 0.3$. The departure of 
the calculations here from the approximate law of \eqref{eq:approx_form} 
arises from the assumption of an isotropic grain model by 
\citet{Takei2009}, whereas the model here has considerable anisotropy 
arising from the differences between the square and hexagonal contacts. 

Figure \ref{fig:cooper_comparison} also shows the expected behaviour from 
the model of \citet{Cooper1989}, which closely 
approximates the calculations presented here for a wide range of 
porosities. The model of \citet{Cooper1989} is also based on 
tetrakaidecahedrons, but the geometry is simplified by assuming that melt 
lies in uniform tubes along the grain edges.  \citet{Cooper1989} argue that 
the dominant contribution to the average shear viscosity comes from the 
hexagonal faces, and thus approximate that the viscosity scales with the 
fourth power of the distance from the centre of a hexagonal face to the 
melt. This is a reasonable approximation to have made -- while we have 
shown here that the average shear viscosity depends on a weighted sum of 
squares of the areas of both the square and hexagonal contacts, it is 
indeed the hexagonal contacts that form the dominant contribution. For 
example, in the absence of melt, the hexagonal faces contribute 88\% of 
the 
total sum in the estimation of the average shear viscosity. The remaining 
differences between the \citet{Cooper1989} calculations and the model 
presented here arise due to the more accurate computation of the melt 
geometry here (i.e. not approximating by tubes), and the direct calculation 
of the contributions from both the square and hexagonal faces. 

\section{Discussion}

Perhaps the most useful outcome of this work is the expressions relating 
the bulk and 
shear 
viscosities to porosity, given for Coble creep by \eqref{eq:approx_form}, 
and 
for Nabarro-Herring creep by \eqref{eq:nhapprox_shear} and 
\eqref{eq:nhapprox_bulk} in appendix \ref{sec:nh_param}. These expressions 
can be used in larger-scale 
geodynamic calculations, such as those models based on the compaction 
equations 
of \citet{McKenzie1984}.

There are important differences between the rheological laws proposed here 
and 
those in common use by the geodynamics community. Perhaps the most 
important 
difference concerns the bulk viscosity. Here, the bulk viscosity either 
has 
a 
logarithmic singularity at small porosity (Nabarro-Herring creep), or is a 
constant multiple of the shear viscosity (Coble creep). In large-scale 
geodynamic models, bulk viscosity is usually given by a law with a 
singularity 
proportional to $1/\phi$ (e.g.  
\citet{Scott1986,Sleep1988,Bercovici2001,Hewitt2008,Simpson2010a,
Schmeling2012}). The reason these studies have a different singular 
behaviour is 
ultimately due to a different conception of the physics at the scale of 
individual grains. In these studies, the grain itself is treated as a 
Newtonian 
viscous fluid, so that the governing equations at the microscale are those 
of 
Stokes flow. In the present study, the physics of the individual grain is 
described in terms of diffusion. As shown in section \ref{sec:governing}, 
only at the macroscale does the medium 
act like a Newtonian viscous fluid. The different singular behaviour at 
small porosity is ultimately a consequence of the different PDEs that 
are 
solved at the grain scale: here, Laplace's or Poisson's equation; in the 
previous studies, Stokes flow. The rheological laws determined here are 
likely to better represent the viscosities of partially molten rocks 
because they better reflect the true microscale physics.  

In fact, a logarithmic singularity of the bulk viscosity was 
originally advocated by \citet{McKenzie1984} (see his Figure 6 and (C12)). 
His 
equation (C12) arises from a model of Coble creep with spherical pores, as 
discussed 
by 
\citet{Arzt1983} and \citet{Pan1994}. Here, the results for Coble creep do 
not 
show a logarithmic singularity, and this is due to the different melt 
geometry -- here melt lies in tubes along the grain edges for dihedral 
angles less than 
$60^\circ$, 
whereas \citet{Arzt1983}'s model assumes spherical pores.

Another key point of difference from previous studies is the effect of a 
very 
small amount of melt on the shear viscosity. \citet{Takei2009} argued that 
during Coble creep the shear viscosity is 20\% that of the 
melt-free case when a small amount of melt is present, due to the 
short-circuiting effect of the melt pathways. The results here 
suggest 
the effect of the short-circuiting may be much more modest, since even 
in the absence of melt, diffusion may be able to take place through the 
complete 
network of grain boundaries rather than being restricted to a single grain 
as 
\citet{Takei2009} assume. For Coble creep, the shear viscosity with a very 
small amount of melt is then 71\% of the melt-free value.
 During 
Nabarro-Herring creep there is no sudden drop in the shear viscosity at 
the 
onset of melting, but there is a very rapid decrease as the 
porosity 
increases: At just 0.1\% porosity the shear viscosity is halved from the 
melt-free value. Experimental studies have produced mixed results as to 
whether 
there is a sudden 
drop in the shear viscosity at the onset of melting. Drops in the shear 
viscosity have been reported for both the organic borneol system 
\citep{McCarthy2011a} and olivine-basaltic melt systems \citep{Faul2007}, 
but 
more recent work on borneol has not identified a drop \citep{Yamauchi2016}.

It is important to bear in mind the simplifications that have made in this 
study. Importantly, there has been no exploration here of the effects of 
chemistry, which is likely to play an important role in the polyphase, 
polycrystalline, partially molten mantle. Here all crystals have been 
treated as being the same phase, and all grain-boundaries are those 
between identical grains. In a polyphase system, there are a series of 
different grain boundaries depending on the combinations of different 
mineral phases that meet \citep{Ford2004}. Thus matter may be transported 
more easily across some boundaries than others. This may mean that the 
short-circuiting effect of melt might be greater in a polyphase system 
than 
one with a single solid phase. Moreover, each grain-boundary has been 
treated as freely slipping, and the calculations here could be extended by 
considering a finite viscosity for the grain boundaries 
\citep{Lifshitz1963}. The inclusion of grain boundary viscosity will be 
particularly important for modelling short-timescale transient deformation 
(e.g. that associated with seismic attenuation). However, the 
freely-slipping assumption is likely to be good for modelling the 
long-timescale deformation associated with mantle convection and melt 
transport.

The melt geometries on which this study is based are textural 
equilibrium geometries assuming isotropic surface energies. Anisotropy is 
likely to play an important role in wetting some grain boundaries, and in 
turn reducing the shear viscosity. Moreover, the geometry of the melt 
network must evolve with finite deformation, and more work needs to be 
done 
to explore with interactions between effective rheology and finite 
deformation.

This study is also based on an assumption of an infinite diffusivity in 
the melt 
phase. Finite melt diffusivity will mean the effect of melt-weakening 
is not as strong as calculated here. The effect of finite diffusivity has 
been explored by \citet{Takei2009a}, and similar calculations could be 
performed for the more complex geometries considered here. As 
\citet{Takei2009a} point out, the effects of finite melt diffusivity are 
most important at very small porosities. Moreover, at large porosities, 
the 
rheological laws produced here will also breakdown as they do not consider 
the finite viscosity of the melt phase. At very large porosities one has a 
suspension of isolated crystals within a melt, which is described by a 
very 
different rheological law.  

The precise behaviour of bulk viscosity as a function of porosity can have 
important consequences for larger-scale dynamics. An important example of 
this 
has been given recently by \citet{ReesJones2018} and concerns the 
reaction-infiltration instability (RII), an instability that may be 
responsible for localising melt transport during mantle melting. 
\citet{ReesJones2018} have demonstrated that a bulk viscosity which 
varies 
strongly with porosity is more likely to suppress the RII than one which 
does 
not.

\begin{figure}
 \includegraphics[width=\columnwidth]{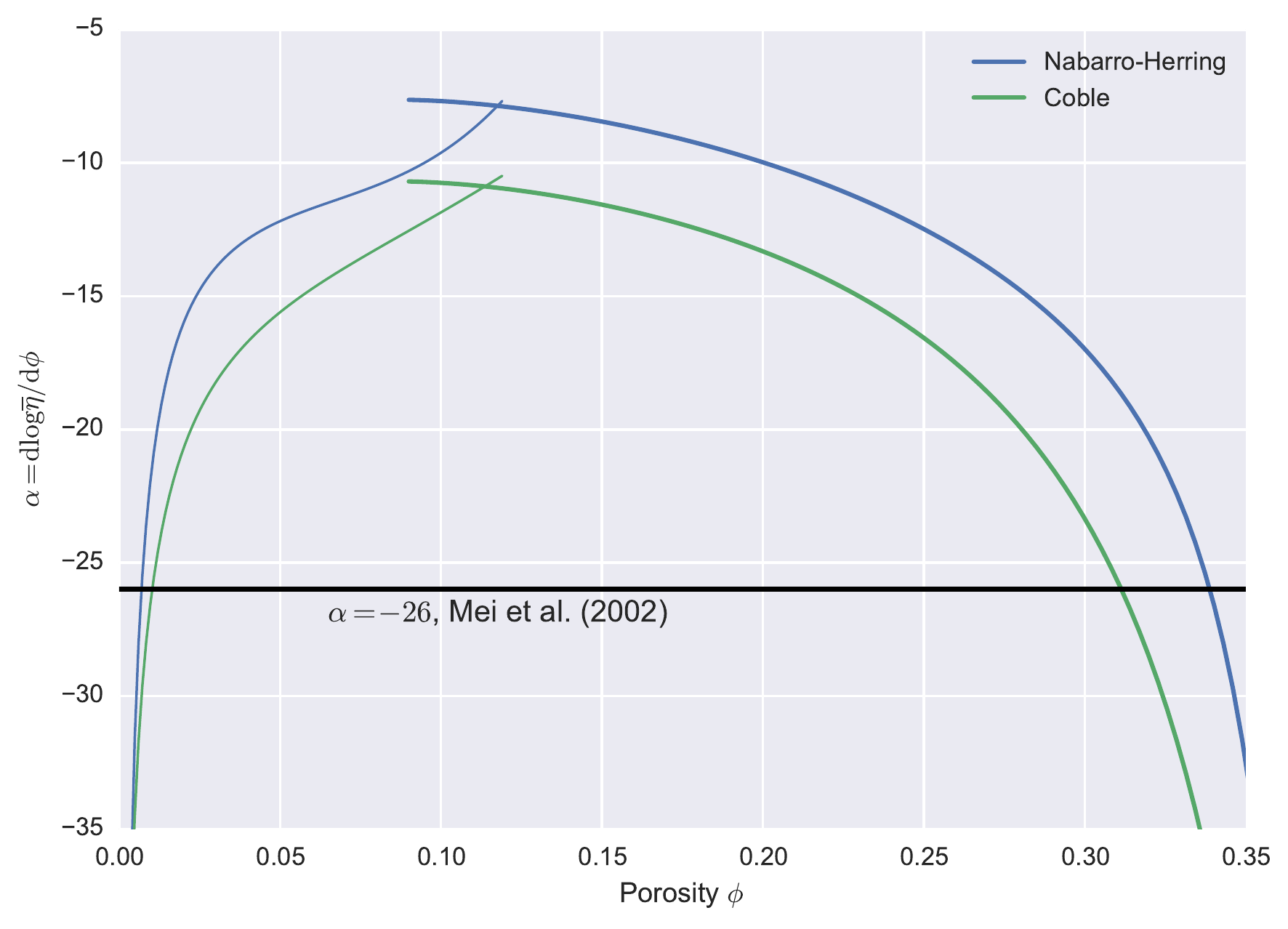}
\caption{A plot of the porosity-weakening exponent $\alpha = \d \log 
\overline{\eta} / 
\d \phi$ against porosity for Nabarro-Herring and Coble creep and a 
dihedral angle of $40^\circ$. The thin lines shows the connected ``c'' 
topology and the thick lines the square-wetted ``s'' topology.  Shown as a 
black horizontal line is the 
corresponding result for the empirical exponential law proposed by 
\citet{Mei2002}.}
\label{fig:weakening}
\end{figure}

The rheological laws derived here have consequences for the rate at which 
melt bands develop during shear 
\citep{Stevenson1989,Spiegelman2003,Holtzman2003}. Linear 
stability analysis shows that melt bands are expected to initially grow at 
a 
rate 
proportional to a weakening factor $\alpha = \d \log \eta / \d \phi$ 
\citep{Stevenson1989,Spiegelman2003}, which is plotted as a function of 
porosity for the rheologies considered here in Figure \ref{fig:weakening}. 
Modelling work on this instability (e.g. 
\citet{Katz2006,Takei2015,Rudge2015, Bercovici2016}) has typically used an 
empirical 
rheological law of the form $\eta \propto \exp(\alpha \phi)$, with 
constant 
$\alpha \sim -26$, based on a fit to experimental data by \citet{Mei2002}. 
As pointed out by \citet{Takei2009} for Coble creep, the microscale 
calculations predict a weakening factor $\alpha$ which varies with 
porosity. For porosities around 3\% as used in experiments 
\citep{Holtzman2003}, a weakening factor around $\alpha = -17$ is 
predicted 
for Coble creep (Figure \ref{fig:weakening}). The growth-rate of melt 
bands 
also depends on the ratio of 
bulk to shear viscosity. The rheological laws here suggest the bulk 
viscosity is comparable in magnitude to the shear viscosity, as also 
recently argued by \citet{Alisic2016} on the basis of a comparison of 
numerical models of melt bands with the laboratory experiments by 
\citet{Qi2013}. A bulk viscosity comparable to the shear viscosity was 
also inferred in four-point bending experiments by \citet{Cooper1990}, 
who found $\zeta/\eta=1.9$, not too dissimilar the value of 
$\zeta/\eta=\tfrac{5}{3}=1.67$ predicted for Coble creep.

\section{Conclusions}

This study represents one step towards a better understanding of the 
rheology of 
partially molten materials. The effective viscosity tensor has been 
obtained as 
a function of porosity and dihedral angle for diffusion creep, both 
assuming 
body diffusion (Nabarro-Herring creep) and surface diffusion (Coble 
creep). The 
3D calculations were based on an assumption of a melt geometry that is in 
textural equilibrium, with a tessellation of identical tetrakaidecahedral 
(truncated octahedral) unit cells. For Coble creep, the bulk viscosity was 
shown to be a constant multiple of the shear viscosity, whereas 
Nabarro-Herring 
creep has a logarithmic singularity in the bulk viscosity for small 
porosities. An important challenge for the future is to develop a model of 
finite deformation which allows the melt geometry to evolve with time, and 
to be 
out of textural equilibrium. An additional challenge is to explore other 
forms 
of deformation, such as dislocation creep. Such models will undoubtedly 
provide 
important insights into the dynamics of melt extraction from the Earth's 
mantle.

\subsection*{Acknowledgements}

The ideas for this work arose out of the 2016 ``Melt in the Mantle'' 
programme at the Isaac Newton Institute for Mathematical Sciences. I am 
grateful to many of the participants of that programme for useful 
discussions, but particularly grateful to Yasuko Takei and Dan McKenzie 
for introducing me to the 
microscale physics of diffusion creep. I am also grateful to David Rees 
Jones, David Kohlstedt, Richard Katz, and an anonymous reviewer for their 
comments that helped improve this manuscript. ``Melt in the Mantle'' 
was 
supported 
by EPSRC Grant Number 
EP/K032208/1. I am also very grateful to the Leverhulme Trust for 
support. Data tables of the computed viscosities can be found in the 
supporting information.  
Code for generating the textural equilibrium geometries is available at 
\url{https://www.johnrudge.com/melt/}.

\clearpage

\appendix
\section{Nabarro-Herring creep analytical solutions}

\subsection{Decomposition into trace and symmetric trace-free 
components}\label{sec:decom}

It is helpful to decompose equations (\ref{eq:g1}-\ref{eq:g4}) for 
Nabarro-Herring creep into trace and symmetric trace-free components, by 
writing $\gamma_{ij}$ as

\begin{equation}
 \gamma_{ij} = \frac{\gamma}{N} \delta_{ij} + \tilde{\gamma}_{ij}
\end{equation}
where the symbol $\gamma$ without subscripts is used to represent the 
trace 
part ($\gamma \equiv \gamma_{kk}$), and $\tilde{\gamma}_{ij}$ is the 
symmetric trace-free part ($\tilde{\gamma}_{kk}=0$). $N$ is the dimension 
of the space ($N=2$ for the hexagons, $N=3$ for the tetrakaidecahedrons). 
The trace part of the decomposition is associated with bulk deformation, 
and the symmetric trace-free part is associated with shear.

The viscosity tensor can be decomposed as
\begin{equation}
 C_{ijkl} = B_{ij} \delta_{kl} + \tilde{C}_{ijkl},
\end{equation}
where the tensors $B_{ij}$ and $\tilde{C}_{ijkl}$ can be obtained from two 
separate problems, one for bulk deformation:
\begin{gather}
 \nabla^2 \gamma = 0 \text{ in }V_s, \label{eq:gb1} \\
\gamma = 0 \text{ on } S_\text{sl}, \label{eq:gb2} \\
\pdd{\gamma}{n} = x_p n_p
\text{ on } S_\text{ss}, \label{eq:gb3}\\
B_{ij} = \frac{1}{N \Vc} \int \gamma  x_i n_j\; \d 
S, \label{eq:gb4}
\end{gather}
and one for shear:
\begin{gather}
 \nabla^2 \tilde{\gamma}_{ij} = 0 \text{ in }V_s, \label{eq:gs1} \\
\tilde{\gamma}_{ij} = 0 \text{ on } S_\text{sl}, \label{eq:gs2} \\
\pdd{\tilde{\gamma}_{ij}}{n} = x_p n_p \tilde{N}_{ij} 
\text{ on } S_\text{ss}, \label{eq:gs3}\\
\tilde{C}_{ijkl} = \frac{1}{2 \Vc} \int \tilde{\gamma}_{kl} 
\pdd{\tilde{X}_{ij}}{n} \; \d 
S. \label{eq:gs4}
\end{gather}
The tensors $\tilde{N}_{ij}$ and $\tilde{X}_{ij}$ are the trace-free parts 
of the outer product of the normal vector $\mathbf{n}$ with itself, and 
the 
position vector $\mathbf{x}$ with itself respectively, i.e.
\begin{gather}
 \tilde{N}_{ij} \equiv n_i n_j - \frac{\delta_{ij}}{N}, \label{eq:Ndef}\\
\tilde{X}_{ij} \equiv x_i x_j - x_p x_p \frac{\delta_{ij}}{N}. 
\label{eq:Xdef}
\end{gather}
Note that \eqref{eq:g4} has been simplified to \eqref{eq:gs4} using the 
identity
\begin{equation}
 n_i x_j + n_j x_i = \pdd{}{n}\left(x_i x_j\right).
\end{equation}

\subsubsection{Isotropy}

For situations where the fourth-rank tensor $C_{ijkl}$ is isotropic, the 
expressions in \eqref{eq:gb4} and \eqref{eq:gs4} can be simplified further 
and written in terms of bulk ($\zeta$) and shear ($\eta$) viscosities. 
Assuming isotropy, the tensors $B_{ij}$ and $\tilde{C}_{ijkl}$ can be 
written as
\begin{gather}
 B_{ij} = \zeta \delta_{ij}, \\
\tilde{C}_{ijkl} =\eta \left( \delta_{ik} \delta_{jl} + \delta_{il} 
\delta_{jk}- 
\frac{2}{N} \delta_{ij} \delta_{kl} \right), \label{eq:ctilde_iso}
\end{gather}
where expressions for the bulk and shear viscosities can be simplified to
\begin{gather}
\zeta = \frac{1}{N^2 \Vc} \int \gamma  x_k n_k\; \d 
S, \label{eq:gbp4} \\
\eta = \frac{1}{2\Vc (N+2)(N-1)} \int \tilde{\gamma}_{kl} 
\pdd{\tilde{X}_{kl}}{n} \; \d 
S, \label{eq:gsp4}
\end{gather}
using the relationship  $\tilde{C}_{klkl} = \eta (N+2) (N-1)$ from 
\eqref{eq:ctilde_iso}.

\subsection{Pure solid}\label{sec:nhpure}

When the melt phase is absent, the equations of Nabarro-Herring creep 
simplify considerably. Only shear deformation is permissible, and hence it 
suffices to consider only the problem for $\tilde{C}_{ijkl}$, which is 
given by (\ref{eq:gs1}-\ref{eq:gs4}). The boundary condition 
\eqref{eq:gs2} 
is no longer needed as there is no melt phase. Since both 
$\tilde{\gamma}_{ij}$ and $\tilde{X}_{ij}$ satisfy Laplace's equation, 
Green's second identity allows us to rewrite \eqref{eq:gs4} as
\begin{equation}
\tilde{C}_{ijkl} =  \frac{1}{2 \Vc} \int \tilde{X}_{ij}  
\pdd{\tilde{\gamma}_{kl}}{n}
 \; \d S.  \label{eq:almostcijkl}
\end{equation}
For a pure solid, the boundary condition \eqref{eq:gs3} applies to the 
whole boundary of the grain. Substituting \eqref{eq:gs3} into 
\eqref{eq:almostcijkl} yields
\begin{equation}
 \tilde{C}_{ijkl} = \frac{1}{2 \Vc} \int x_p n_p \tilde{X}_{ij} 
\tilde{N}_{kl} \; 
\d 
S, \label{eq:cijkl} 
\end{equation}
where $\tilde{N}_{kl}$  and $\tilde{X}_{ij}$ are given by \eqref{eq:Ndef} 
and \eqref{eq:Xdef}. The above expression is a key result, as it enables 
one to calculate the 
viscosity tensor for Nabarro-Herring creep for a pure solid without 
explicitly solving Laplace's equation: all one needs to do is evaluate the 
geometric integral in \eqref{eq:cijkl}.

\subsubsection{Isotropic examples}

If the fourth rank tensor $C_{ijkl}$ is isotropic, then the result in 
\eqref{eq:cijkl} can be written in terms of a shear viscosity as
\begin{equation}
 \eta = \frac{1}{2 (N+2)(N-1) \Vc} \int \left( \left(\x \cdot \n\right)^2 
- 
\frac{1}{N} \left(\x \cdot \x \right) \right) \x \cdot \n  \; \d S. 
\label{eq:shearvisc}
\end{equation}

As a simple example of the use of this formula, consider the unit sphere 
embedded in $N$ dimensional space ($N=3$ corresponds to an ordinary 
sphere, 
$N=2$ corresponds to a circle). In this case 
$\x 
= \n$ on the surface of the sphere, and $\x \cdot \x = 1$. The integral 
reduces 
to
\begin{equation}
 \eta = \frac{S}{2 \Vc N (N+2)},
\end{equation}
where $S$ is the surface area. $S/\Vc$ for the unit sphere is $N$, and 
hence
\begin{equation}
 \eta = \frac{1}{2 (N+2)}.
\end{equation}
For an ordinary sphere ($N=3$), $\eta = 1/10$, which is 
\citet{Herring1950}'s classic result.  The corresponding result for a 
circle ($N=2$) is $\eta = 1/8$. The formula in \eqref{eq:shearvisc} can be 
used to directly obtain the shear viscosity for any grain geometry that 
leads to an isotropic viscosity tensor, including the case of hexagonal 
grains. 

\subsubsection{An anisotropic example: orthorhombic grains}

For geometries where the viscosity tensor is anisotropic (such as 
tetrakaidecahedrons), the expression in \eqref{eq:cijkl} should be used 
to calculate the viscosity tensor. As an example, consider an orthorhombic 
grain (a cuboid), with side lengths $A_1$, $A_2$, and $A_3$. The fourth 
rank tensor $\tilde{C}_{ijkl}$ has orthotropic symmetry. The integrals in 
\eqref{eq:cijkl} yield 
\begin{gather}
\tilde{C}_{1111} = \frac{1}{108} \left(4 A_1^2 + A_2^2 + A_3^2 \right), \\
\tilde{C}_{1122} = \frac{1}{108} \left(-2 A_1^2 -2 A_2^2 + A_3^2 \right), 
\\
\tilde{C}_{1212} =  0, 
\end{gather}
where the remaining components of $\tilde{C}_{ijkl}$ can be found by 
appropriate relabelling of indices and the orthotropic symmetry. The above 
results agree with those found previously by \citet{Lifshitz1963} and 
\citet{Greenwood1985}, but were determined without explicitly solving 
Laplace's equation. An important special case of the orthorhombic grain is 
the unit cube, which has $A_1 = A_2 = A_3 = 1$.  As in \eqref{eq:Voigt}, 
the tensor $\tilde{C}_{ijkl}$ with cubic symmetry can then be written in 
terms of two shear viscosities, given by
\begin{gather}
 \eta_1 = \frac{1}{24}, \quad \quad \eta_2 = 0.
\end{gather}
As discussed by \citet{Lifshitz1963}, the vanishing of $\eta_2$ arises 
from a degeneracy in the assumed packing of grains, which allows the array 
of grains to be freely sheared in some directions without the need for 
diffusive transport of matter. Note that \citet{Greenwood1992} argues for 
a non-zero $\eta_2$ for cubic grains, but this is based on an erroneous 
assumption that no minima in the creep strength occurs for principal axes 
of shear between those parallel to the principal axes of the orthorhombic 
grain. The packing of tetrakaidecahedral grains considered here has no 
such degeneracy, and both $\eta_1$ and $\eta_2$ are non-zero.

\subsection{Asymptotics for small porosity and hexagonal grains}

%
%
%

\subsubsection{Bulk viscosity}\label{sec:nh_bulk_asm}

For the case of hexagonal grains, it is possible to make further 
analytical 
progress when the porosity is non-zero, but small. This can be done 
through 
matched asymptotics, and the analysis presented here closely follows a 
related analysis for the Poisson equation by \citet{Ward2010}.

To determine an asymptotic expansion for bulk viscosity, we need to 
analyse 
\eqref{eq:gb1}, \eqref{eq:gb2}, \eqref{eq:gb3} and \eqref{eq:gbp4} in the 
limit of infinitesimally small pores (vanishing porosity). The problem can 
be divided into two parts: First, an outer problem which gives a 
description at the grain scale, where the pores can be treated as point 
sources/ point sinks at the vertices of the hexagon. Second, an inner 
problem, which zooms in on the pore scale behaviour. A matching procedure 
then joins the two problems in an asymptotically consistent manner.

Dimensionless units are chosen so that the distance between opposite sides 
of the hexagon is 1. This leads to an area $\Vc = \sqrt{3}/2$, perimeter 
$S=2 \sqrt{3}$, side length $a= 1/\sqrt{3}$, and perpendicular distance 
$\mathbf{x} \cdot 
\mathbf{n} = 1/2$ around the edges of the hexagon. The outer problem can 
be 
stated as
\begin{gather}
 \nabla^2 \psi = \sum_{k = 1}^{6} q \, \delta \left(\mathbf{x} - 
\mathbf{x}^{(k)} \right) \text{ in }V_\text{hex}, \label{eq:gbo1} \\
\pdd{\psi}{n} = \frac{1}{2}
\text{ on } S_\text{hex}, \label{eq:gbo2}\\
\zeta \sim \frac{1}{4 \sqrt{3}} \int \psi \; \d 
S. \label{eq:gbo3}
\end{gather}
\eqref{eq:gbo1} is the equivalent of \eqref{eq:gb1}, except that point 
sources have been introduced at the vertices of the hexagon to represent 
the infinitesimally small pores. Each source has strength $q$ (to be 
determined), and the vector $\mathbf{x}^{(k)}$ gives the position vector 
of 
the $k^\text{th}$ vertex of the hexagon ($k = 1, 2, \ldots, 6$). 
\eqref{eq:gbo2} follows directly from \eqref{eq:gb3}, and \eqref{eq:gbo3} 
follows directly from \eqref{eq:gbp4}. 

$q$ can be determined from a balance of flux. The total flux out of the 
sides of the hexagon is given by integrating \eqref{eq:gbo2} over the 
boundary. This flux must be matched by the flux produced by the point 
sources at the vertices, and hence $q = \sqrt{3}/2$. Since the outer 
problem involves only Neumann boundary conditions, its solution is unique 
only up to a constant. This constant will be determined by an appropriate 
matching to the inner problem, by considering the asymptotic behaviour in 
the neighbourhood of the point sources. This asymptotic behavior is given 
for the outer problem by
\begin{equation}
 \psi(\mathbf{x}^{(k)}) \sim \frac{q}{2 \pi} \log \left\vert \mathbf{x} - 
\mathbf{x}^{(k)} \right \vert + q r \label{eq:outerexp}
\end{equation}
where the constant $r$ will be determined by matching to the inner 
solution. The problem in (\ref{eq:gbo1}-\ref{eq:gbo3}) can be solved 
numerically to yield
\begin{equation}
 \zeta \sim \frac{q r}{2} + B \label{eq:zeta_asym}
\end{equation}
where the numerical constant $B = -0.177431$.

The inner problem arises from a rescaling of the governing equations. Near 
$\mathbf{x} = \mathbf{x}^{(k)}$ we introduce the inner variable 
\begin{equation}
 \mathbf{y} = \frac{\mathbf{x} - \mathbf{x}^{(k)}}{\epsilon} \label{eq:y}
\end{equation}
where the length scale $\epsilon$ is chosen such that the area of an 
individual pore is $\pi$ in the inner co-ordinates. For a dihedral angle 
of 
$180^\circ$, $\epsilon$ is equal to the radius of the circular pore. The 
length scale $\epsilon$ is related to the porosity by
\begin{equation}
 \epsilon = \left(\frac{\sqrt{3} \phi}{4 \pi} \right)^{1/2}. 
\label{eq:epstopor}
\end{equation}
The inner problem is
\begin{gather}
 \nabla_\mathbf{y}^2 \varphi = 0 \text{ in }V, \label{eq:gbi1} \\
\varphi = 0 \text{ on } S_\text{sl}, \label{eq:gbi2} \\
\varphi \sim \alpha \log \left\vert \mathbf{y} \right\vert, \quad 
\text{as} 
\; \left\vert \mathbf{y} \right\vert \rightarrow \infty. \label{eq:gbi3}
\end{gather}
\eqref{eq:gbi1} follows from \eqref{eq:gb1}, and \eqref{eq:gbi2} from 
\eqref{eq:gb2}. We seek solutions that have a logarithmic singularity in 
the far-field \eqref{eq:gbi3} in order to match with the outer solution. 
The unique solution of inner problem has the following far-field 
asymptotic 
behaviour:
\begin{equation}
 \varphi \sim \alpha \log \left\vert \mathbf{y} \right\vert - \alpha \log 
\logcap + \cdots \quad \text{as} \; \left\vert \mathbf{y} 
\right\vert \rightarrow \infty \label{eq:innerexp}
\end{equation}
where $\logcap$ is known as the logarithmic capacity. The 
logarithmic capacity is a function of the shape of the pore, and hence 
here 
is a function of the dihedral angle $\theta$ (Figure \ref{fig:logcap}).

Matching inner \eqref{eq:innerexp} and outer \eqref{eq:outerexp} solutions 
implies that $\alpha = q/(2 \pi)$ and 
\begin{equation}
 r = - \frac{1}{2 \pi} \log \epsilon \logcap. \label{eq:r}
\end{equation}
Substituting \eqref{eq:r} into \eqref{eq:zeta_asym} yields the small 
$\epsilon$ asymptotic behaviour of the bulk viscosity,
\begin{equation}
 \zeta \sim  - \frac{\sqrt{3}}{8 \pi} \log \epsilon \logcap
-0.177431. \label{eq:zeta_e_asym}
\end{equation}

Using the relationship \eqref{eq:epstopor} between $\epsilon$ and porosity 
$\phi$, \eqref{eq:zeta_e_asym} can also be rewritten in terms of porosity 
as
\begin{equation}
 \zeta \sim  - \frac{\sqrt{3}}{16 \pi} \log \phi  - 0.109145 - 
\frac{\sqrt{3}}{8 \pi} \log \logcap.
\end{equation}

\subsubsection{Shear viscosity}\label{sec:nh_shear_asm}

A similar matched asymptotic analysis as performed above for the bulk 
viscosity 
can be performed for the shear viscosity. A key difference in the analysis 
is that while the bulk viscosity is singular as the porosity vanishes, the 
shear viscosity is finite. Indeed the shear viscosity in the absence of 
melt can be obtained directly from the integral expression 
\eqref{eq:shearvisc} to give a value without melt of $\eta_0 = 1/36$. In 
this section we obtain the next term in the asymptotic expansion, for a 
small, but finite, amount of melt.

The analysis proceeds as in the previous section by dividing the problem 
up 
into two parts: an outer problem for a tensor $\tilde{\psi}_{ij}$ and and 
inner problem for a tensor $\tilde{\varphi}_{ij}$. Furthermore, it is 
helpful to subdivide the outer problem as $\tilde{\psi}_{ij} = 
\tilde{W}_{ij} + \tilde{S}_{ij}$, where $\tilde{W}_{ij}$ is the solution 
for a pure solid, and $\tilde{S}_{ij}$ is the singular perturbation due to 
the melt pores. Similarly, the shear viscosity can be subdivided as $\eta 
= 
\eta_0 + \eta_S$, where $\eta_S$ represents the singular perturbation to 
the shear viscosity.

The problem for the pure solid is 
\begin{gather}
 \nabla^2 \tilde{W}_{ij} = 0 \text{ in }V_\text{hex}, \label{eq:gszero1} \\
\pdd{\tilde{W}_{ij}}{n} = x_p n_p \tilde{N}_{ij} 
\text{ on } S_\text{hex}, \label{eq:gszero2}
\end{gather}
where \eqref{eq:gszero1} follows from \eqref{eq:gs1}, and 
\eqref{eq:gszero2} follows from \eqref{eq:gs3}. As mentioned above, to 
determine $\eta_0$ is not necessary to solve for $\tilde{W}_{ij}$, but for 
the asymptotics that follows it is necessary to determine the values of 
$\tilde{W}_{ij}$ at the vertices of the hexagon. The values at the 
$k^\text{th}$ vertex $\mathbf{x}^{(k)}$ can be obtained numerically as  
\begin{equation}
 \tilde{W}_{ij} \left(\mathbf{x}^{(k)} \right) = F \tilde{X}^{(k)}_{ij} 
\label{eq:wvert}
\end{equation}
where
\begin{equation}
 \tilde{X}^{(k)}_{ij} \equiv x_i^{(k)} x_j^{(k)} - x_p^{(k)} x_p^{(k)} 
\frac{\delta_{ij}}{N}
\end{equation}
and the numerical constant $F = 0.319889078$.

The singular part of the outer problem is
\begin{gather}
 \nabla^2 \tilde{S}_{ij} = \sum_{k = 1}^{6} q \tilde{X}^{(k)}_{ij} \, 
\delta \left(\mathbf{x} - \mathbf{x}^{(k)} \right) \text{ in 
}V_\text{hex}, 
\label{eq:gso1} \\
\pdd{\tilde{S}_{ij}}{n} = 0
\text{ on } S_\text{hex}, \label{eq:gso2}\\
\eta_S = \frac{1}{4 \sqrt{3}} \int \tilde{S}_{kl} \pdd{\tilde{X}_{kl}}{n} 
\; \d 
S. \label{eq:gso3}
\end{gather}
Similar to the bulk viscosity problem, \eqref{eq:gso1} is simply 
\eqref{eq:gs1} with the addition of a series of point sources of strength 
$q \tilde{X}^{(k)}_{ij}$ at the vertices of the hexagon. The choice of 
tensorial form $\tilde{X}^{(k)}_{ij}$ is motivated by solution of the 
regular part of the problem at the vertices. \eqref{eq:gso2} follows from 
\eqref{eq:gs3}, and \eqref{eq:gso3} from \eqref{eq:gsp4}. Unlike the outer 
problem for the bulk viscosity, the problem for $\tilde{S}_{ij}$ is unique 
if $q$ is given, since the tensor must have zero trace 
($\tilde{S}_{kk}=0$). But in this case, the prefactor $q$ must be 
determined by matching to the inner solution.

Green's second identity can be used to simplify \eqref{eq:gso3} to
\begin{equation}
 \eta_S = - \frac{q}{12 \sqrt{3}} \sum_{k=1}^{6} \tilde{X}_{kl} 
\tilde{X}_{kl} = - \frac{q}{36 \sqrt{3}}, \label{eq:visc_q}
\end{equation}
and hence the viscosity can be determined once $q$ is determined. The 
asymptotic behaviour of $\tilde{S}_{ij}$ near the vertices is
\begin{equation}
 \tilde{S}_{ij}(\mathbf{x}^{(k)}) \sim \frac{q \tilde{X}^{(k)}_{ij}}{2 
\pi} 
\log \left\vert \mathbf{x} - \mathbf{x}^{(k)} \right \vert + q R 
\tilde{X}^{(k)}_{ij} \label{eq:shear_outer_asym}
\end{equation}
where $R$ is determined by numerical computation as $R=0.150237305$. 

The inner problem uses the same scaled variable $\mathbf{y}$ given in 
\eqref{eq:y}, and is
\begin{gather}
 \nabla_\mathbf{y}^2 \tilde{\varphi}_{ij} = 0 \text{ in }V, 
\label{eq:gsi1} 
\\
\tilde{\varphi}_{ij} = 0 \text{ on } S_\text{sl}, \label{eq:gsi2} \\
\tilde{\varphi}_{ij} \sim \alpha \tilde{X}^{(k)}_{ij} \log \left\vert 
\mathbf{y} \right\vert, \quad \text{as} \; \left\vert \mathbf{y} 
\right\vert \rightarrow \infty.
\end{gather}
The unique solution of inner problem has far-field asymptotic behaviour
\begin{equation}
\tilde{\varphi}_{ij} \sim \alpha \tilde{X}^{(k)}_{ij} \log \left\vert 
\mathbf{y} \right\vert - \alpha \tilde{X}^{(k)}_{ij} \log \logcap  + 
\cdots \quad \text{as} \; \left\vert \mathbf{y} \right\vert 
\rightarrow \infty \label{eq:shear_inner_asym}
\end{equation}
Matching between the inner \eqref{eq:shear_inner_asym} and outer 
(\eqref{eq:wvert} and \eqref{eq:shear_outer_asym}) solutions yields
\begin{gather}
 \alpha = \frac{q}{2 \pi}, \\
F + q R = - \alpha \log \epsilon \logcap,
\end{gather}
and hence
\begin{equation}
 q = - \frac{2 \pi F}{2 \pi R + \log \epsilon \logcap}. \label{eq:q}
\end{equation}
Substituting \eqref{eq:q} into \eqref{eq:visc_q} yields the asymptotic 
expression for the shear viscosity for small $\epsilon$,
\begin{equation}
 \eta \sim \frac{1}{36} + \frac{2 \pi F}{36 \sqrt{3} \left(2 \pi R + \log 
\epsilon \logcap \right)},
\end{equation}
where the numerical constants are $F = 0.319889078$ and $R=0.150237305$.

\subsection{Leading-order bulk viscosity asymptotics for general grain 
shapes}\label{sec:leadingorder}

Part of the asymptotic analysis given in section \ref{sec:nh_bulk_asm} for 
hexagonal grains can be applied in three-dimensions to obtain the 
leading-order behaviour of the bulk viscosity for more general grain 
shapes. Instead of the outer problem consisting of point sources of 
strength $q$ at the vertices of the hexagon, we now consider line sources 
of strength $q$ along grain edges in 3D. This assumes melt forms a 
connected network along the grain edges at vanishing porosity, which will 
only be true for dihedral angles less than 60$^\circ$. Integration of 
\eqref{eq:gb3} around the surface of the grain gives the net flux out of 
the grain faces as $Q=N V_\text{cell}$. This must match the flux produced 
by the line sources, $Q=q L$, where $L$ is the total length of edges for 
the unit cell (an effective length accounting for the fact that each edge 
is shared by more than one grain). Hence
\begin{equation}
 q = \frac{N}{\lambda},
\end{equation}
where $\lambda\equiv L/V_\text{cell}$ is the edge length per unit volume 
in 
the tiling of grains. For tetrakaidecahedrons, $\lambda = 6 \sqrt{2}$ in 
units where the distance between opposite square faces of the 
tetrakaidecahedrons is 1. The equivalent measure for hexagonal grains in 
2D 
is the density of vertices per unit area, where $\lambda=4/\sqrt{3}$ in 
units where the distance between opposite sides of the hexagons is 1.

The asymptotic behaviour in the neighbourhood of the line sources in 3D 
will take the same form as \eqref{eq:outerexp}, with bulk viscosity given 
in $N$-dimensions as
\begin{equation}
 \zeta \sim \frac{q r}{N} \label{eq:zetaN}
\end{equation}
to leading order (this is the $N$-dimensional counterpart to 
\eqref{eq:zeta_asym}).  A similar matching to that in \eqref{eq:r} yields 
\begin{equation}
 r \sim - \frac{1}{4 \pi} \log \phi \label{eq:newr}
\end{equation}
at leading order. Substituting \eqref{eq:newr} into \eqref{eq:zetaN} gives 
the leading order asymptotics
\begin{equation}
 \zeta \sim - \frac{1}{4 \pi \lambda} \log \phi 
\end{equation}
where the next term in the asymptotic expansion will depend on the exact 
geometry of the melt. For the specific case of tetrakaidecahedrons with 
distance 1 between square faces,
\begin{equation}
  \zeta \sim - \frac{1}{24 \pi \sqrt{2}} \log \phi.
\end{equation}

\subsection{Parametrisation of Nabarro-Herring creep} \label{sec:nh_param}

It is useful to have a simple parametrisation of viscosity as a function 
of 
porosity which can be used in larger-scale models. What follows is a 
suggested parametrisation for Nabarro-Herring creep, which provides a 
close 
approximation to the tetrakaidecahedron calculations here for a dihedral 
angle of $40^\circ$ (appropriate for basaltic melts) and porosities up to 
around 10\%, where melt exists as 
a connected network along the grain edges. The formulas below are 
least-squares polynomial fits in terms of a variable $\nu \equiv -1/\log 
\phi$, where this choice of variable has been motivated by the asymptotic 
analysis above. The leading-order term in the fits is provided the 
leading-order asymptotics for small porosity. The higher-order terms have 
been found by fitting the numerical results shown in Figures 
\ref{fig:herring_bulk} and \ref{fig:herring_shear}. The parametrisation is
\begin{gather}
 \frac{\eta}{\eta_0} = 1 + a_1 \nu + a_2 \nu^2 + a_3 \nu^3, 
\label{eq:nhapprox_shear}\\
\frac{\zeta}{\eta} = \frac{160 \sqrt{2}}{139 \pi} \frac{1}{\nu} + b_1 + 
b_2 
\nu 
+ b_3 \nu^2, \label{eq:nhapprox_bulk}
\end{gather}
where the fitting constants are
\begin{gather}
a_1= -4.28207265, \quad a_2=  7.36988663, \quad a_3 = -4.98396638,\\ 
b_1 =  0.97736898, \quad b_2 =  -1.76154195, \quad b_3 = 2.63720462.
\end{gather}

\section{Coble creep analytical solutions}

\subsection{Pure solid, Hexagonal grains}\label{sec:coble_hex_pure}

Simple analytical solutions exist for Coble creep in arrays of regular 
hexagons 
\citep{Spingarn1978,Cocks1990}. The Poisson equation for the tensor $\g$ 
in 
\eqref{eq:gc1} can be written in co-ordinates along a single edge of the 
hexagon as
\begin{equation}
 - \nablas^2 \g = \begin{pmatrix}
                            0 & 0 \\
                            0 & 1 
                           \end{pmatrix} \label{eq:component_eq}
\end{equation}
where the horizontal co-ordinate is along an edge and the vertical 
coordinate 
is perpendicular to the edge (i.e. the normal vector is $\mathbf{n}= 
(0,1)$). 
Since the edge is straight the surface Laplacian operator $\nablas^2$ 
becomes 
$\d^2/\d s^2$ where $s$ measures distance along an edge. $s$ is chosen to 
be 
zero in the middle of an edge, and equal to $h$ at one end and $-h$ at the 
other end. In the absence of melt $h=1/(2 \sqrt{3})$ with lengths scaled 
such that the distance between opposite sides of the hexagon is 1. 
\eqref{eq:component_eq} can be 
integrated 
as
\begin{equation}
 \g = \begin{pmatrix}
                            a_{11} + b_{11} s & a_{12} + 
b_{12} s \\
                            a_{12} + b_{12} s & a_{22} + 
b_{22} s - \tfrac{1}{2} s^2 
                           \end{pmatrix} \label{eq:gam_com}
\end{equation}
for constants $a_{ij}$ and $b_{ij}$ that will be determined by 
the 
boundary conditions. 

\eqref{eq:gam_com} can be simplified immediately by 
the 
mirror symmetry of the hexagon along a plane through the middle of an 
edge, 
which yields
\begin{equation}
 \g = \begin{pmatrix}
                            a_{11}  &  
b_{12} s \\
                             b_{12} s & a_{22} - \tfrac{1}{2} s^2 
                           \end{pmatrix}. \label{eq:gmat}
\end{equation} 
By the rotational symmetry of the hexagon, the solution for $\g$ on the 
other 
5 edges is the same, up to a suitable rotation of the coordinates. Thus in 
what 
follows we find the solution just for one edge of the hexagon, exploiting 
the symmetry to match this solution to the solutions on the other edges.

\subsubsection{Isolated grain}

In the absence of melt, what the effective viscosity is depends very much 
on 
the nature of the assumed boundary conditions. Since there is no melt 
present, 
the medium is incompressible, and this restricts the discussion to the 
trace-free part of $\g$, which is denoted as $\tilde{\g}$. Following on 
from  
\eqref{eq:gmat}, this can be written in terms of just two constants, 
$a$ 
and $b$,
\begin{equation}
 \tilde{\g} = \begin{pmatrix}
                            -a + \tfrac{1}{4} s^2  &  
b s \\
                             b s & a - \tfrac{1}{4} s^2 
\label{eq:gtmat}
                           \end{pmatrix}
\end{equation}
where the constants will be determined by the choice of boundary 
condition. 
Given the rotational symmetry of the hexagon, it is helpful to describe 
the 
boundary conditions using the rotation matrix
\begin{equation}
 \tensor{R}_\theta = \begin{pmatrix}
             \cos \theta & -\sin \theta \\
	    \sin \theta & \cos \theta
            \end{pmatrix}.
\end{equation}
If one considers each individual 
grain to be isolated, and grain-boundary diffusion to be restricted to 
each 
individual grain (as \citet{Takei2009} do), then at each corner of the 
grain 
we must have continuity of vacancies (chemical potential) and continuity 
of 
flux. Continuity of vacancies can be written as
\begin{equation}
\tilde{\g}({s= h}) =  \tensor{R}_{-\pi/3} \cdot
\tilde{\g}({s= -h}) \cdot \tensor{R}_{-\pi/3}^T \label{eq:isobc1}
\end{equation}
which simply ensures that the concentration at the end of one edge matches 
the 
start of the next. By symmetry the solution on one edge is related to the 
solution on another edge simply by a rotation of co-ordinates.
Continuity of flux is
\begin{equation}
\dd{\tilde{\g}}{s}({s= h}) =  \tensor{R}_{-\pi/3} \cdot
\dd{\tilde{\g}}{s}({s= -h}) \cdot \tensor{R}_{-\pi/3}^T, \label{eq:isobc2}
\end{equation}
since the flux is given by the derivative of concentration along each 
edge. 
\eqref{eq:isobc1} and \eqref{eq:isobc2} determine the constants 
in \eqref{eq:gtmat} as
\begin{equation}
 a = \frac{5}{144}, \quad \quad b = \frac{1}{12}.
\end{equation}

\subsubsection{Networked}

Alternatively, if the grain boundaries form a connected network along 
which 
diffusion can occur, then the boundary conditions are slightly different. 
Continuity of vacancies is unchanged and given by \eqref{eq:isobc1}, but 
continuity of flux becomes
\begin{equation}
\dd{\tilde{\g}}{s}({s= h}) =  \tensor{R}_{-\pi/3} \cdot
\dd{\tilde{\g}}{s}({s= -h}) \cdot \tensor{R}_{-\pi/3}^T + 
\tensor{R}_{\pi/3} 
\cdot
\dd{\tilde{\g}}{s}({s= -h}) \cdot \tensor{R}_{\pi/3}^T \label{eq:netbc2}
\end{equation}
because the flux balance is now between the three edges that meet at each 
triple junction (Figure \ref{fig:hex_coble_pure}). \eqref{eq:isobc1} and 
\eqref{eq:netbc2} determine the 
constants in \eqref{eq:gtmat} as
\begin{equation}
 a = \frac{1}{48}, \quad \quad b = 0.
\end{equation}

Once $\tilde{\g}$ is determined it is straightforward to integrate 
\eqref{eq:gc2} to find the relevant viscosities. For the matrix 
given in \eqref{eq:gtmat}, the shear viscosity is
\begin{equation}
 \eta = \frac{a}{2} + \frac{b}{36} - \frac{1}{288},
\end{equation}
and hence $\eta = 7/432$ for a pure solid under the isolated assumption 
and 
$\eta = 1/144$ for a pure solid under the networked assumption.

%

\subsection{Coble creep with melt}\label{sec:coble_melt}

In the presence of melt, the equations of Coble creep can be further 
simplified. The solution to \eqref{eq:gc1} can be broken down into a 
series 
of isolated Poisson problems on each individual grain--grain contact. 
Moreover, since we are assuming the boundaries are planar, we can write 
these individual Poisson problems in terms of a single scalar variable 
$\varphi$, since the normal vector $\mathbf{n}$ is a constant over each 
grain--grain contact. Writing $\gamma_{ij} = \varphi n_i n_j$, 
\eqref{eq:gc1} and \eqref{eq:gc2} become
\begin{gather}
 -\nablas^2 \varphi= 2 x_p n_p, \label{eq:gc3} \\
C_{ijkl} = \frac{1}{\Vc} \int \varphi  x_i n_j  n_k n_l \; \d 
S. \label{eq:gc4} 
\end{gather}
The tensor $C_{ijkl}$ satisfies the following symmetry, known as the 
Cauchy 
relation,
\begin{equation}
 C_{ijkl} = C_{ikjl}.
\end{equation}
As a consequence of the above symmetry, for isotropic cases the following 
relationship holds between the bulk and shear viscosities,
\begin{equation}
 \frac{\zeta}{\eta} = \frac{N+2}{N}, \label{eq:Cauchy_result}
\end{equation}
so $\zeta = 2 \eta$ in 2D, and $\zeta = \tfrac{5}{3} \eta$ in 3D.

The integral in \eqref{eq:gc4} can be further simplified by considering 
the 
integral as a sum over the contributions from each grain--grain contact,
\begin{equation}
 C_{ijkl} = \frac{1}{\Vc} \sum_{f} n^f_i n^f_j n^f_k n^f_l d_f \int 
\varphi \; \d 
S, \label{eq:gc4b}
\end{equation}
where $f$ is an index used to denote the individual grain--grain contacts, 
and $\mathbf{n}^f$ are the corresponding normals to those contacts. $d_f$ 
are the perpendicular distances of the contacts from the origin. 
\eqref{eq:gc4b} has been obtained from \eqref{eq:gc4} by pulling out the 
$n_j n_k n_l$ factor in front of the integral sign, which can be done 
because $\mathbf{n}$ is constant on each contact. What remains in the 
integrand of \eqref{eq:gc4} is then $\varphi x_i$, which yields a vector 
quantity for the integral. However, provided the contact face has at least 
one non-trivial symmetry (e.g. symmetric under reflection or rotation), 
this vector quantity must be proportional to the normal vector, and hence 
a 
factor of $n_i$ can also be pulled in front of the integral to yield the 
final scalar integral expression in \eqref{eq:gc4b}.     

The right hand side of \eqref{eq:gc3} is constant over each contact, equal 
to twice the perpendicular distance from the origin. Hence the procedure 
for calculating $C_{ijkl}$ consists of solving a series of Poisson 
problems 
with constant right-hand side for each face, and then integrating the 
resulting solution over the contact. This is identical to the problem of 
finding the hydraulic resistance in Hagen-Poiseuille flow in a pipe of 
given cross-sectional shape \citep{Mortensen2005,Bazant2016}. The 
hydraulic 
resistance depends both on the cross-sectional area and the particular 
cross-sectional shape of the pipe. The same is true here of the effective 
viscosity tensor for Coble creep, which depends both on the areas of the 
contact patches and their shapes. The two effects can be separated by 
introducing the following scaled problem on each contact face,
\begin{gather}
 -\nablas^2 w_f = 1, \label{eq:s1} \\
\chi_f = \frac{1}{A_f^{(N+1)/(N-1)}} \int w_f \; \d 
S, \label{eq:s2} 
\end{gather}
where $\chi_f$ is a shape factor, defined such that it is independent of 
the area of the contact. On each contact $\varphi = 2 d_f w_f$. The 
viscosity tensor can then be written
\begin{equation}
 C_{ijkl} = \frac{2}{\Vc} \sum_{f} n_i^f n_j^f n_k^f n_l^f d_{f}^2 
\chi_f A_f^{(N+1)/(N-1)}. \label{eq:coblecmelt}
\end{equation}
For the tetrakaidecahedral geometry considered here the contact faces go 
from being squares and hexagons at zero porosity, to being more circular 
at 
higher porosity. For a circle $\chi = 1/(8\pi) = 0.0397887$, for a 
hexagon $\chi=0.0383503$, and for a square $\chi= 0.0351443$. $\chi$ 
is very similar for all three end-member shapes, so the shape effects on 
the viscosity are likely to be modest. The principal control on the 
viscosity tensor is contact area. In 3D, the viscosity is proportional to 
$A_f^2$ and in 2D, the viscosity is proportional to $A_f^3$. This agrees 
with the results of \citet{Cooper1984} and \citet{Takei2009}, and indeed, 
\eqref{eq:coblecmelt} is 
a generalisation of (42) of \citet{Takei2009} to a more general geometry. 

The bulk viscosity can be obtained from \eqref{eq:coblecmelt} as
\begin{equation}
 \zeta = \frac{2}{N^2 \Vc} \sum_{f}  d_{f}^2 \chi_f 
A_f^{(N+1)/(N-1)}. \label{eq:coblebulkmelt}
\end{equation}

\subsubsection{Hexagonal grains}\label{sec:hex_melt}

For the hexagon in 2D, $N=2$, $\Vc = \sqrt{3}/2$, $d_f=1/2$, 
$\chi_f=1/12$, 
and $A_f = \rho/\sqrt{3}$, where $\rho \equiv A_\text{ss}/A_\text{cell}$ 
is the fraction of the unit cell 
boundary 
which is grain--grain contact. The bulk viscosity is given by 
\eqref{eq:coblebulkmelt} as $\zeta = \rho^3 / 72$, in agreement with 
\citet{Cocks1990}. From \eqref{eq:Cauchy_result}, the shear viscosity 
$\eta 
= 
\zeta/2 = \rho^3/144$. Some straightforward trigonometry can be used to 
relate 
$\rho$ 
to $\theta$ and $\phi$ by
\begin{equation}
 \phi = (1-\rho)^2 \phi_d \left( \theta \right) 
\end{equation}
where
\begin{equation}
 \phi_d \left( \theta \right) = \frac{1}{8 \sqrt{3}}  
\csc\left(\tfrac{\pi}{6}-\tfrac{\theta}{2}\right) \left( 4 \sqrt{3} \cos 
\tfrac{\theta}{2} - (\pi - 3 \theta) 
\csc\left(\tfrac{\pi}{6}-\tfrac{\theta}{2}\right) \right).
\end{equation}

\subsubsection{Tetrakaidecahedral grains}\label{sec:tetra_melt}

For tetrakaidecahedrons, \eqref{eq:coblecmelt} can be simplified to give 
the two shear viscosities as
\begin{gather}
 \eta_1 = \frac{1}{3 \Vc} \sum_\text{squares}  d_{f}^2 \chi_f A_f^2, \\
 \eta_2 = \frac{2}{9 \Vc} \sum_\text{hexagons}  d_{f}^2 \chi_f A_f^2,
\end{gather}
where $\eta_1$ is given in terms of a sum over the square faces only, and 
$\eta_2$ is terms of a sum over the hexagonal faces only. The bulk 
viscosity is given by $\zeta = \tfrac{5}{3} \overline{\eta} = \tfrac{2}{3} 
\eta_1 + \eta_2$. In coordinates where the distance between opposite 
square 
faces is 1, $\Vc = 1/2$, $d_\text{sq} = 1/2$, and 
$d_\text{hex}=\sqrt{3}/4$. When there is only an infinitesimal amount of 
melt present,  $A_\text{sq} = 1/8$ and $A_\text{hex} = 3 \sqrt{3}/16$, and 
hence
\begin{gather}
 \eta_{10} = \frac{\chi_{\text{sq}}}{64}, \\
 \eta_{20} = \frac{9\chi_{\text{hex}}}{128},
\end{gather}
and these are the values given in the shorted row of 
Table \ref{tab:tetvisc}.


\begin{thebibliography}{}

\bibitem [\protect \citeauthoryear {%
Alisic%
, Rhebergen%
, Rudge%
, Katz%
\BCBL {}\ \BBA {} Wells%
}{%
Alisic%
\ \protect \BOthers {.}}{%
{\protect \APACyear {2016}}%
}]{%
Alisic2016}
\APACinsertmetastar {%
Alisic2016}%
\begin{APACrefauthors}%
Alisic, L.%
, Rhebergen, S.%
, Rudge, J\BPBI F.%
, Katz, R\BPBI F.%
\BCBL {}\ \BBA {} Wells, G\BPBI N.%
\end{APACrefauthors}%
\unskip\
\newblock
\APACrefYearMonthDay{2016}{}{}.
\newblock
{\BBOQ}\APACrefatitle {{Torsion of a cylinder of partially molten rock with 
a
  spherical inclusion: Theory and simulation}} {{Torsion of a cylinder of
  partially molten rock with a spherical inclusion: Theory and
  simulation}}.{\BBCQ}
\newblock
\APACjournalVolNumPages{Geochemistry, Geophys. 
Geosystems}{17}{1}{143--161}.
\newblock
\begin{APACrefDOI} \doi{10.1002/2015GC006061} \end{APACrefDOI}
\PrintBackRefs{\CurrentBib}

\bibitem [\protect \citeauthoryear {%
Arzt%
, Ashby%
\BCBL {}\ \BBA {} Easterling%
}{%
Arzt%
\ \protect \BOthers {.}}{%
{\protect \APACyear {1983}}%
}]{%
Arzt1983}
\APACinsertmetastar {%
Arzt1983}%
\begin{APACrefauthors}%
Arzt, E.%
, Ashby, M\BPBI F.%
\BCBL {}\ \BBA {} Easterling, K\BPBI E.%
\end{APACrefauthors}%
\unskip\
\newblock
\APACrefYearMonthDay{1983}{}{}.
\newblock
{\BBOQ}\APACrefatitle {{Practical Applications of Hot-lsostatic Pressing
  Diagrams : Four Case Studies}} {{Practical Applications of Hot-lsostatic
  Pressing Diagrams : Four Case Studies}}.{\BBCQ}
\newblock
\APACjournalVolNumPages{Metall. Trans. A}{14A}{}{211--221}.
\newblock
\begin{APACrefDOI} \doi{10.1007/BF02651618} \end{APACrefDOI}
\PrintBackRefs{\CurrentBib}

\bibitem [\protect \citeauthoryear {%
Bazant%
}{%
Bazant%
}{%
{\protect \APACyear {2016}}%
}]{%
Bazant2016}
\APACinsertmetastar {%
Bazant2016}%
\begin{APACrefauthors}%
Bazant, M\BPBI Z.%
\end{APACrefauthors}%
\unskip\
\newblock
\APACrefYearMonthDay{2016}{}{}.
\newblock
{\BBOQ}\APACrefatitle {{Exact solutions and physical analogies for
  unidirectional flows}} {{Exact solutions and physical analogies for
  unidirectional flows}}.{\BBCQ}
\newblock
\APACjournalVolNumPages{Phys. Rev. Fluids}{1}{2}{024001}.
\newblock
\begin{APACrefDOI} \doi{10.1103/PhysRevFluids.1.024001} \end{APACrefDOI}
\PrintBackRefs{\CurrentBib}

\bibitem [\protect \citeauthoryear {%
Beer{\'{e}}%
}{%
Beer{\'{e}}%
}{%
{\protect \APACyear {1976}}%
}]{%
Beere1976}
\APACinsertmetastar {%
Beere1976}%
\begin{APACrefauthors}%
Beer{\'{e}}, W.%
\end{APACrefauthors}%
\unskip\
\newblock
\APACrefYearMonthDay{1976}{}{}.
\newblock
{\BBOQ}\APACrefatitle {{Stress redistribution during Nabarro-Herring and
  superplastic creep}} {{Stress redistribution during Nabarro-Herring and
  superplastic creep}}.{\BBCQ}
\newblock
\APACjournalVolNumPages{Met. Sci.}{10}{4}{133--139}.
\newblock
\begin{APACrefDOI} \doi{doi:10.1179/030634576790431958} \end{APACrefDOI}
\PrintBackRefs{\CurrentBib}

\bibitem [\protect \citeauthoryear {%
Bercovici%
, Ricard%
\BCBL {}\ \BBA {} Schubert%
}{%
Bercovici%
\ \protect \BOthers {.}}{%
{\protect \APACyear {2001}}%
}]{%
Bercovici2001}
\APACinsertmetastar {%
Bercovici2001}%
\begin{APACrefauthors}%
Bercovici, D.%
, Ricard, Y.%
\BCBL {}\ \BBA {} Schubert, G.%
\end{APACrefauthors}%
\unskip\
\newblock
\APACrefYearMonthDay{2001}{}{}.
\newblock
{\BBOQ}\APACrefatitle {{A two-phase model for compaction and damage 1. 
General
  theory}} {{A two-phase model for compaction and damage 1. General
  theory}}.{\BBCQ}
\newblock
\APACjournalVolNumPages{J. Geophys. Res.}{106}{B5}{8887--8906}.
\newblock
\begin{APACrefDOI} \doi{10.1029/2000JB900430} \end{APACrefDOI}
\PrintBackRefs{\CurrentBib}

\bibitem [\protect \citeauthoryear {%
Bercovici%
\ \BBA {} Rudge%
}{%
Bercovici%
\ \BBA {} Rudge%
}{%
{\protect \APACyear {2016}}%
}]{%
Bercovici2016}
\APACinsertmetastar {%
Bercovici2016}%
\begin{APACrefauthors}%
Bercovici, D.%
\BCBT {}\ \BBA {} Rudge, J\BPBI F.%
\end{APACrefauthors}%
\unskip\
\newblock
\APACrefYearMonthDay{2016}{}{}.
\newblock
{\BBOQ}\APACrefatitle {{A mechanism for mode selection in melt band
  instabilities}} {{A mechanism for mode selection in melt band
  instabilities}}.{\BBCQ}
\newblock
\APACjournalVolNumPages{Earth Planet. Sci. Lett.}{433}{}{139--145}.
\newblock
\begin{APACrefDOI} \doi{10.1016/j.epsl.2015.10.051} \end{APACrefDOI}
\PrintBackRefs{\CurrentBib}

\bibitem [\protect \citeauthoryear {%
Coble%
}{%
Coble%
}{%
{\protect \APACyear {1963}}%
}]{%
Coble1963}
\APACinsertmetastar {%
Coble1963}%
\begin{APACrefauthors}%
Coble, R\BPBI L.%
\end{APACrefauthors}%
\unskip\
\newblock
\APACrefYearMonthDay{1963}{}{}.
\newblock
{\BBOQ}\APACrefatitle {{A Model for Boundary Diffusion Controlled Creep in
  Polycrystalline Materials}} {{A Model for Boundary Diffusion Controlled 
Creep
  in Polycrystalline Materials}}.{\BBCQ}
\newblock
\APACjournalVolNumPages{J. Appl. Phys.}{34}{6}{1679--1682}.
\newblock
\begin{APACrefDOI} \doi{10.1063/1.1702656} \end{APACrefDOI}
\PrintBackRefs{\CurrentBib}

\bibitem [\protect \citeauthoryear {%
Cocks%
}{%
Cocks%
}{%
{\protect \APACyear {1996}}%
}]{%
Cocks1996}
\APACinsertmetastar {%
Cocks1996}%
\begin{APACrefauthors}%
Cocks, A\BPBI C\BPBI F.%
\end{APACrefauthors}%
\unskip\
\newblock
\APACrefYearMonthDay{1996}{}{}.
\newblock
{\BBOQ}\APACrefatitle {{Variational principles, numerical schemes and 
bounding
  theorems for deformation by Nabarro-Herring creep}} {{Variational 
principles,
  numerical schemes and bounding theorems for deformation by 
Nabarro-Herring
  creep}}.{\BBCQ}
\newblock
\APACjournalVolNumPages{J. Mech. Phys. Solids}{44}{9}{1429--1452}.
\newblock
\begin{APACrefDOI} \doi{10.1016/0022-5096(96)00040-3} \end{APACrefDOI}
\PrintBackRefs{\CurrentBib}

\bibitem [\protect \citeauthoryear {%
Cocks%
\ \BBA {} Searle%
}{%
Cocks%
\ \BBA {} Searle%
}{%
{\protect \APACyear {1990}}%
}]{%
Cocks1990}
\APACinsertmetastar {%
Cocks1990}%
\begin{APACrefauthors}%
Cocks, A\BPBI C\BPBI F.%
\BCBT {}\ \BBA {} Searle, A\BPBI A.%
\end{APACrefauthors}%
\unskip\
\newblock
\APACrefYearMonthDay{1990}{}{}.
\newblock
{\BBOQ}\APACrefatitle {{Cavity growth in ceramic materials under multiaxial
  stress states}} {{Cavity growth in ceramic materials under multiaxial 
stress
  states}}.{\BBCQ}
\newblock
\APACjournalVolNumPages{Acta Metall. Mater.}{38}{12}{2493--2505}.
\newblock
\begin{APACrefDOI} \doi{10.1016/0956-7151(90)90261-E} \end{APACrefDOI}
\PrintBackRefs{\CurrentBib}

\bibitem [\protect \citeauthoryear {%
Cooper%
}{%
Cooper%
}{%
{\protect \APACyear {1990}}%
}]{%
Cooper1990}
\APACinsertmetastar {%
Cooper1990}%
\begin{APACrefauthors}%
Cooper, R\BPBI F.%
\end{APACrefauthors}%
\unskip\
\newblock
\APACrefYearMonthDay{1990}{}{}.
\newblock
{\BBOQ}\APACrefatitle {{Differential stress-induced melt migration: An
  experimental approach}} {{Differential stress-induced melt migration: An
  experimental approach}}.{\BBCQ}
\newblock
\APACjournalVolNumPages{J. Geophys. Res.}{95}{B5}{6979}.
\newblock
\begin{APACrefDOI} \doi{10.1029/JB095iB05p06979} \end{APACrefDOI}
\PrintBackRefs{\CurrentBib}

\bibitem [\protect \citeauthoryear {%
Cooper%
\ \BBA {} Kohlstedt%
}{%
Cooper%
\ \BBA {} Kohlstedt%
}{%
{\protect \APACyear {1984}}%
}]{%
Cooper1984}
\APACinsertmetastar {%
Cooper1984}%
\begin{APACrefauthors}%
Cooper, R\BPBI F.%
\BCBT {}\ \BBA {} Kohlstedt, D\BPBI L.%
\end{APACrefauthors}%
\unskip\
\newblock
\APACrefYearMonthDay{1984}{}{}.
\newblock
{\BBOQ}\APACrefatitle {{Solution-precipitation enhanced diffusional creep 
of
  partially molten olivine-basalt aggregates during hot-pressing}}
  {{Solution-precipitation enhanced diffusional creep of partially molten
  olivine-basalt aggregates during hot-pressing}}.{\BBCQ}
\newblock
\APACjournalVolNumPages{Tectonophysics}{107}{3-4}{207--233}.
\newblock
\begin{APACrefDOI} \doi{10.1016/0040-1951(84)90252-X} \end{APACrefDOI}
\PrintBackRefs{\CurrentBib}

\bibitem [\protect \citeauthoryear {%
Cooper%
\ \BBA {} Kohlstedt%
}{%
Cooper%
\ \BBA {} Kohlstedt%
}{%
{\protect \APACyear {1986}}%
}]{%
Cooper1986}
\APACinsertmetastar {%
Cooper1986}%
\begin{APACrefauthors}%
Cooper, R\BPBI F.%
\BCBT {}\ \BBA {} Kohlstedt, D\BPBI L.%
\end{APACrefauthors}%
\unskip\
\newblock
\APACrefYearMonthDay{1986}{}{}.
\newblock
{\BBOQ}\APACrefatitle {{Rheology and structure of olivine-basalt partial
  melts}} {{Rheology and structure of olivine-basalt partial 
melts}}.{\BBCQ}
\newblock
\APACjournalVolNumPages{J. Geophys. Res.}{91}{B9}{9315}.
\newblock
\begin{APACrefDOI} \doi{10.1029/JB091iB09p09315} \end{APACrefDOI}
\PrintBackRefs{\CurrentBib}

\bibitem [\protect \citeauthoryear {%
Cooper%
, Kohlstedt%
\BCBL {}\ \BBA {} Chyung%
}{%
Cooper%
\ \protect \BOthers {.}}{%
{\protect \APACyear {1989}}%
}]{%
Cooper1989}
\APACinsertmetastar {%
Cooper1989}%
\begin{APACrefauthors}%
Cooper, R\BPBI F.%
, Kohlstedt, D\BPBI L.%
\BCBL {}\ \BBA {} Chyung, K.%
\end{APACrefauthors}%
\unskip\
\newblock
\APACrefYearMonthDay{1989}{}{}.
\newblock
{\BBOQ}\APACrefatitle {{Solution-precipitation enhanced creep in 
solid-liquid
  aggregates which display a non-zero dihedral angle}} 
{{Solution-precipitation
  enhanced creep in solid-liquid aggregates which display a non-zero 
dihedral
  angle}}.{\BBCQ}
\newblock
\APACjournalVolNumPages{Acta Metall.}{37}{7}{1759--1771}.
\newblock
\begin{APACrefDOI} \doi{10.1016/0001-6160(89)90061-8} \end{APACrefDOI}
\PrintBackRefs{\CurrentBib}

\bibitem [\protect \citeauthoryear {%
Driscoll%
}{%
Driscoll%
}{%
{\protect \APACyear {2002}}%
}]{%
Driscoll2002}
\APACinsertmetastar {%
Driscoll2002}%
\begin{APACrefauthors}%
Driscoll, T\BPBI A.%
\end{APACrefauthors}%
\unskip\
\newblock
\APACrefYearMonthDay{2002}{}{}.
\newblock
\APACrefbtitle {{Schwarz-Christoffel Toolbox User's Guide, Version 2.3}.}
  {{Schwarz-Christoffel Toolbox User's Guide, Version 2.3}.}
\newblock
\begin{APACrefURL} \url{http://www.math.udel.edu/{~}driscoll/SC/}
  \end{APACrefURL}
\PrintBackRefs{\CurrentBib}

\bibitem [\protect \citeauthoryear {%
Driscoll%
\ \BBA {} Trefethen%
}{%
Driscoll%
\ \BBA {} Trefethen%
}{%
{\protect \APACyear {2002}}%
}]{%
Driscoll2002a}
\APACinsertmetastar {%
Driscoll2002a}%
\begin{APACrefauthors}%
Driscoll, T\BPBI A.%
\BCBT {}\ \BBA {} Trefethen, L\BPBI N.%
\end{APACrefauthors}%
\unskip\
\newblock
\APACrefYear{2002}.
\newblock
\APACrefbtitle {{Schwarz-Christoffel Mapping}} {{Schwarz-Christoffel 
Mapping}}.
\newblock
\APACaddressPublisher{}{Cambridge University Press}.
\PrintBackRefs{\CurrentBib}

\bibitem [\protect \citeauthoryear {%
Faul%
\ \BBA {} Jackson%
}{%
Faul%
\ \BBA {} Jackson%
}{%
{\protect \APACyear {2007}}%
}]{%
Faul2007}
\APACinsertmetastar {%
Faul2007}%
\begin{APACrefauthors}%
Faul, U\BPBI H.%
\BCBT {}\ \BBA {} Jackson, I.%
\end{APACrefauthors}%
\unskip\
\newblock
\APACrefYearMonthDay{2007}{}{}.
\newblock
{\BBOQ}\APACrefatitle {{Diffusion creep of dry, melt-free olivine}} 
{{Diffusion
  creep of dry, melt-free olivine}}.{\BBCQ}
\newblock
\APACjournalVolNumPages{J. Geophys. Res. Solid Earth}{112}{B4}{1--14}.
\newblock
\begin{APACrefDOI} \doi{10.1029/2006JB004586} \end{APACrefDOI}
\PrintBackRefs{\CurrentBib}

\bibitem [\protect \citeauthoryear {%
Ford%
\ \BBA {} Wheeler%
}{%
Ford%
\ \BBA {} Wheeler%
}{%
{\protect \APACyear {2004}}%
}]{%
Ford2004}
\APACinsertmetastar {%
Ford2004}%
\begin{APACrefauthors}%
Ford, J\BPBI M.%
\BCBT {}\ \BBA {} Wheeler, J.%
\end{APACrefauthors}%
\unskip\
\newblock
\APACrefYearMonthDay{2004}{}{}.
\newblock
{\BBOQ}\APACrefatitle {{Modelling interface diffusion creep in two-phase
  materials}} {{Modelling interface diffusion creep in two-phase
  materials}}.{\BBCQ}
\newblock
\APACjournalVolNumPages{Acta Mater.}{52}{8}{2365--2376}.
\newblock
\begin{APACrefDOI} \doi{10.1016/j.actamat.2004.01.045} \end{APACrefDOI}
\PrintBackRefs{\CurrentBib}

\bibitem [\protect \citeauthoryear {%
German%
, Suri%
\BCBL {}\ \BBA {} Park%
}{%
German%
\ \protect \BOthers {.}}{%
{\protect \APACyear {2009}}%
}]{%
German2009}
\APACinsertmetastar {%
German2009}%
\begin{APACrefauthors}%
German, R\BPBI M.%
, Suri, P.%
\BCBL {}\ \BBA {} Park, S\BPBI J.%
\end{APACrefauthors}%
\unskip\
\newblock
\APACrefYearMonthDay{2009}{}{}.
\newblock
{\BBOQ}\APACrefatitle {{Review: Liquid phase sintering}} {{Review: Liquid 
phase
  sintering}}.{\BBCQ}
\newblock
\APACjournalVolNumPages{J. Mater. Sci.}{44}{1}{1--39}.
\newblock
\begin{APACrefDOI} \doi{10.1007/s10853-008-3008-0} \end{APACrefDOI}
\PrintBackRefs{\CurrentBib}

\bibitem [\protect \citeauthoryear {%
Greenwood%
}{%
Greenwood%
}{%
{\protect \APACyear {1985}}%
}]{%
Greenwood1985}
\APACinsertmetastar {%
Greenwood1985}%
\begin{APACrefauthors}%
Greenwood, G\BPBI W.%
\end{APACrefauthors}%
\unskip\
\newblock
\APACrefYearMonthDay{1985}{}{}.
\newblock
{\BBOQ}\APACrefatitle {{An analysis of the effect of multiaxial stresses 
and
  grain shape on Nabarro-Herring creep}} {{An analysis of the effect of
  multiaxial stresses and grain shape on Nabarro-Herring creep}}.{\BBCQ}
\newblock
\APACjournalVolNumPages{Philos. Mag. A}{51}{4}{537--542}.
\newblock
\begin{APACrefDOI} \doi{10.1080/01418618508237575} \end{APACrefDOI}
\PrintBackRefs{\CurrentBib}

\bibitem [\protect \citeauthoryear {%
Greenwood%
}{%
Greenwood%
}{%
{\protect \APACyear {1992}}%
}]{%
Greenwood1992}
\APACinsertmetastar {%
Greenwood1992}%
\begin{APACrefauthors}%
Greenwood, G\BPBI W.%
\end{APACrefauthors}%
\unskip\
\newblock
\APACrefYearMonthDay{1992}{}{}.
\newblock
{\BBOQ}\APACrefatitle {{A Formulation for Anisotropy in Diffusional Creep}} 
{{A
  Formulation for Anisotropy in Diffusional Creep}}.{\BBCQ}
\newblock
\APACjournalVolNumPages{Proc. R. Soc. A Math. Phys. Eng.
  Sci.}{436}{1896}{187--196}.
\newblock
\begin{APACrefDOI} \doi{10.1098/rspa.1992.0013} \end{APACrefDOI}
\PrintBackRefs{\CurrentBib}

\bibitem [\protect \citeauthoryear {%
Herring%
}{%
Herring%
}{%
{\protect \APACyear {1950}}%
}]{%
Herring1950}
\APACinsertmetastar {%
Herring1950}%
\begin{APACrefauthors}%
Herring, C.%
\end{APACrefauthors}%
\unskip\
\newblock
\APACrefYearMonthDay{1950}{}{}.
\newblock
{\BBOQ}\APACrefatitle {{Diffusional Viscosity of a Polycrystalline Solid}}
  {{Diffusional Viscosity of a Polycrystalline Solid}}.{\BBCQ}
\newblock
\APACjournalVolNumPages{J. Appl. Phys.}{21}{5}{437--445}.
\newblock
\begin{APACrefDOI} \doi{10.1063/1.1699681} \end{APACrefDOI}
\PrintBackRefs{\CurrentBib}

\bibitem [\protect \citeauthoryear {%
Hewitt%
\ \BBA {} Fowler%
}{%
Hewitt%
\ \BBA {} Fowler%
}{%
{\protect \APACyear {2008}}%
}]{%
Hewitt2008}
\APACinsertmetastar {%
Hewitt2008}%
\begin{APACrefauthors}%
Hewitt, I\BPBI J.%
\BCBT {}\ \BBA {} Fowler, A\BPBI C.%
\end{APACrefauthors}%
\unskip\
\newblock
\APACrefYearMonthDay{2008}{}{}.
\newblock
{\BBOQ}\APACrefatitle {{Partial melting in an upwelling mantle column}}
  {{Partial melting in an upwelling mantle column}}.{\BBCQ}
\newblock
\APACjournalVolNumPages{Proc. Roy. Soc. A}{464}{2097}{2467--2491}.
\newblock
\begin{APACrefDOI} \doi{10.1098/rspa.2008.0045} \end{APACrefDOI}
\PrintBackRefs{\CurrentBib}

\bibitem [\protect \citeauthoryear {%
Holtzman%
}{%
Holtzman%
}{%
{\protect \APACyear {2016}}%
}]{%
Holtzman2016}
\APACinsertmetastar {%
Holtzman2016}%
\begin{APACrefauthors}%
Holtzman, B\BPBI K.%
\end{APACrefauthors}%
\unskip\
\newblock
\APACrefYearMonthDay{2016}{}{}.
\newblock
{\BBOQ}\APACrefatitle {{Questions on the existence, persistence, and 
mechanical
  effects of a very small melt fraction in the asthenosphere}} {{Questions 
on
  the existence, persistence, and mechanical effects of a very small melt
  fraction in the asthenosphere}}.{\BBCQ}
\newblock
\APACjournalVolNumPages{Geochemistry, Geophys. 
Geosystems}{17}{2}{470--484}.
\newblock
\begin{APACrefDOI} \doi{10.1002/2015GC006102} \end{APACrefDOI}
\PrintBackRefs{\CurrentBib}

\bibitem [\protect \citeauthoryear {%
Holtzman%
, Groebner%
, Zimmerman%
, Ginsberg%
\BCBL {}\ \BBA {} Kohlstedt%
}{%
Holtzman%
\ \protect \BOthers {.}}{%
{\protect \APACyear {2003}}%
}]{%
Holtzman2003}
\APACinsertmetastar {%
Holtzman2003}%
\begin{APACrefauthors}%
Holtzman, B\BPBI K.%
, Groebner, N\BPBI J.%
, Zimmerman, M\BPBI E.%
, Ginsberg, S\BPBI B.%
\BCBL {}\ \BBA {} Kohlstedt, D\BPBI L.%
\end{APACrefauthors}%
\unskip\
\newblock
\APACrefYearMonthDay{2003}{}{}.
\newblock
{\BBOQ}\APACrefatitle {{Stress-driven melt segregation in partially molten
  rocks}} {{Stress-driven melt segregation in partially molten 
rocks}}.{\BBCQ}
\newblock
\APACjournalVolNumPages{Geochemistry, Geophys. Geosystems}{4}{5}{8607}.
\newblock
\begin{APACrefDOI} \doi{10.1029/2001GC000258} \end{APACrefDOI}
\PrintBackRefs{\CurrentBib}

\bibitem [\protect \citeauthoryear {%
Katz%
, Spiegelman%
\BCBL {}\ \BBA {} Holtzman%
}{%
Katz%
\ \protect \BOthers {.}}{%
{\protect \APACyear {2006}}%
}]{%
Katz2006}
\APACinsertmetastar {%
Katz2006}%
\begin{APACrefauthors}%
Katz, R\BPBI F.%
, Spiegelman, M.%
\BCBL {}\ \BBA {} Holtzman, B.%
\end{APACrefauthors}%
\unskip\
\newblock
\APACrefYearMonthDay{2006}{}{}.
\newblock
{\BBOQ}\APACrefatitle {{The dynamics of melt and shear localization in
  partially molten aggregates}} {{The dynamics of melt and shear 
localization
  in partially molten aggregates}}.{\BBCQ}
\newblock
\APACjournalVolNumPages{Nature}{442}{7103}{676--679}.
\newblock
\begin{APACrefDOI} \doi{10.1038/nature05039} \end{APACrefDOI}
\PrintBackRefs{\CurrentBib}

\bibitem [\protect \citeauthoryear {%
Lifshitz%
}{%
Lifshitz%
}{%
{\protect \APACyear {1963}}%
}]{%
Lifshitz1963}
\APACinsertmetastar {%
Lifshitz1963}%
\begin{APACrefauthors}%
Lifshitz, I\BPBI M.%
\end{APACrefauthors}%
\unskip\
\newblock
\APACrefYearMonthDay{1963}{}{}.
\newblock
{\BBOQ}\APACrefatitle {{On the theory of diffusion-viscous flow of
  polycrystalline bodies}} {{On the theory of diffusion-viscous flow of
  polycrystalline bodies}}.{\BBCQ}
\newblock
\APACjournalVolNumPages{Sov. Phys. JETP}{17}{44}{1349--1367}.
\PrintBackRefs{\CurrentBib}

\bibitem [\protect \citeauthoryear {%
Logg%
, Mardal%
\BCBL {}\ \BBA {} Wells%
}{%
Logg%
\ \protect \BOthers {.}}{%
{\protect \APACyear {2012}}%
}]{%
Logg2012}
\APACinsertmetastar {%
Logg2012}%
\begin{APACrefauthors}%
Logg, A.%
, Mardal, K\BHBI A.%
\BCBL {}\ \BBA {} Wells, G.%
\end{APACrefauthors}%
\ (\BEDS).
\unskip\
\newblock
\APACrefYear{2012}.
\newblock
\APACrefbtitle {{Automated Solution of Differential Equations by the Finite
  Element Method, Lecture Notes in Computational Science and Engineering}}
  {{Automated Solution of Differential Equations by the Finite Element 
Method,
  Lecture Notes in Computational Science and Engineering}}.
\newblock
\APACaddressPublisher{}{vol 84, Springer, Berlin Heidelberg}.
\PrintBackRefs{\CurrentBib}

\bibitem [\protect \citeauthoryear {%
Logg%
\ \BBA {} Wells%
}{%
Logg%
\ \BBA {} Wells%
}{%
{\protect \APACyear {2010}}%
}]{%
Logg2010}
\APACinsertmetastar {%
Logg2010}%
\begin{APACrefauthors}%
Logg, A.%
\BCBT {}\ \BBA {} Wells, G\BPBI N.%
\end{APACrefauthors}%
\unskip\
\newblock
\APACrefYearMonthDay{2010}{}{}.
\newblock
{\BBOQ}\APACrefatitle {{DOLFIN}} {{DOLFIN}}.{\BBCQ}
\newblock
\APACjournalVolNumPages{ACM Trans. Math. Softw.}{37}{2}{1--28}.
\newblock
\begin{APACrefDOI} \doi{10.1145/1731022.1731030} \end{APACrefDOI}
\PrintBackRefs{\CurrentBib}

\bibitem [\protect \citeauthoryear {%
McCarthy%
\ \BBA {} Takei%
}{%
McCarthy%
\ \BBA {} Takei%
}{%
{\protect \APACyear {2011}}%
}]{%
McCarthy2011a}
\APACinsertmetastar {%
McCarthy2011a}%
\begin{APACrefauthors}%
McCarthy, C.%
\BCBT {}\ \BBA {} Takei, Y.%
\end{APACrefauthors}%
\unskip\
\newblock
\APACrefYearMonthDay{2011}{}{}.
\newblock
{\BBOQ}\APACrefatitle {{Anelasticity and viscosity of partially molten rock
  analogue: Toward seismic detection of small quantities of melt}}
  {{Anelasticity and viscosity of partially molten rock analogue: Toward
  seismic detection of small quantities of melt}}.{\BBCQ}
\newblock
\APACjournalVolNumPages{Geophys. Res. Lett.}{38}{18}{L18306}.
\newblock
\begin{APACrefDOI} \doi{10.1029/2011GL048776} \end{APACrefDOI}
\PrintBackRefs{\CurrentBib}

\bibitem [\protect \citeauthoryear {%
McKenzie%
}{%
McKenzie%
}{%
{\protect \APACyear {1984}}%
}]{%
McKenzie1984}
\APACinsertmetastar {%
McKenzie1984}%
\begin{APACrefauthors}%
McKenzie, D.%
\end{APACrefauthors}%
\unskip\
\newblock
\APACrefYearMonthDay{1984}{}{}.
\newblock
{\BBOQ}\APACrefatitle {{The Generation and Compaction of Partially Molten
  Rock}} {{The Generation and Compaction of Partially Molten Rock}}.{\BBCQ}
\newblock
\APACjournalVolNumPages{J. Pet.}{25}{}{713--765}.
\newblock
\begin{APACrefDOI} \doi{10.1093/petrology/25.3.713} \end{APACrefDOI}
\PrintBackRefs{\CurrentBib}

\bibitem [\protect \citeauthoryear {%
Mei%
, Bai%
, Hiraga%
\BCBL {}\ \BBA {} Kohlstedt%
}{%
Mei%
\ \protect \BOthers {.}}{%
{\protect \APACyear {2002}}%
}]{%
Mei2002}
\APACinsertmetastar {%
Mei2002}%
\begin{APACrefauthors}%
Mei, S.%
, Bai, W.%
, Hiraga, T.%
\BCBL {}\ \BBA {} Kohlstedt, D\BPBI L.%
\end{APACrefauthors}%
\unskip\
\newblock
\APACrefYearMonthDay{2002}{}{}.
\newblock
{\BBOQ}\APACrefatitle {{Influence of melt on the creep behavior of
  olivine-basalt aggregates under hydrous conditions}} {{Influence of melt 
on
  the creep behavior of olivine-basalt aggregates under hydrous
  conditions}}.{\BBCQ}
\newblock
\APACjournalVolNumPages{Earth Planet. Sci. Lett.}{201}{3-4}{491--507}.
\newblock
\begin{APACrefDOI} \doi{10.1016/S0012-821X(02)00745-8} \end{APACrefDOI}
\PrintBackRefs{\CurrentBib}

\bibitem [\protect \citeauthoryear {%
Mortensen%
, Okkels%
\BCBL {}\ \BBA {} Bruus%
}{%
Mortensen%
\ \protect \BOthers {.}}{%
{\protect \APACyear {2005}}%
}]{%
Mortensen2005}
\APACinsertmetastar {%
Mortensen2005}%
\begin{APACrefauthors}%
Mortensen, N\BPBI A.%
, Okkels, F.%
\BCBL {}\ \BBA {} Bruus, H.%
\end{APACrefauthors}%
\unskip\
\newblock
\APACrefYearMonthDay{2005}{}{}.
\newblock
{\BBOQ}\APACrefatitle {{Reexamination of Hagen-Poiseuille flow: Shape
  dependence of the hydraulic resistance in microchannels}} {{Reexamination 
of
  Hagen-Poiseuille flow: Shape dependence of the hydraulic resistance in
  microchannels}}.{\BBCQ}
\newblock
\APACjournalVolNumPages{Phys. Rev. E}{71}{5}{057301}.
\newblock
\begin{APACrefDOI} \doi{10.1103/PhysRevE.71.057301} \end{APACrefDOI}
\PrintBackRefs{\CurrentBib}

\bibitem [\protect \citeauthoryear {%
Nabarro%
}{%
Nabarro%
}{%
{\protect \APACyear {1948}}%
}]{%
Nabarro1948}
\APACinsertmetastar {%
Nabarro1948}%
\begin{APACrefauthors}%
Nabarro, F\BPBI R\BPBI N.%
\end{APACrefauthors}%
\unskip\
\newblock
\APACrefYearMonthDay{1948}{}{}.
\newblock
{\BBOQ}\APACrefatitle {{Deformation of crystals by the motion of single 
ions}}
  {{Deformation of crystals by the motion of single ions}}.{\BBCQ}
\newblock
\BIn{} \APACrefbtitle {Report on Conf. on the Strength of Solids} {Report 
on
  conf. on the strength of solids}\ (\BPGS\ 75--90).
\newblock
\APACaddressPublisher{}{Physical Society, London}.
\PrintBackRefs{\CurrentBib}

\bibitem [\protect \citeauthoryear {%
Pan%
\ \BBA {} Cocks%
}{%
Pan%
\ \BBA {} Cocks%
}{%
{\protect \APACyear {1994}}%
}]{%
Pan1994}
\APACinsertmetastar {%
Pan1994}%
\begin{APACrefauthors}%
Pan, J.%
\BCBT {}\ \BBA {} Cocks, A.%
\end{APACrefauthors}%
\unskip\
\newblock
\APACrefYearMonthDay{1994}{}{}.
\newblock
{\BBOQ}\APACrefatitle {{A constitutive model for stage 2 sintering of fine
  grained materials—I. Grain-boundaries act as perfect sources and sinks 
for
  vacancies}} {{A constitutive model for stage 2 sintering of fine grained
  materials—I. Grain-boundaries act as perfect sources and sinks for
  vacancies}}.{\BBCQ}
\newblock
\APACjournalVolNumPages{Acta Metall. Mater.}{42}{4}{1215--1222}.
\newblock
\begin{APACrefDOI} \doi{10.1016/0956-7151(94)90138-4} \end{APACrefDOI}
\PrintBackRefs{\CurrentBib}

\bibitem [\protect \citeauthoryear {%
Qi%
, Zhao%
\BCBL {}\ \BBA {} Kohlstedt%
}{%
Qi%
\ \protect \BOthers {.}}{%
{\protect \APACyear {2013}}%
}]{%
Qi2013}
\APACinsertmetastar {%
Qi2013}%
\begin{APACrefauthors}%
Qi, C.%
, Zhao, Y\BHBI H.%
\BCBL {}\ \BBA {} Kohlstedt, D\BPBI L.%
\end{APACrefauthors}%
\unskip\
\newblock
\APACrefYearMonthDay{2013}{}{}.
\newblock
{\BBOQ}\APACrefatitle {{An experimental study of pressure shadows in 
partially
  molten rocks}} {{An experimental study of pressure shadows in partially
  molten rocks}}.{\BBCQ}
\newblock
\APACjournalVolNumPages{Earth Planet. Sci. Lett.}{382}{}{77--84}.
\newblock
\begin{APACrefDOI} \doi{10.1016/J.EPSL.2013.09.004} \end{APACrefDOI}
\PrintBackRefs{\CurrentBib}

\bibitem [\protect \citeauthoryear {%
Raj%
\ \BBA {} Ashby%
}{%
Raj%
\ \BBA {} Ashby%
}{%
{\protect \APACyear {1971}}%
}]{%
Raj1971}
\APACinsertmetastar {%
Raj1971}%
\begin{APACrefauthors}%
Raj, R.%
\BCBT {}\ \BBA {} Ashby, M\BPBI F.%
\end{APACrefauthors}%
\unskip\
\newblock
\APACrefYearMonthDay{1971}{}{}.
\newblock
{\BBOQ}\APACrefatitle {{On grain boundary sliding and diffusional creep}} 
{{On
  grain boundary sliding and diffusional creep}}.{\BBCQ}
\newblock
\APACjournalVolNumPages{Metall. Trans.}{2}{4}{1113--1127}.
\newblock
\begin{APACrefDOI} \doi{10.1007/BF02664244} \end{APACrefDOI}
\PrintBackRefs{\CurrentBib}

\bibitem [\protect \citeauthoryear {%
{Rees Jones}%
\ \BBA {} Katz%
}{%
{Rees Jones}%
\ \BBA {} Katz%
}{%
{\protect \APACyear {2018}}%
}]{%
ReesJones2018}
\APACinsertmetastar {%
ReesJones2018}%
\begin{APACrefauthors}%
{Rees Jones}, D\BPBI W.%
\BCBT {}\ \BBA {} Katz, R\BPBI F.%
\end{APACrefauthors}%
\unskip\
\newblock
\APACrefYearMonthDay{2018}{}{}.
\newblock
{\BBOQ}\APACrefatitle {{Reaction-infiltration instability in a compacting
  porous medium}} {{Reaction-infiltration instability in a compacting 
porous
  medium}}.{\BBCQ}
\newblock
\APACjournalVolNumPages{J. Fluid Mech.}{852}{}{5--36}.
\newblock
\begin{APACrefDOI} \doi{10.1017/jfm.2018.524} \end{APACrefDOI}
\PrintBackRefs{\CurrentBib}

\bibitem [\protect \citeauthoryear {%
Rudge%
}{%
Rudge%
}{%
{\protect \APACyear {2018}}%
}]{%
Rudge2018}
\APACinsertmetastar {%
Rudge2018}%
\begin{APACrefauthors}%
Rudge, J\BPBI F.%
\end{APACrefauthors}%
\unskip\
\newblock
\APACrefYearMonthDay{2018}{}{}.
\newblock
{\BBOQ}\APACrefatitle {{Textural equilibrium melt geometries around
  tetrakaidecahedral grains}} {{Textural equilibrium melt geometries around
  tetrakaidecahedral grains}}.{\BBCQ}
\newblock
\APACjournalVolNumPages{Proc. R. Soc. A Math. Phys. Eng.
  Sci.}{474}{2212}{20170639}.
\newblock
\begin{APACrefDOI} \doi{10.1098/rspa.2017.0639} \end{APACrefDOI}
\PrintBackRefs{\CurrentBib}

\bibitem [\protect \citeauthoryear {%
Rudge%
\ \BBA {} Bercovici%
}{%
Rudge%
\ \BBA {} Bercovici%
}{%
{\protect \APACyear {2015}}%
}]{%
Rudge2015}
\APACinsertmetastar {%
Rudge2015}%
\begin{APACrefauthors}%
Rudge, J\BPBI F.%
\BCBT {}\ \BBA {} Bercovici, D.%
\end{APACrefauthors}%
\unskip\
\newblock
\APACrefYearMonthDay{2015}{}{}.
\newblock
{\BBOQ}\APACrefatitle {{Melt-band instabilities with two-phase damage}}
  {{Melt-band instabilities with two-phase damage}}.{\BBCQ}
\newblock
\APACjournalVolNumPages{Geophys. J. Int.}{201}{2}{640--651}.
\newblock
\begin{APACrefDOI} \doi{10.1093/gji/ggv040} \end{APACrefDOI}
\PrintBackRefs{\CurrentBib}

\bibitem [\protect \citeauthoryear {%
Schmeling%
, Kruse%
\BCBL {}\ \BBA {} Richard%
}{%
Schmeling%
\ \protect \BOthers {.}}{%
{\protect \APACyear {2012}}%
}]{%
Schmeling2012}
\APACinsertmetastar {%
Schmeling2012}%
\begin{APACrefauthors}%
Schmeling, H.%
, Kruse, J\BPBI P.%
\BCBL {}\ \BBA {} Richard, G.%
\end{APACrefauthors}%
\unskip\
\newblock
\APACrefYearMonthDay{2012}{}{}.
\newblock
{\BBOQ}\APACrefatitle {{Effective shear and bulk viscosity of partially 
molten
  rock based on elastic moduli theory of a fluid filled poroelastic 
medium}}
  {{Effective shear and bulk viscosity of partially molten rock based on
  elastic moduli theory of a fluid filled poroelastic medium}}.{\BBCQ}
\newblock
\APACjournalVolNumPages{Geophys. J. Int.}{190}{3}{1571--1578}.
\newblock
\begin{APACrefDOI} \doi{10.1111/j.1365-246X.2012.05596.x} \end{APACrefDOI}
\PrintBackRefs{\CurrentBib}

\bibitem [\protect \citeauthoryear {%
Scott%
\ \BBA {} Stevenson%
}{%
Scott%
\ \BBA {} Stevenson%
}{%
{\protect \APACyear {1986}}%
}]{%
Scott1986}
\APACinsertmetastar {%
Scott1986}%
\begin{APACrefauthors}%
Scott, D\BPBI R.%
\BCBT {}\ \BBA {} Stevenson, D\BPBI J.%
\end{APACrefauthors}%
\unskip\
\newblock
\APACrefYearMonthDay{1986}{}{}.
\newblock
{\BBOQ}\APACrefatitle {{Magma ascent by porous flow}} {{Magma ascent by 
porous
  flow}}.{\BBCQ}
\newblock
\APACjournalVolNumPages{J. Geophys. Res.}{91}{B9}{9283--9296}.
\newblock
\begin{APACrefDOI} \doi{10.1029/JB091iB09p09283} \end{APACrefDOI}
\PrintBackRefs{\CurrentBib}

\bibitem [\protect \citeauthoryear {%
Shah%
\ \BBA {} Chokshi%
}{%
Shah%
\ \BBA {} Chokshi%
}{%
{\protect \APACyear {1998}}%
}]{%
Shah1998}
\APACinsertmetastar {%
Shah1998}%
\begin{APACrefauthors}%
Shah, S.%
\BCBT {}\ \BBA {} Chokshi, A\BPBI H.%
\end{APACrefauthors}%
\unskip\
\newblock
\APACrefYearMonthDay{1998}{}{}.
\newblock
{\BBOQ}\APACrefatitle {{The significance of diffusional flow in
  ultrafine-grained materials}} {{The significance of diffusional flow in
  ultrafine-grained materials}}.{\BBCQ}
\newblock
\APACjournalVolNumPages{Colloids Surfaces A Physicochem. Eng.
  Asp.}{133}{1-2}{57--61}.
\newblock
\begin{APACrefDOI} \doi{10.1016/S0927-7757(97)00124-6} \end{APACrefDOI}
\PrintBackRefs{\CurrentBib}

\bibitem [\protect \citeauthoryear {%
Simpson%
, Spiegelman%
\BCBL {}\ \BBA {} Weinstein%
}{%
Simpson%
\ \protect \BOthers {.}}{%
{\protect \APACyear {2010}}%
}]{%
Simpson2010a}
\APACinsertmetastar {%
Simpson2010a}%
\begin{APACrefauthors}%
Simpson, G.%
, Spiegelman, M.%
\BCBL {}\ \BBA {} Weinstein, M\BPBI I.%
\end{APACrefauthors}%
\unskip\
\newblock
\APACrefYearMonthDay{2010}{}{}.
\newblock
{\BBOQ}\APACrefatitle {{A multiscale model of partial melts: 2. Numerical
  results}} {{A multiscale model of partial melts: 2. Numerical
  results}}.{\BBCQ}
\newblock
\APACjournalVolNumPages{J. Geophys. Res.}{115}{B4}{B04411}.
\newblock
\begin{APACrefDOI} \doi{10.1029/2009JB006376} \end{APACrefDOI}
\PrintBackRefs{\CurrentBib}

\bibitem [\protect \citeauthoryear {%
Sleep%
}{%
Sleep%
}{%
{\protect \APACyear {1988}}%
}]{%
Sleep1988}
\APACinsertmetastar {%
Sleep1988}%
\begin{APACrefauthors}%
Sleep, N\BPBI H.%
\end{APACrefauthors}%
\unskip\
\newblock
\APACrefYearMonthDay{1988}{}{}.
\newblock
{\BBOQ}\APACrefatitle {{Tapping of melt by veins and dikes}} {{Tapping of 
melt
  by veins and dikes}}.{\BBCQ}
\newblock
\APACjournalVolNumPages{J. Geophys. Res. Solid 
Earth}{93}{B9}{10255--10272}.
\newblock
\begin{APACrefDOI} \doi{10.1029/JB093iB09p10255} \end{APACrefDOI}
\PrintBackRefs{\CurrentBib}

\bibitem [\protect \citeauthoryear {%
Spiegelman%
}{%
Spiegelman%
}{%
{\protect \APACyear {2003}}%
}]{%
Spiegelman2003}
\APACinsertmetastar {%
Spiegelman2003}%
\begin{APACrefauthors}%
Spiegelman, M.%
\end{APACrefauthors}%
\unskip\
\newblock
\APACrefYearMonthDay{2003}{}{}.
\newblock
{\BBOQ}\APACrefatitle {{Linear analysis of melt band formation by simple
  shear}} {{Linear analysis of melt band formation by simple 
shear}}.{\BBCQ}
\newblock
\APACjournalVolNumPages{Geochemistry, Geophys. Geosystems}{4}{9}{8615}.
\newblock
\begin{APACrefDOI} \doi{10.1029/2002GC000499} \end{APACrefDOI}
\PrintBackRefs{\CurrentBib}

\bibitem [\protect \citeauthoryear {%
Spingarn%
\ \BBA {} Nix%
}{%
Spingarn%
\ \BBA {} Nix%
}{%
{\protect \APACyear {1978}}%
}]{%
Spingarn1978}
\APACinsertmetastar {%
Spingarn1978}%
\begin{APACrefauthors}%
Spingarn, J\BPBI R.%
\BCBT {}\ \BBA {} Nix, W\BPBI D.%
\end{APACrefauthors}%
\unskip\
\newblock
\APACrefYearMonthDay{1978}{}{}.
\newblock
{\BBOQ}\APACrefatitle {{Diffusional creep and diffusionally accommodated 
grain
  rearrangement}} {{Diffusional creep and diffusionally accommodated grain
  rearrangement}}.{\BBCQ}
\newblock
\APACjournalVolNumPages{Acta Metall.}{26}{9}{1389--1398}.
\newblock
\begin{APACrefDOI} \doi{10.1016/0001-6160(78)90154-2} \end{APACrefDOI}
\PrintBackRefs{\CurrentBib}

\bibitem [\protect \citeauthoryear {%
Stevenson%
}{%
Stevenson%
}{%
{\protect \APACyear {1989}}%
}]{%
Stevenson1989}
\APACinsertmetastar {%
Stevenson1989}%
\begin{APACrefauthors}%
Stevenson, D\BPBI J.%
\end{APACrefauthors}%
\unskip\
\newblock
\APACrefYearMonthDay{1989}{}{}.
\newblock
{\BBOQ}\APACrefatitle {{Spontaneous small-scale melt segregation in partial
  melts undergoing deformation}} {{Spontaneous small-scale melt segregation 
in
  partial melts undergoing deformation}}.{\BBCQ}
\newblock
\APACjournalVolNumPages{Geophys. Res. Lett.}{16}{9}{1067--1070}.
\newblock
\begin{APACrefDOI} \doi{10.1029/GL016i009p01067} \end{APACrefDOI}
\PrintBackRefs{\CurrentBib}

\bibitem [\protect \citeauthoryear {%
Takei%
\ \BBA {} Holtzman%
}{%
Takei%
\ \BBA {} Holtzman%
}{%
{\protect \APACyear {2009}}%
{\protect \APACexlab {{\protect \BCnt {1}}}}}]{%
Takei2009}
\APACinsertmetastar {%
Takei2009}%
\begin{APACrefauthors}%
Takei, Y.%
\BCBT {}\ \BBA {} Holtzman, B\BPBI K.%
\end{APACrefauthors}%
\unskip\
\newblock
\APACrefYearMonthDay{2009{\protect \BCnt {1}}}{}{}.
\newblock
{\BBOQ}\APACrefatitle {{Viscous constitutive relations of solid-liquid
  composites in terms of grain boundary contiguity: 1. Grain boundary 
diffusion
  control model}} {{Viscous constitutive relations of solid-liquid 
composites
  in terms of grain boundary contiguity: 1. Grain boundary diffusion 
control
  model}}.{\BBCQ}
\newblock
\APACjournalVolNumPages{J. Geophys. Res. Solid Earth}{114}{6}{1--19}.
\newblock
\begin{APACrefDOI} \doi{10.1029/2008JB005850} \end{APACrefDOI}
\PrintBackRefs{\CurrentBib}

\bibitem [\protect \citeauthoryear {%
Takei%
\ \BBA {} Holtzman%
}{%
Takei%
\ \BBA {} Holtzman%
}{%
{\protect \APACyear {2009}}%
{\protect \APACexlab {{\protect \BCnt {2}}}}}]{%
Takei2009a}
\APACinsertmetastar {%
Takei2009a}%
\begin{APACrefauthors}%
Takei, Y.%
\BCBT {}\ \BBA {} Holtzman, B\BPBI K.%
\end{APACrefauthors}%
\unskip\
\newblock
\APACrefYearMonthDay{2009{\protect \BCnt {2}}}{}{}.
\newblock
{\BBOQ}\APACrefatitle {{Viscous constitutive relations of solid-liquid
  composites in terms of grain boundary contiguity: 2. Compositional model 
for
  small melt fractions}} {{Viscous constitutive relations of solid-liquid
  composites in terms of grain boundary contiguity: 2. Compositional model 
for
  small melt fractions}}.{\BBCQ}
\newblock
\APACjournalVolNumPages{J. Geophys. Res.}{114}{B6}{B06206}.
\newblock
\begin{APACrefDOI} \doi{10.1029/2008JB005851} \end{APACrefDOI}
\PrintBackRefs{\CurrentBib}

\bibitem [\protect \citeauthoryear {%
Takei%
\ \BBA {} Katz%
}{%
Takei%
\ \BBA {} Katz%
}{%
{\protect \APACyear {2015}}%
}]{%
Takei2015}
\APACinsertmetastar {%
Takei2015}%
\begin{APACrefauthors}%
Takei, Y.%
\BCBT {}\ \BBA {} Katz, R\BPBI F.%
\end{APACrefauthors}%
\unskip\
\newblock
\APACrefYearMonthDay{2015}{}{}.
\newblock
{\BBOQ}\APACrefatitle {{Consequences of viscous anisotropy in a deforming,
  two-phase aggregate. Why is porosity-band angle lowered by viscous
  anisotropy?}} {{Consequences of viscous anisotropy in a deforming, 
two-phase
  aggregate. Why is porosity-band angle lowered by viscous 
anisotropy?}}{\BBCQ}
\newblock
\APACjournalVolNumPages{J. Fluid Mech.}{784}{}{199--224}.
\newblock
\begin{APACrefDOI} \doi{10.1017/jfm.2015.592} \end{APACrefDOI}
\PrintBackRefs{\CurrentBib}

\bibitem [\protect \citeauthoryear {%
von Bargen%
\ \BBA {} Waff%
}{%
von Bargen%
\ \BBA {} Waff%
}{%
{\protect \APACyear {1986}}%
}]{%
vonBargen1986}
\APACinsertmetastar {%
vonBargen1986}%
\begin{APACrefauthors}%
von Bargen, N.%
\BCBT {}\ \BBA {} Waff, H\BPBI S.%
\end{APACrefauthors}%
\unskip\
\newblock
\APACrefYearMonthDay{1986}{}{}.
\newblock
{\BBOQ}\APACrefatitle {{Permeabilities, interfacial-areas and curvatures of
  partially molten systems - results of numerical computation of 
equilibrium
  microstructures}} {{Permeabilities, interfacial-areas and curvatures of
  partially molten systems - results of numerical computation of 
equilibrium
  microstructures}}.{\BBCQ}
\newblock
\APACjournalVolNumPages{J. Geophys. Res.}{91}{}{9261--9276}.
\PrintBackRefs{\CurrentBib}

\bibitem [\protect \citeauthoryear {%
Ward%
\ \BBA {} Kropinski%
}{%
Ward%
\ \BBA {} Kropinski%
}{%
{\protect \APACyear {2010}}%
}]{%
Ward2010}
\APACinsertmetastar {%
Ward2010}%
\begin{APACrefauthors}%
Ward, M\BPBI J\BPBI M.%
\BCBT {}\ \BBA {} Kropinski, M\BHBI C\BPBI M.%
\end{APACrefauthors}%
\unskip\
\newblock
\APACrefYearMonthDay{2010}{}{}.
\newblock
{\BBOQ}\APACrefatitle {{Asymptotic Methods for PDE Problems in Fluid 
Mechanics
  and Related Systems with Strong Localized Perturbations in 
Two-Dimensional
  Domains}} {{Asymptotic Methods for PDE Problems in Fluid Mechanics and
  Related Systems with Strong Localized Perturbations in Two-Dimensional
  Domains}}.{\BBCQ}
\newblock
\BIn{} \APACrefbtitle {Asymptotic Methods in Fluid Mechanics: Survey and 
Recent
  Advances. CISM Courses and Lectures, vol 523} {Asymptotic methods in 
fluid
  mechanics: Survey and recent advances. cism courses and lectures, vol 
523}\
  (\BPGS\ 1--48).
\newblock
\APACaddressPublisher{Vienna}{Springer, Vienna}.
\newblock
\begin{APACrefDOI} \doi{10.1007/978-3-7091-0408-8_2} \end{APACrefDOI}
\PrintBackRefs{\CurrentBib}

\bibitem [\protect \citeauthoryear {%
Yamauchi%
\ \BBA {} Takei%
}{%
Yamauchi%
\ \BBA {} Takei%
}{%
{\protect \APACyear {2016}}%
}]{%
Yamauchi2016}
\APACinsertmetastar {%
Yamauchi2016}%
\begin{APACrefauthors}%
Yamauchi, H.%
\BCBT {}\ \BBA {} Takei, Y.%
\end{APACrefauthors}%
\unskip\
\newblock
\APACrefYearMonthDay{2016}{}{}.
\newblock
{\BBOQ}\APACrefatitle {{Polycrystal anelasticity at near-solidus 
temperatures}}
  {{Polycrystal anelasticity at near-solidus temperatures}}.{\BBCQ}
\newblock
\APACjournalVolNumPages{J. Geophys. Res. Solid Earth}{121}{11}{7790--7820}.
\newblock
\begin{APACrefDOI} \doi{10.1002/2016JB013316} \end{APACrefDOI}
\PrintBackRefs{\CurrentBib}

\end{thebibliography}
\end{document}